\documentclass[journal]{IEEEtran}

\usepackage{xcolor,soul,framed} 

\colorlet{shadecolor}{yellow}
\usepackage[pdftex]{graphicx}
\graphicspath{{../pdf/}{../jpeg/}}
\DeclareGraphicsExtensions{.pdf,.jpeg,.png}

\usepackage[cmex10]{amsmath}
\usepackage{array}
\usepackage{mdwmath}
\usepackage{mdwtab}
\usepackage{eqparbox}
\usepackage{url}
\usepackage{cite}
\usepackage{xcolor}
\usepackage{breqn}

\usepackage{algorithm}
\usepackage{algorithmic}
\makeatletter
\newcommand\fs@norules{\def\@fs@cfont{\bfseries}\let\@fs@capt\floatc@ruled
  \def\@fs@pre{}%
  \def\@fs@post{}%
  \def\@fs@mid{\kern3pt}%
  \let\@fs@iftopcapt\iftrue}
\makeatother
\restylefloat{algorithm}

\ifCLASSINFOpdf
\else
\fi
\begin{document}
\bstctlcite{IEEEexample:BSTcontrol}
    \title{
    Metaverse Communications, Networking, Security, and Applications: Research Issues, State-of-the-Art, and Future Directions}
 \author{Mansoor Ali,~\IEEEmembership{ Member,~IEEE}, Faisal Naeem,~\IEEEmembership{ Member,~IEEE}, Georges Kaddoum,~\IEEEmembership{ Senior Member,~IEEE}, and Ekram Hossain,~\IEEEmembership{ Fellow,~IEEE}
 
     \thanks
		{Mansoor Ali, Faisal Naeem and Georges Kaddoum are with the Electrical Engineering Department, ETS, University of Quebec, Montreal, Canada. (e-mail: mansoor.ali.1@etsmtl.net, faisal.naeem.1@ens.etsmtl.ca georges.kaddoum@etsmtl.ca). 
		}
  \thanks
  {Ekram Hossain is with Department of Computer and Electrical Engineering, University of Manitoba, Canada. (email: Ekram.Hossain@umanitoba.ca).}
  \thanks
  {Corresponding Authors: Mansoor Ali (email:mansoor.ali.1@etsmtl.net)}
}
%
%


\maketitle

\begin{abstract}
Metaverse is an evolving orchestrator of the next-generation Internet architecture that produces an immersive and self-adapting virtual world in which humans perform activities similar to those in the real world, such as playing sports, doing work, and socializing. It is becoming a reality and is driven by ever-evolving advanced technologies such as extended reality, artificial intelligence, and blockchain. In this context, Metaverse will play an essential role in developing smart cities, which becomes more evident in the post-COVID-19-pandemic metropolitan setting. However, the new paradigm imposes new challenges, such as developing novel privacy and security threats that can emerge in the digital Metaverse ecosystem. Moreover, it requires the convergence of several media types with the capability to quickly process massive amounts of data to keep the residents safe and well-informed, which can raise issues related to scalability and interoperability. In light of this, this research study aims to review the literature on the state of the art of integrating the Metaverse architecture concepts in smart cities. First, this paper presents the theoretical architecture of Metaverse and discusses international companies' interest in this emerging technology. It also examines the notion of Metaverse relevant to virtual reality, identifies the prevalent threats, and determines the importance of communication infrastructure in information gathering for efficient Metaverse operation. Next, the notion of blockchain technologies is discussed regarding privacy preservation and how it can provide tamper-proof content sharing among Metaverse users. Finally, the application of distributed Metaverse for social good is highlighted. Most importantly, the paper explores the reﬂections of this cutting-edge technology on the smart city, talks about the role and impact of the Metaverse in the production of urban policies, and eventually identifies the research gaps and the future research directions in this domain.
\end{abstract}


%
\IEEEpeerreviewmaketitle

\section{Introduction}

\subsection{Background}
METAVERSE, which is a combination of the word ``universe" and the prefix ``meta" (meaning transforming), depicts an imaginary, artificial environment connected with the physical world. Neal Stephenson coined the term ``Metaverse" in 1992 in his exploratory fiction \textit{novel Snow Crash} \cite{joshua2017information}. Stephenson defined Metaverse as a huge virtual world that exists alongside the physical world wherein users interact through digital avatars. Since then, Metaverse has been defined from a variety of very different points of view, such as "an omniverse: a venue of simulation and collaboration \cite{cheng2022will}, a mirror world,collective space in virtuality \cite{burns2018everything}, such as lifelogging \cite{bruun2019lifelogging}, and  embodied Internet/ spatial Internet \cite{chayka2021facebook}". 

Metaverse is an advanced virtual world that merges digital and physical entities through the convergence of extended reality (XR), web technologies, and the Internet. According to Milgram and Kishino's Reality-Virtuality Continuum \cite{milgram1995augmented}, the physical and digital worlds can be integrated by XR to varying degrees, e.g., virtual reality (VR), mixed reality (MR), and augmented reality (AR). In the Metaverse, users virtually experience an alternate life through their avatars (analogous to their respective user's personality in the physical world) \cite{lee2021all}. As mentioned above, Metaverse integrates a variety of technologies. For instance, VR and AR make it possible to develop high-quality 3D models, 6G, beyond-5G and other advanced communication technologies to provide reliable low-latency connections among different Metaverse devices ranging from wearable health devices to personally used equipment. In addition, artificial intelligence (AI) helps to create intelligent Metaverse environments, while for asset management, blockchain and non-fungible tokens are considered essential tools \cite{9}. In recent years, technological advancements and our dependence on smart devices in day-to-day life have matured the Metaverse terminology for implementation in the near future \cite{8}. Metaverse has attracted the attention of many high-tech companies, including NVIDIA, Facebook, and Tencent, its realistic and feasible applications drive that. Furthermore, to show its dedication to the Metaverse, Facebook has gone one step further and re-branded itself as "Meta".     

\begin{figure}
  \begin{center}
  \includegraphics[width=3.5in]{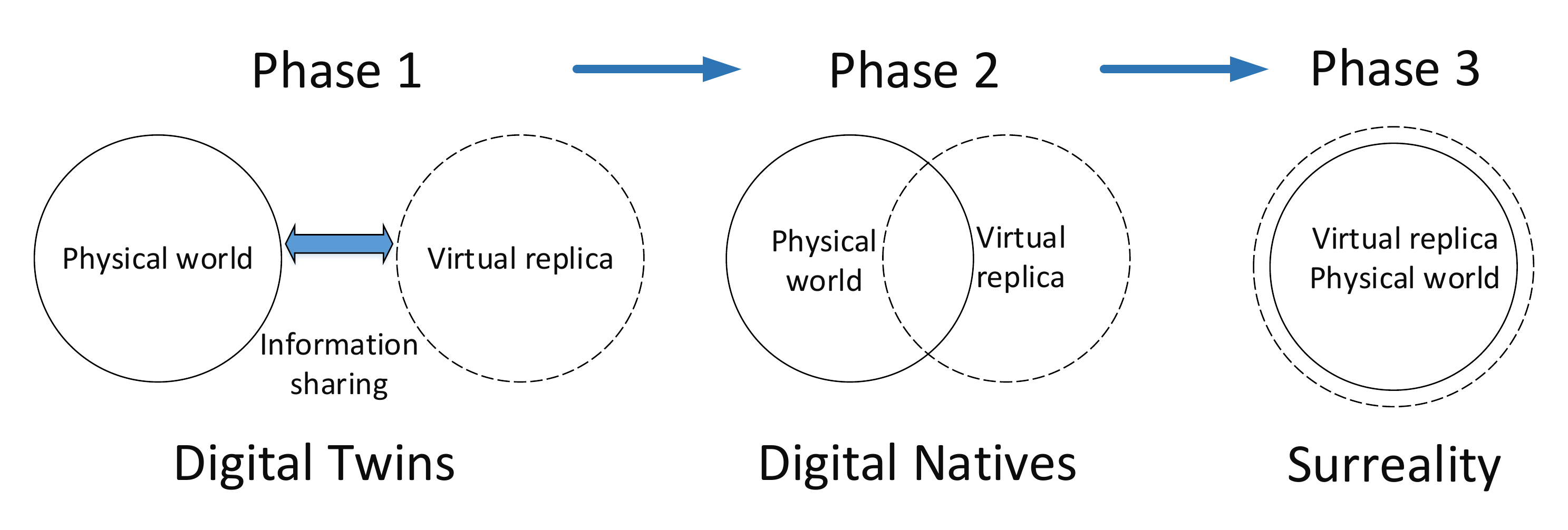}\\
  \caption{Developmental phases of the metaverse 
  }\label{phases}
  \end{center}
\end{figure}

From a macro perspective, Metaverse development generally includes three consecutive phases: Digital Twins (DTs), digital natives, and surreality, as shown in Fig.~\ref{phases}. In the first phase, a mirror world is produced that involves large-scale high-fidelity DTs of things and humans in virtual environments to produce a meaningful depiction of the physical reality. The next phase is focused on the creation of native content, with innovations being produced by digital natives in virtual realms. The Metaverse is considered mature in the final phase and develops into a self-sustainable and persistent surreality world that absorbs reality in itself. The smooth incorporation and mutual symbiosis of the virtual and physical worlds are achieved at this step. The virtual world will have a broader scope compared to the real world. Moreover, lives and scenes that are not present in the real world might exist in the virtual worlds \cite{wang2022survey}.

The concept of Metaverse has applications in different aspects of city living (daily life, the economy, art, culture, health and science) of the city. According to Thomas Alsop's report \cite{statista_2020},  improvements in urban planning and navigation solutions are the two well-known applications of MR/VR/AR/XR in a smart city. These applications include smart systems for building management, parking, and traffic signalling.

Realtors use XR to enable potential buyers to view and experience properties without needing to visit physically, which is believed to be a huge breakthrough for the real estate sector \cite{kemecreality}. It also makes it easier for interior designers and architects to design house interiors efficiently without mistakes. One practical example of XR can be found in IKEA's mobile shopping app. App users can choose any products from the catalogue and virtually place them in their homes to decide which product best suits the specific area \cite{11}. In addition, the Metaverse offers users/customers an entire change experience to co-create experiences in the retail industry, strengthen communication, increase customer engagement, and enhance design services \cite{gadalla2013metaverse}. Moreover, complicated interactions between different entities (such as digital content creators and users) involved in the Metaverse can be managed and automated by the Metaverse service providers through blockchain-based smart contract mechanisms \cite{nguyen2021metachain}. 

The digital economy environment is another important dimension the  Metaverse offers that allows users to create, sell and buy various items. Cryptocurrencies are usable on the Metaverse, and several coins are listed by the cryptocurrency exchanges specifically for doing trades in Metaverse society. Metaverse coins include Starlink (STARL), WEMIX (WEMIX), Enjin Coin (ENJ), UFO Gaming (UFO), THETA (THETA), The Sandbox (SAND), Axie Infinity (AXS), Decentraland (MANA) \cite{coinmarket_2022}. These Ethereum blockchain-enabled platforms serve as VR platforms to enable users to create applications and content; create, experience, and monetize digital assets as a game; and interact without disclosing personal information \cite{kemecreality}.

\subsection {Challenges For Metaverse}
Research work is being conducted into developing and implementing 'Metaverse' in smart urban planning, from constructing DTs to designing virtually habitable cities. However, this fundamental idea is not new and was previously envisioned by researchers at least three decades ago. As a combination of a control‐freak, DTs visions of 'Mirror Worlds' \cite{gelernter1993mirror}, constraints and pollution-free living \cite{benedikt1991cyberspace}, interactive abilities and free-form "liquid architecture" designs, most of the Metaverse rhetoric echoes the primarily cyberspace-hailing, anti‐urban hype of the 1990s. These ideas point towards forming novel digital spaces, an innovative futuristic frontier, to overcome shortcomings in today's smart city design architecture. Still, several risks and challenges are associated with jettisoning places, some of which are discussed below. 

\textbf{Relevance :} Modelling may help to eliminate the impractical factors of the hypothetical "model" of reality or simplify the designing process. The formation of the contemporary, digitally-enabled realms could easily result in shortcomings related to framing the social environments within them and rendering or excluding unnoticeable places, living practices, cultures, and social groups. Various concerns expressed by the empowered middle class and elite class populations regarding the smart city vision, such as residential security, car‐based mobility and remote working, have been addressed in \cite{cardullo2019right}. However, some basic requirements of low‐income local communities, such as lack of access to education, poor living conditions, and water poverty, were ignored. These issues can be addressed through information and communications technology (ICT) provided that the modelling or virtualization of functions is done inclusively using a well-grounded, multi‐dimensional, strategic, and complex strategy rather than a more generic one.

\textbf{Communication and networking:}
The deployment of metaverse will be based on novel applications such as VR, XR, MR, and AR that have stringent QoS performance metrics. For example, the requirement of real-time views of video streaming in VR systems demands a bitrate of up to 1 Gbps. Moreover, the motion-to-photon (MPT) latency delay requirement is less than 20 ms to provide smooth movements in the virtual space. Thus, high visual quality and low MTP latency are essential requirements of Metaverse networks to provide a high-quality user experience. The XR application should physically allow the users to control and steer both virtual and real objects. Such tactile experiences and control would require real-time transmission and feedback of haptic information with less than 1 ms end-to-end latency and 99.999\% reliability. To achieve such physical and visual experiences with heterogeneous networks, Metaverse system requires an architectural transformation with high computation and communication capacity.

\textbf{Agency:} Suppose that smart cities can be envisioned and projected in a pertinent way and address several concerns by reading the places and their dimensions. However, it is observed that there could be challenges related to social groups and communities that are not only recognized but actively changing futuristic perspectives of the smart cities, i.e, their agency, their inability to take certain actions, and "right to the city" \cite{cardullo2019right}, \cite{foth2015citizen}. Furthermore, there are concerns regarding the transparency of smart cities and their Metaverse, and the possibility of actually owning something in the Metaverse. In addition to these issues, the public raises some more questions: would it be possible to rearrange things and introduce new opportunities proactively? Do users have the right to intervene in the environment in any way, or only the right to use the preset gamified space (only act as a player in the game) \cite{graham1997virtual}? These concerns regarding agency and public rights to the city deeply affect the nature of the smart cities being shaped.

\textbf{Socio‐spatial polarization:} Another inevitable challenge is that the greater the separation between the physical and digital spaces, the higher the chances of encouraging a socio‐spatially polarized, two‐tier urban environment. For instance, one may be travelling virtually to exotic places while finding it difficult to adjust to and live in the physical city, find good job options, and have a hard time developing a real social life. Moreover, one might be dressing up their avatars in the digital world but possibly finding it hard to pay their energy bills or apartment rent. Focusing on virtually enhancing the city experience does not guarantee improvements in the physical world; instead, it might result in a more polarized world in which the rich can maintain or further improve their ability to interact and play with real places and spaces;. In contrast, others are demoted to a lower‐agency digital surrogate \cite{graham1997virtual}.

First of all, since the Metaverse involves the integration of various modern technologies, it has a high likelihood of inheriting the intrinsic flaws and vulnerabilities associated with these technologies.

\textbf{ Privacy and security:} Despite the Metaverse's potential, privacy and security are the key concerns hindering its further development. Several privacy invasions and security breaches might occur in the Metaverse through pervasive user-profiling activities, the mismanagement of massive volumes of data, and inequitable results of AI algorithms for the safety of human bodies and physical infrastructures. Second, the impact of existing threats may become more intense in the virtual world due to the intertwining of several technologies. Moreover, there is a possibility of new risks, such as virtual spying and virtual stalking \cite{leenes2007privacy}. In particular, the personal information used in the Metaverse might be remarkably ubiquitous and more granular for building a real-world's digital copy, thus opening new options for abuse of confidential big data \cite{falchuk2018social}. Finally, the loopholes in the system can be exploited by hackers to compromise devices and gain access to real-world equipment. For instance, they might get access to home appliances for threatening personal security or essential setups like water supply systems, high-speed rail systems, and power grid systems through advanced persistent threat (APT) attacks \cite{hu2015dynamic}.

The existing countermeasures might need to be more adaptable and effective, which is required for secured Metaverse applications. More specifically, the Metaverse's innate attributes, including heterogeneity, scalability, interoperability, sustainability, hyper-spatiotemporality, and immersiveness, may create several hurdles for ensuring effective security. These include: 
\begin{enumerate}
    \item The fact that the fully developed immersive virtual Metaverse environment brings real world enjoyments to the virtual world. However, in order to do so, a massive amount of data must be continuously exchanged between real users and their respective avatars, which ultimately increases security and privacy concerns.
    \item The Metaverse needs to be constructed on a decentralized architecture to be persistent and self-sustainable to eliminate a situation in which only a few entities have influence and the risk of having a single point of failure (SPoF) \cite{nguyen2021metachain}. In order to convince the big entities to agree to this newly proposed solution is also one of the big challenges that need to be tackled in the Metaverse.
    \item Incorporating the ternary world results in the Metaverse being hyper-spatiotemporal \cite{nevelsteen2018virtual}, which makes trust management much more difficult and complex. The Metaverse may make it more difficult to differentiate between fiction and fact because of the blurred boundary between the virtual and real worlds.
    \item The Metaverse scalability and interoperability point to users being able to easily and concurrently move across several sub-Metaverses easily under different interaction modes and scenes \cite{yoon2021interfacing}, which also presents challenges in terms of ensuring accountability enforcement, compliance auditing, and fast service authorization for multi-source data blending.
\end{enumerate}

\subsection {Related Work And Contributions}
The topic of the Metaverse has gained substantial research attention over the past few years. As of now, several survey papers have been published addressing various aspects of this ground-breaking concept. For instance, Dionisio et al. \cite{dionisio20133d} previously identified four attributes of workable virtual spaces (or Metaverse), i.e., interoperability, scalability, realism, and ubiquity, and talked about the current developments in fundamental virtual world technologies. Lee et al. \cite{lee2021all} evaluated eight elementary technologies for building the Metaverse in line with its prospects from six user-based factors. Ning et al. \cite{ning2021survey} discussed the Metaverse development status with reference to the social Metaverse, virtual reality, supporting technologies, infrastructures, industrial projects, and national policies. Yang et al. \cite{yang2022fusing} highlighted the potential of blockchain technology and AI for future Metaverse construction. Huynh-The et al. \cite{huynh2022artificial} also examined the role of AI-based schemes in founding and developing the Metaverse. In another recent study, Park et al. \cite{park2022metaverse} talked about the three components of the Metaverse including content, software, and hardware. They also reviewed representative applications, implementation, and user interactions within the Metaverse. Xu et al. \cite{xu2022full} provided a detailed survey on the edge-enabled Metaverse from the blockchain technology, computation, networking, and communication perspectives. From the privacy and security perspective, Leenes \cite{leenes2007privacy} explored the possible privacy threats in the online game 
'Second Life" from the legal and social angles. Very recently, Wang et al. \cite{wang2022survey} investigated the potential privacy/security risks, crucial privacy/security-related challenges, and contemporary defense mechanisms pertaining to the Metaverse. They conducted a thorough survey of the primary challenges and solutions invloved in building a privacy-preserving and secure Metaverse. 

Contrary to the above mentioned survey papers on the generic Metaverse concept \cite{lee2021all}, \cite{leenes2007privacy},  \cite{dionisio20133d},  \cite{ning2021survey}, \cite{park2022metaverse}; AI-enabled Metaverse \cite{yang2022fusing}, \cite{huynh2022artificial}; edge-enabled Metaverse  \cite{xu2022full}; and in social goods \cite{duan2021metaverse}, computational arts \cite{lee2021creators}, education \cite{diaz2020virtual}, retailing \cite{bourlakis2009retail}, and social AR/VR game \cite{falchuk2018social} applications this paper comprehensively discusses the significant role that could be played by the Metaverse in enabling technologies in the smart cities. The major contributions of this work are as follows:
\begin{figure*}
  \begin{center}
  \includegraphics[width=18cm, height=12cm]{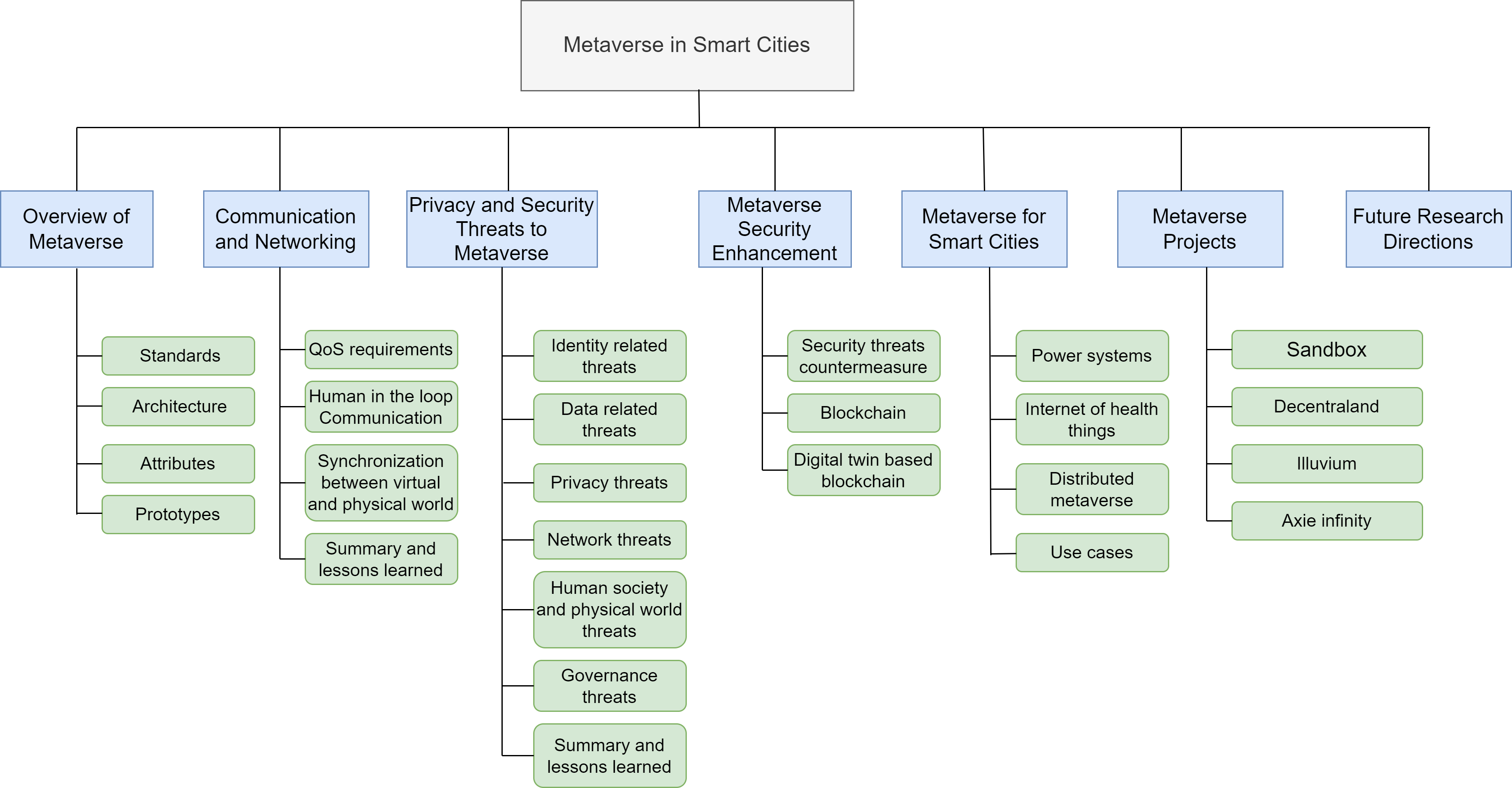}
  \caption{Outline of the paper 
  }\label{arch1}
  \end{center}
\end{figure*}

\begin{itemize}
   \item It explains the basic concepts of the Metaverse, including its architecture, primary attributes, enabling technologies, and the current state-of-the-art prototypes of the Metaverse application.
   \item It identifies the major computation, networking, communication, and distributed ledger/blockchain technologies, which would contribute to achieving the ubiquitous Metaverse and addressing the above-mentioned challenges. 
   \item It discusses how blockchain technology can play a vital part in developing the Metaverse. In particular, it explains how the networking and communication aspects of the Metaverse can be effectively supported by blockchain technology, e.g., through cross-chain and scalable block-chain sharing for reliable Metaverse applications, decentralized and secure storage, and edge data sharing for the secure Metaverse. This complies with the beyond 5G (BG5) / 6G vision \cite{saad2019vision} that a major part will be played by distributed ledger technologies to ensure reliable and secure networking and communication. 
    \item It explores Metaverse privacy and security threats from various aspects (i.e., social/physical, governance, economy, network, privacy, data, and identity) and identifies the potential countermeasures.
    \item It highlights the significant role played by Metaverse for smart cities in terms of power systems, the health sector, social good, etc., and discusses the contemporary Metaverse projects.
    \item It outlines open future research directions for the effective, and secure integration of Metaverse concepts in the designing and development of smart cities while preserving users' privacy. 
\end{itemize}

This research study is believed to be the first work that discusses in detail Metaverse-enabled smart urbanization. A comparison of available review articles and this survey paper is provided in Table I. The rest of the paper is organized as follows: Section II provides a comprehensive overview of the Metaverse concept, covering the standards, architecture, key characteristics, enabling technologies, and modern prototypes. Section III discusses the communication and networking procedures that are related to this evolving concept. The existing threats to the privacy and security of the Metaverse and their possible countermeasures are covered in Section IV. Section V sheds light on the modern concept of Metaverse for smart cities. Existing Metaverse projects are discussed in Section VI and open research challenges and future research directions are identified in Section VII. Section VIII includes future direction and Section IX concludes this paper.

\begin{table*}\centering
\caption{Comparison with available survey papers}
 \begin{tabular}{||p{2cm}|p{2cm}|p{10cm}||}  \hline
    Reference & Year & Contributions \\ [0.5ex] 
     \hline
     \cite{dionisio20133d} & 2013 & Discussed main attributes of Metaverse and current
developments of the fundamental virtual world technologies \\ \hline
 \cite{lee2021all} & 2021 & Reviewed the elementary technologies for building the Metaverse \\ \hline
  \cite{ning2021survey} & 2021 & Evaluated the development status of Metaverse with respect to social good, virtual reality, supporting technologies, infrastructures, industrial and national projects \\ \hline
  \cite{yang2022fusing} & 2022 & Discussed the potential of blockchain and AI technologies
in constructing the future Metaverse \\ \hline
 \cite{huynh2022artificial}  & 2022 & Discussed the technical role of artificial intelligence from different  aspects in developing Metaverse \\ \hline
\cite{park2022metaverse} & 2022 & Discussed the main components of the Metaverse, including content, software, and hardware, and reviewed the representative applications, implementation, and  users interaction in the Metaverse \\ \hline
 \cite{xu2022full} & 2022 & Conducted a comprehensive survey on the edge-enabled Metaverse with regards to  computation, networking, and communication \\ \hline
\cite{leenes2007privacy} & 2007 & Discussed the legal and social threats to the privacy in the game Second Life \\ \hline
\cite{wang2022survey} & 2022 & Comprehensively reviewed the fundamentals, privacy, and security of Metaverse, open challenges, potential solutions, and future research directions for securing Metaverse \\ \hline
\cite{duan2021metaverse} & 2021 & Discussed the Metaverse architecture and its role in promoting social good \\ \hline
\cite{diaz2020virtual} & 2020 & Discussed Metaverse applications in the education sector \\ \hline
\cite{falchuk2018social} & 2018 & Identified the digital footprints-related privacy risks and countermeasures  in social Metaverse games \\ \hline
 \cite{lee2021creators} & 2021 & Reviewed the Metaverse applications with reference to digital arts \\ \hline
 \cite{bourlakis2009retail} & 2009 & Discussed the Metaverse applications in the retailing sector \\ \hline
\textbf{Our Survey} & \textbf{2022} & \textbf{Thoroughly analyzes the significant role played by the groundbreaking Metaverse concept in enabling novel technologies in smart cities}  \\ \hline
 \end{tabular}
\end{table*}

\section{An Overview of Metaverse}
In this section we provides an overview of the Metaverse concept from different aspects, including its current standards, generic architecture, primary attributes, enabling technologies, state-of-the-art prototypes, and potential applications.



\subsection {Metaverse Architecture}
The Metaverse is a synthetic space that consists of computer-generated elements, virtual environments, digital things, and user-controlled avatars, wherein human beings can use any smart device to have their avatar socialize, collaborate, and communicate with other avatars belonging to other users. The Metaverse construction involves a blending of the ternary digital, human, and physical worlds. It's generic architecture is shown in Fig.~\ref{architect}. The connection between the three worlds, the main elements in the Metaverse, and the flow of information are discussed below.
 
\subsubsection {The Human World} Human users and their social communications and personal psychologies, make up the human world \cite{heller2016avatars}. The human users (equipped with smart devices, such as AR/VR helmets) can direct their digital avatars to interact, socialize, work, and play with other virtual entities or other avatars in the Metaverse through XR and human-computer interaction (HCI) technologies \cite{genay2021being}.

\subsubsection {The Physical World} It provides the Metaverse with supporting infrastructures (including storage, computation, communication, and control/sensing frameworks) for multi-sensory data perception, processing, and transmission in addition to physical control, thereby enabling effective interactions between the human and digital worlds. The communication framework, which is comprised of several heterogeneous wired or wireless networks (e.g., satellite, unmanned aerial vehicle (UAV), and cellular communication), provides networking connectivity. Moreover, the storage and computation infrastructure offers the storage capacity and powerful computations with the help of cloud-edge-end computing \cite{kai2020collaborative}, as massive number of frames are generated per seconds in the virtual world. 

\subsubsection {The Digital World} According to the IEEE 2888 \cite{2888} and ISO/IEC 23005 standards \cite{MPEG-V}, a digital world can be comprised of several mature distributed virtual worlds that are linked together. Each virtual world (or sub-Metaverse) offers users, different types of virtual environments (such as virtual cities and game scenes) and virtual services/goods (e.g., online concerts, online museums, social dating, and gaming) \cite{wang2022survey}.

The two main sources of information in the Metaverse, are \cite{wang2022survey}:
\begin{itemize}
\item Real space input (i.e., knowledge acquired and information obtained from the real world is digitally exhibited in the virtual world)
\item Virtual worlds output (i.e., the data produced by Metaverse services, digital objects, and avatars in the virtual world) 
\end{itemize}

\begin{figure}
  \begin{center}
  \includegraphics[width=3.5in]{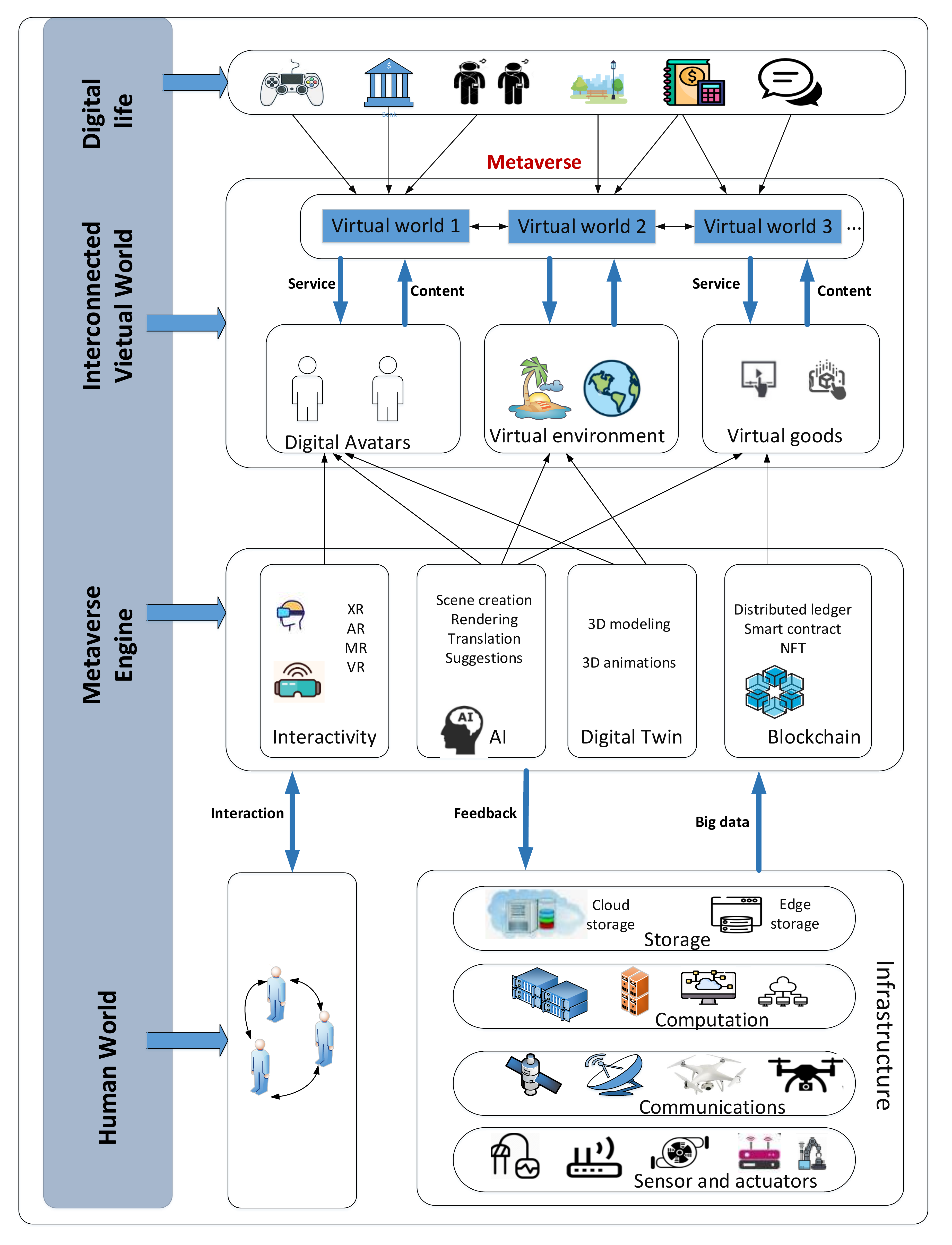}\\
  \caption{Metavserse architecture involving the integration of the digital, physical, and human worlds}\label{architect}
  \end{center}
\end{figure}

\subsubsection {Metaverse Engine} In the Metaverse, physical world data is used as input to generate, maintain, and update the virtual space via blockchain, DTs, and AI technologies \cite{xu2022full}. The HCI and XR techniques enable users in the physical worlds to manage their digital avatars through their bodies and senses for varied social and collective activities, e.g., virtual item trading, dating, and car racing. AI-based algorithms create customized content/avatars and offer intelligent services for enriching the ecology of the Metaverse. Furthermore, incorporating AI techniques in big data analysis can help to acquire knowledge for simulating, digitizing, and mirroring of the real space using DT technology for producing a vivid virtual space experience for the users. Lastly, adopting the blockchain technology make it possible for trustworthy and transparent transactions for different social services can be made among digital avatars, thus enabling the creation of a value based economic system in the Metaverse  \cite{wang2022survey}.

To summarize, the Metaverse main resource is the information flow in the ternary worlds. This makes the creation of digital ecology and helps to integrate  the real and virtual worlds. The information flow is described in the two subsections below.

\subsubsection {Intra-World Flow of Information} The human world is interlinked through social networks and is developed based on the mutual communications and shared activities of human beings. In the physical world, the IoT-assisted control/sensing framework plays a vital part in the digitization or transformation of the physical world through pervasive actuators and sensors, while the computation and networking infrastructures transmit and process the big data generated. In the digital world, the Metaverse engine is responsible for processing and managing the digital information produced by the human and physical worlds to support the rendering or creation of Metaverse and facilitate different services in it. Furthermore, the digital creations can be produced and distributed by the users across different sub-Metaverses by user to promote Metaverse creativity. 
  
\subsubsection{Inter-World Flow of Information}Fig.~\ref{architect} shows that the IoT, Internet, and subjective consciousness are the primary channels linking the three worlds. The objective information is acquired by humans from the physical world, which is transformed into intelligence and knowledge using subjective consciousness, and subsequently used as a guide for changing the objective world. Moreover, the HCI technology can help humans in interacting with physical objects and experience virtual AR  through the XR technology. The digital world and human worlds are linked via the Internet for knowledge acquisition. Besides that, the digital and physical worlds are also inter connected through the IoT infrastructure that uses inter-linked smart devices for a smooth flow of information between the two worlds  \cite{jayasinghe2018machine}. Furthermore, the processed information feedback from the digital world (e.g, intelligent decisions based on processed data) facilitates the physical world transformation \cite{wang2022survey}.

\subsection {Major Attributes of Metaverse}
Web 1.0 Internet users were simply content users, with websites providing the fundamental contents for users. In Web 2.0, i.e., mobile Internet, the content is not only used but also produced by the users, and the websites become platforms for providing services. Some common service provision platforms includes TikTok, WeChat, and Wikipedia. However, Metaverse is known to be an emerging Web 3.0 paradigm. 
Metaverse users can smoothly travel between different sub-Metaverses to experience a virtual life along with making economic interactions and digital creations, which are assisted by the Metaverse engine and physical infrastructures. More specifically, the Metaverse exhibits the following unique attributes. 


\subsubsection {Heterogeneity} This includes heterogeneous modes of communication (e.g., satellite and cellular communications), heterogeneous types of data (e.g., structured and unstructured data), heterogeneous physical devices having different interfaces, and heterogeneous virtual environments with diverse implementations, and the variation in the human psyche. 

\subsubsection {Scalability} This refers to the capability of Metaverse to remain effective with numerous concurrent active avatars/users,scenario complexity levels, and modes of interaction among avatars/users (in terms of range, scope, and type) \cite{dionisio20133d}.

\subsubsection {Interoperability} This implies that (1) users can travel between sub-Metaverses with a smooth immersive experience \cite{lee2021all}, and (2) the digital assets used to construct or render sub-Metaverses can be interchanged across different platforms \cite{dionisio20133d}.

\subsubsection {Sustainability} This shows that the Metaverse maintains reliable value system and a closed economic loop infrastructure with a high level of independence. On one hand, it must be formed on decentralized framework to maintain persistence, eliminate SPoF threats, and prevent some influential entities from taking control \cite{wang2022survey}. On the other hand, it should continuously arouse the users' enthusiasm for creating novel digital content and innovations.

\subsubsection {Hyper-Spatiotemporality} Irreversible time and finite space are the two main limitations of the real world. Since Metaverse is a virtual time-space perpetuity parallel to the real world, therefore, breaking restrictions of space and time is referred to as hyper-spatiotemporality  \cite{ning2021survey}. Users can therefore seamlessly move between different worlds that have varied spatiotemporal dimensions for experiencing an alternative life with the smooth transformation of scenes. 

\subsubsection{Immersiveness} This implies that the computer-generated virtual environment is adequately realistic for users to feel emotionally and psychologically immersed \cite{han2010user}. It may also be termed "immersive realism" \cite{dionisio20133d}.

\subsection {Enabling Metaverse Technologies}
\begin{figure}
  \begin{center}
  \includegraphics[width=8.5cm, height=7cm]{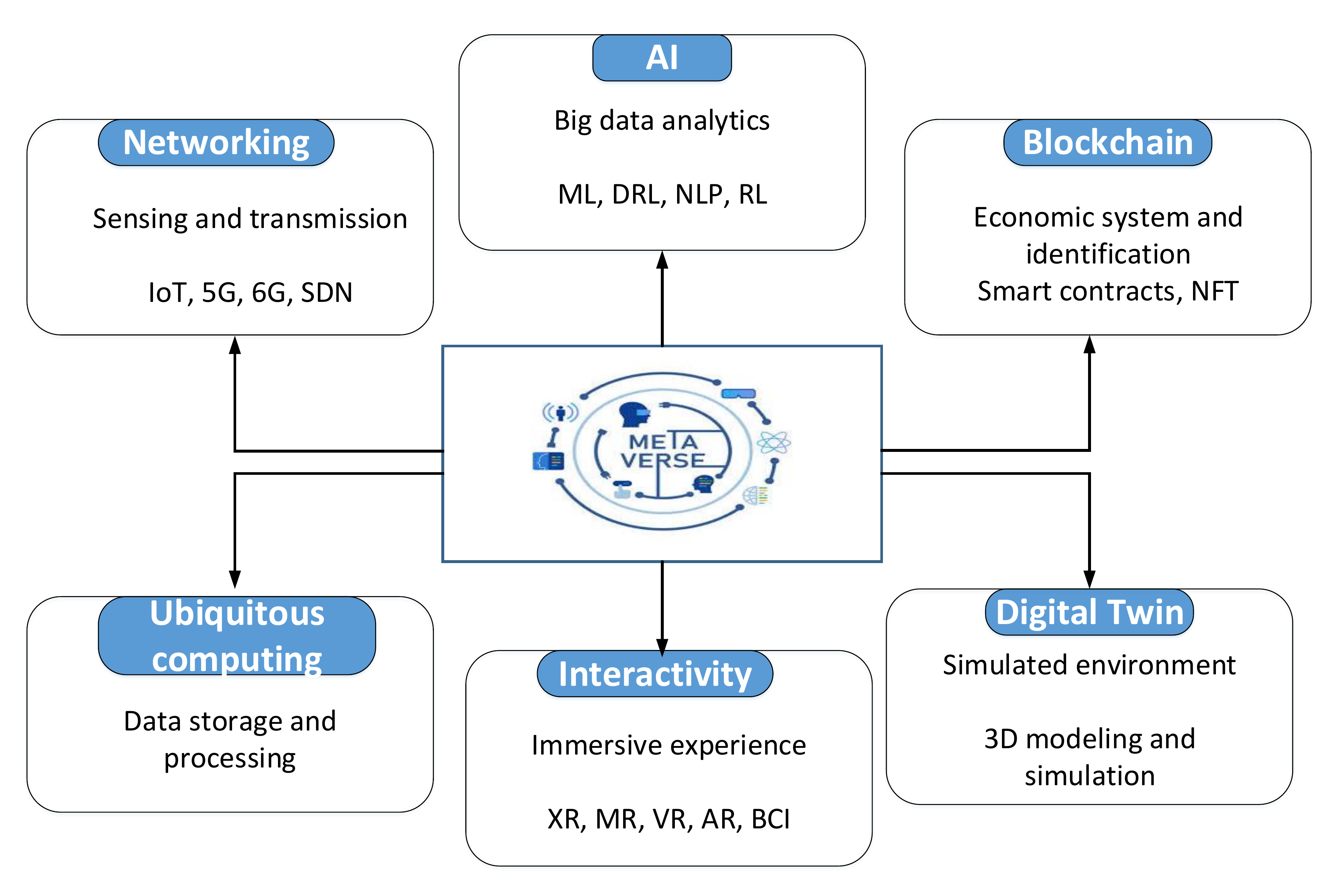}\\
  \caption{Six fundamental technologies and their role in the Metaverse}\label{tech}
  \end{center}
\end{figure}

The Metaverse is built on the following six fundamental technologies, as depicted in Fig.~\ref{tech}. 

\subsubsection {Interactivity} With the maturation of XR technology, embedded technology, and miniaturized sensors, XR devices are expected to become the major terminal through which to enter the Metaverse \cite{sugimoto2021extended}. MR/AR/VR technologies are deeply incorporated in XR to offer an augmented experience,  multi-sensory immersiveness, and real-time interactions between the environment, the avatar, and user through large-scale 3D modeling and HCI (especially the brain-computer interface (BCI)) \cite{jaynes2003metaverse}. In particular, MR provides an experience for transitioning between AR and VR.  AR is responsible for the experience of the actual presence of virtual videos, graphics, and holograms in the real world, while VR delivers the experience of immersiveness in the virtual world. Moreover, the wearable XR equipment carries out the perceiving of explicit, human-specific details and the pervasive sensing of surroundings and objects using indoor smart devices.  In this way, user-avatar interactivity is not be restricted to mobile inputs (for instance, laptops and mobile phones) only, but possible with every type of interactive device connected to the Metaverse. 

\subsubsection {Digital Twin (DT)} 
DTs are the digital clones of real-world systems and objects with high consciousness and fidelity \cite{wu2021digital}. This technology makes it possible for data that was input through physical elements to be employed for self-adaptation and self-learning in the mirrored world. Furthermore, DTs may offer precise digital models of objects with the requisite features in the Metaverse with the assistance of advanced AI techniques and be capable of simulating the complex models. Moreover,DTs facilitates accident traceability and predictive maintenance for physical security because of the bidirectional link between virtual entities and physical counterparts, thus reducing risks and improving efficiency in the physical world.  

\subsubsection {Ubiquitous Computing}
\begin{figure}
  \begin{center}
  \includegraphics[width=3.5in]{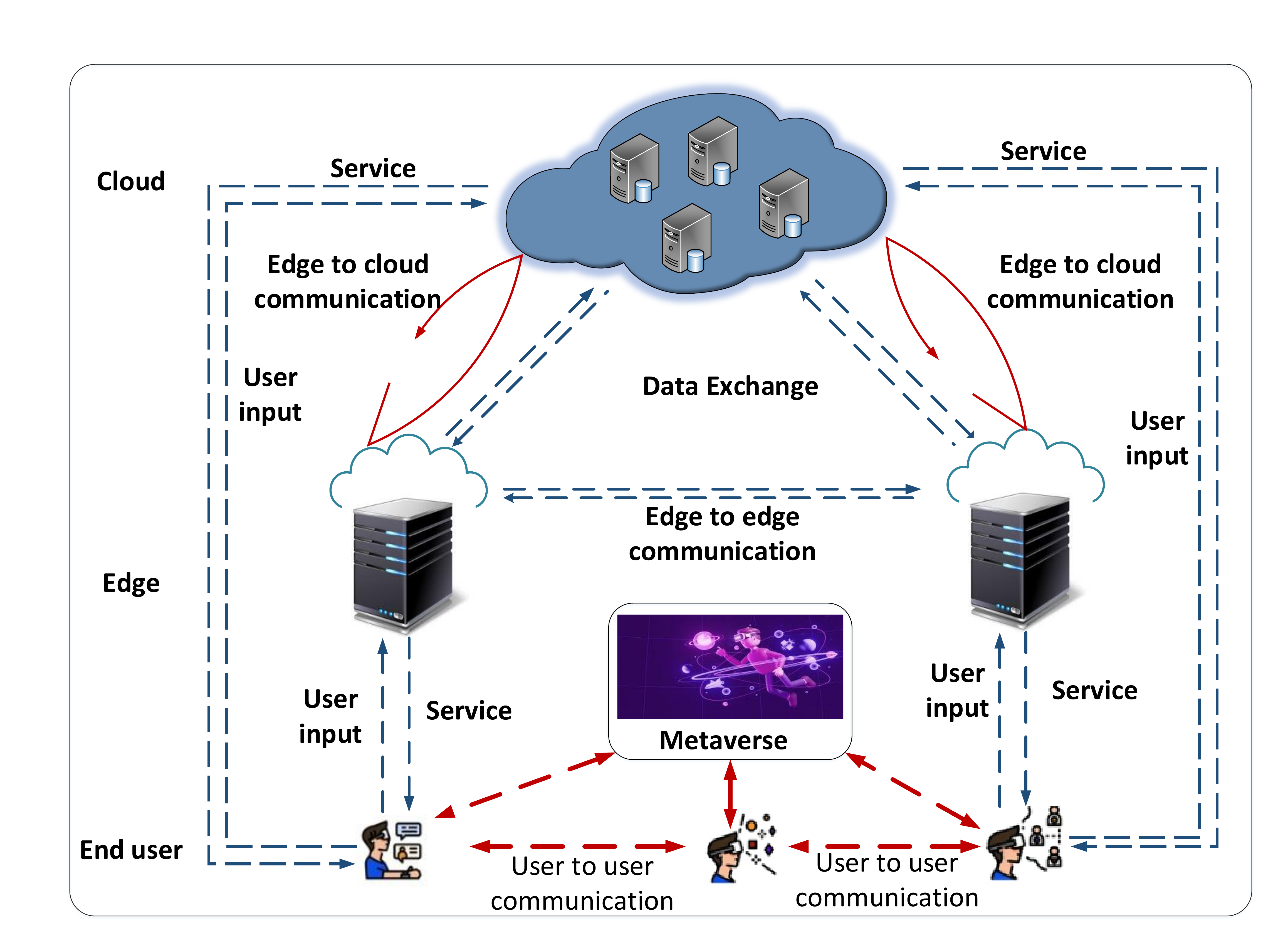}\\
  \caption{Illustration of cloud-edge-end computing in Metaverse service offering.}\label{cloud}
  \end{center}
\end{figure}

The purpose behind ubiquitous computing is to create an environment where computing is accessible for users everywhere all the time \cite{vural2012survey}. Ubiquitous computing makes it possible to seamlessly adapt to interactions between the physical environment and human users through pervasive smart devices carried by humans or embedded in the surroundings. With the help of ubiquitous computing, the users can easily interact with their avatar and enjoy immersive Metaverse devices in real-time. For a better quality user experience in ubiquitous computing,  the heterogeneous edge computing infrastructures and highly scalable cloud infrastructures are orchestrated by the cloud-edge-end computing \cite{kai2020collaborative} through sophisticated inter/inner-layer cooperation paradigms, as depicted in Fig.~\ref{cloud}. As such, these advanced infrastructures provides effective real-time resource allocation to Metaverse applications that are in use as required to meet user demand. 

\subsubsection {Networking} In the Metaverse, ubiquitous network access and the real-time transmission of massive amounts of data between the virtual and real worlds and within sub-Metaverses is empowered by the networking technologies including the IoT, the software-defined network (SDN), B5G and 6G \cite{du2021optimal}. In 6G, space-air-ground integrated network (SAGIN) \cite{wang2021blockchain} is a potential paradigm for ubiquitous and smooth network access to the Metaverse applications. Moreover, SDN allows the scalable and flexible control of massive Metaverse networks by separating the data plane and the control plane. A logically centralized controller uses a standardized interface for managing the resources and physical devices in an SDN-enabled Metaverse and thus enable the dynamic allocation of virtualized bandwidth, storage, and computation resources in response to the real-time requirements of different sub-Metaverses  \cite{wu2020interactive}. Furthermore, IoT sensors in the Metaverse serve as an extension of human senses. 

\subsubsection{AI} Artificial intelligence serves as the “brain” of the Metaverse that makes it possible to have customized Metaverse services (such as personalized and vivid avatars creation) and multilingual support in the Metaverse \cite{huynh2022artificial}. Furthermore, AI technology allows intelligent interactions (for instance, smart user movements predictions and shopping guides) between the avatar and user through intelligent decision-making. Further details regarding AI integration in the Metaverse are available in \cite{huynh2022artificial}.

\subsubsection{Blockchain} 
\begin{figure}
  \begin{center}
   \includegraphics[width=3.5in]{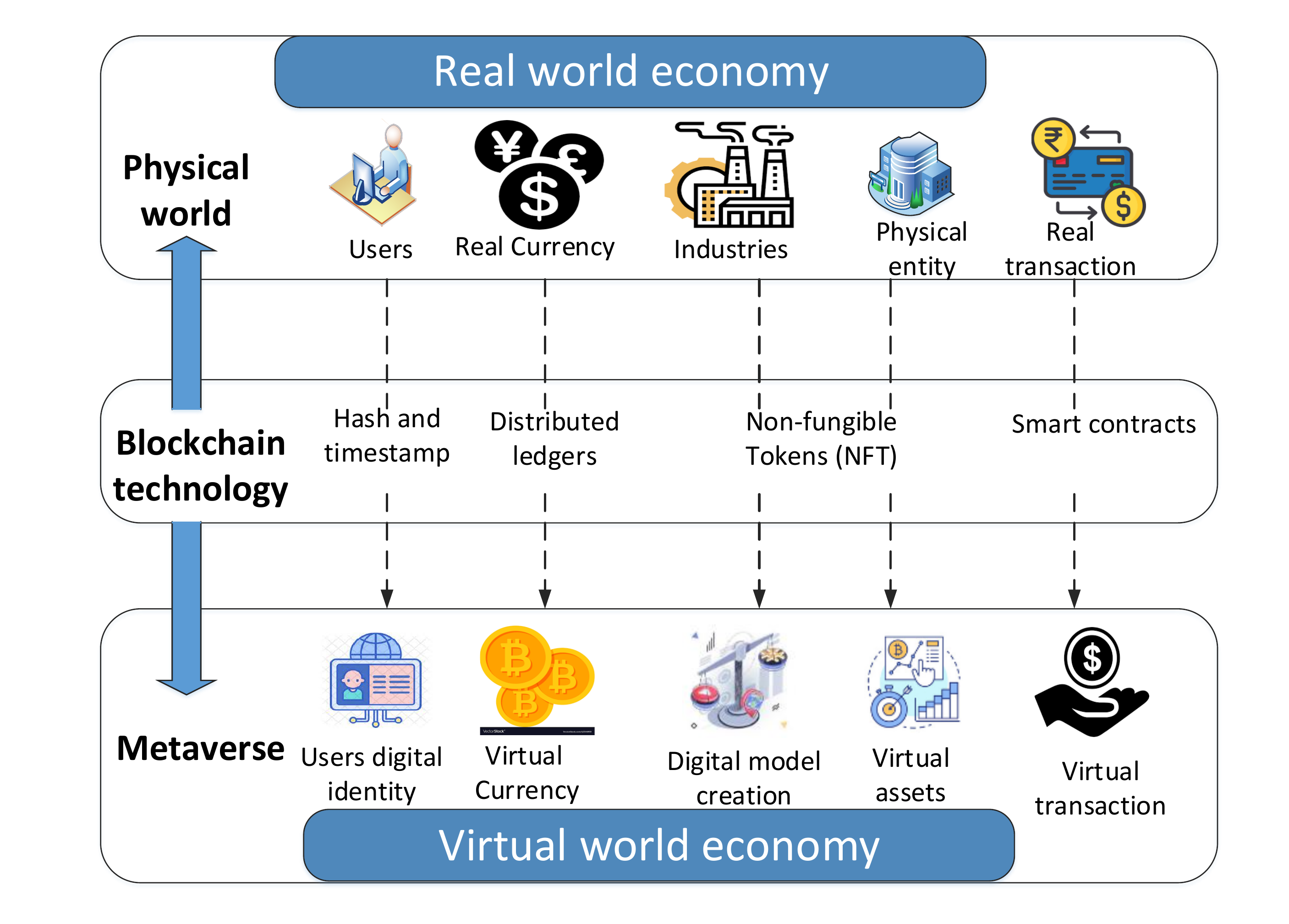}\\
  \caption{Role played by blockchain technology in bridging the Metaverse economy and traditional economy}\label{blockchain}
  \end{center}
\end{figure}

The Metaverse needs to be constructed on a decentralized framework to be persistent and avoid few entities having control, low transparency, and centralization risks \cite{nguyen2021metachain}. Furthermore, the blockchain-assisted virtual value and economy system are the fundamental elements of the Metaverse. Fig.~\ref{blockchain} shows that blockchain technologies offer a decentralized and open solution  for developing a persistent virtual economy in the Metaverse. In the blockchain, the data is structured in hash-chained blocks and attributed with auditability, transparency, immutability, and decentralization  \cite{wang2021blockchain}. There are three main categories of blockchains-private, consortium, and public that differ in their level of  decentralization \cite{wang2021blockchain}. The scalability of the system and consistency of the ledger are determined by the consensus protocol. Moreover, blockchain makes it possible to use smart contracts to automate the performance of functions within disloyal parties in an arbitrary manner. Non-fungible token (NFT) denotes indivisible and irreplaceable tokens \cite{wang2021non} that can assist in determining ownership provenance and asset identification in the blockchain. In addition, ``De-Fi"  which stands for ``decentralized finance" aims to deliver complex, transparent, and secure monetary services (such as currency conversion or stock exchange) in the Metaverse. 

\subsection {Existing Metaverse-Related Standards}
This subsection briefly introduces two existing standards related to the Metaverse-ISO/IEC 23005 (MPEG-V) \cite{MPEG-V} and IEEE 2888 \cite{2888}.
\begin{enumerate}
    \item ISO/IEC 23005 (MPEG-V) is aimed at standardizing the interfaces between the virtual and real worlds, and between two virtual worlds for interoperability,  the smooth exchange of information and response synchronization \cite{MPEG-V}. Fig.~\ref{arch1} shows the generic ISO/IEC 23005  architecture that involves three scenarios for exchanging information.    
    \begin{itemize}
    \item In the first scenario, information is sent from the real world to the virtual one (i.e., R → V adaptation) and once there, the sensory data taken from the real world is used as input to create the virtual world object characteristics (VWOCs) by taking into consideration both users sensing preferences and sensor capacity.
    \item In the second scenario, media is sent from the virtual world to the real one (i.e., V → R adaptation). In this phase VWOCs and sensory effects are used as inputs, which in turn generate the commands for the actuators placed in the real world by taking into consideration both the maximum limits of the actuators and users’ actuation preferences.
    \item The third scenario involves the adaptation of data between the virtual worlds by converting exclusive VWOCs to normatively defined VWOCs using the V→V adaptation engine.
    \end{itemize}
    \item IEEE 2888 \cite{2888} was introduced in 2019 to supplement the ISO/IEC 23005 standard because the latter focuses heavily on sensory effects and does not have enough potential to offer general-purpose interfaces between the real and virtual worlds. IEEE 2888 is aimed at defining standardized interfaces for synchronizing the physical and cyber worlds. It provides a base on which to build Metaverse systems by specifying the application program interfaces (APIs) and information formats to use to  control actuators and acquire sensory data. It has four major components: (1) sensor interface specifications (IEEE 2888.1), (2) actuator interface standard (IEEE 2888.2), (3) standard for orchestrating digital synchronization (IEEE 2888.3); (4) standard for VR disaster response training system architecture (IEEE 2888.4) \cite{yoon2021interfacing}. 

\begin{figure}
  \begin{center}
  \includegraphics[width=7cm, height=10cm]{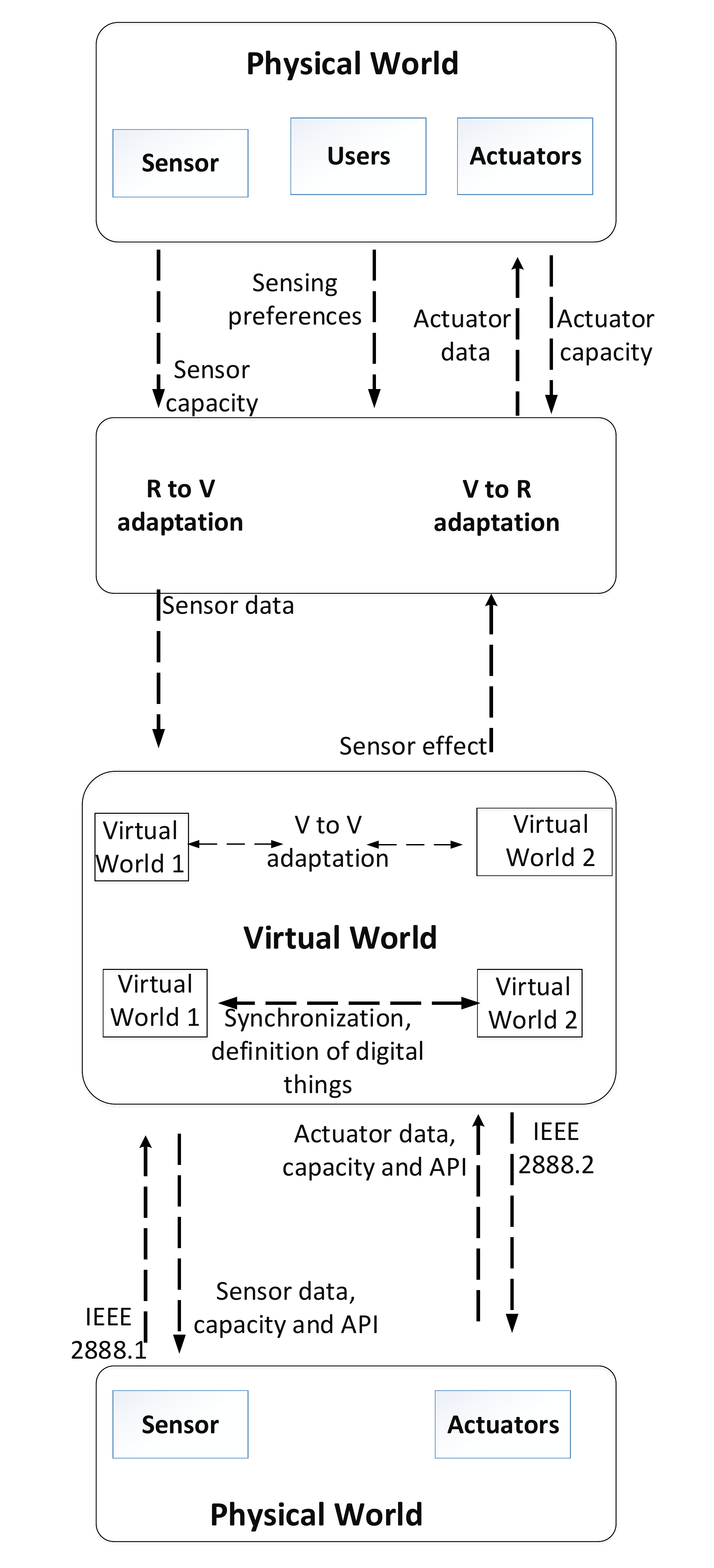}
  \caption{The architecture of ISO/IEC 23005 and 2888 (MPEG-V) standards 
  }\label{arch1}
  \end{center}
\end{figure}
The IEEE 2888 architecture is depicted in Fig.~\ref{arch1}, which also shows that the IEEE 2888.1 and IEEE 2888.2 are used to exchange actuator-related and sensory information between the real and the virtual worlds. Moreover, the IEEE 2888.3 is used to define the mission control data and synchronization for virtual objects or digital things. 
\end{enumerate} 

\subsection{Tool and Platforms for Metaverse}
Different types of tools and platforms used for the development of Metaverse are discussed in this section. Some of the modern tools that plays a significant role in making the concept of Metaverse true are as follows:

\subsubsection{Unity}
Unity is used for the designing 3D models and  environments, as it is integrated with the 3D engine \cite{59}. A different experience for Metaverse, VR, and AR can be created using the advanced features of Unity. In addition, decentralized options are also available and include features of advanced algorithms that works on decentralization, such as blockchain, edge computing, and AI agents.  
\subsubsection{Unreal Engine}
For generating other assets such as Metahuman creator and marketplace assets, for Metaverse environment, another designing studio platform named Unreal Engine is used \cite{59}. Furthermore, by using Metahuman creator functionality, the time taken for the creation of digitized human avatars can be significantly reduced.
\subsubsection{Roblox}
Roblox has been known as a game, but with passage of time, it is evolved into a platform, that provides tools for developing the avatars and teleportation mechanism \cite{59}. Roblox also provides a proprietary 3D engine that is connected with the design studio and Metaverse marketplace, where the avatars can perform trading.
\subsubsection{Meta Avatars}
Meta Avatars is a system that is used to support Unity developers for Quest and different VR platforms \cite{60}. Through this users can customize their avatars and ensure that functionality of Meta Avatars reflect them consistently. Moreover, different tracking controllers like hand tracking, motion tracking, and audio inputs are used by Meta Avatars to produce high quality expressive avatars representations. 
\subsubsection{Hololens2}
It is an MR head-mounted device with built-in WiFi and computer and has more computation power and battery life \cite{62}. Hololens2 can be connected to the users worldwide through Microsoft Mesh, enabling them to communicate, maintain eye contact, observe other users' behaviors located in other places.
\subsubsection{Oculus Quest2}
It is a head-mounted VR device introduced by Meta \cite{63}. It is used to run different VR applications and is connected to the computer either by USB or WiFi.

Different tools and platforms are used to develop the Metaverse prototypes, but many of them such as Grand Theft Auto Online, require huge computation resources to run. Moreover, it will require intensive rendering, updating the DTs in real time, and low latency communication infrastructure. Adding to these implementation requirements, each industry is building the Metaverse based on different protocols, tools and platform. One standardized protocol, and platforms is required for developing the Metaverse, so the avatars can easily create content and move from one virtual space to another without difficulty.

\subsection {Existing Modern Prototypes of Metaverse Applications}

This section introduces the state of the art of Metaverse application prototypes. 

\subsubsection {Online Collaboration} Metaverse offers novel opportunities for virtual collaborations in the form of meetings and panel discussions in virtual conference rooms, studying in virtual classrooms, and telecommuting in virtual workplaces. For instance, Microsoft Mesh is an Azure-supported MR platform that enables users working in different locations to collaborate virtually through shared experiences and holographic presence from anywhere in a digital copy of their office \cite{wang2022survey}.

\subsubsection{Social Experiences}
Metaverse has the power to revolutionize society by enabling several social applications, including time/space travel, global travel, virtual chatting, virtual dating, virtual shopping, and virtual lives \cite{wang2022survey}. For instance, in COVID-19 pandemic the virtual graduation festival was arranged for UC Berkeley by developing the virtual campus using Mincraft platform. Tencent designed a virtual museum in 2018, that allowed users to visit its collections and exhibits from any where using VR helmets.  

\subsubsection {Game} Gaming is currently the most well-known application of Metaverse. Considering content adaptability, user matching, and technological maturity, games are considered to be an efficient approach to explore the Metaverse. Fortnite, Roblox, and Second Life are a few representative examples of Metaverse games. 

\subsubsection {Simulation and Design} 3D design, modeling, and simulation are other promising Metaverse applications. FNVIDIA built the open platform Omniverse to support the real-time 3D visualization and simulation of physical attributes and objects in a shared virtual environment for industry-level applications, such as automotive design. 

\subsubsection {Creator Economy} The content creation process in Metaverse is broadly classified into four modes, which are: AI-generated content (AIGC), user-generated content (UGC), professional- and user-generated content (PUGC), and professional-generated content (PGC). In the Metaverse, it is expected that the content consumers will be in larger amount then the content producers; thus, the AIGC mode might facilitate virtual service providers (VSPs) in creating massive customized and qualified content with lesser cost and higher efficiency. The AIGC is further classified into two categories: (i) content made fully by AI, and (ii) content made jointly by AI and users. One example of AIGC is Epic Games's MetaHuman project \cite{44}, which uses advanced AI algorithms to create real virtual profile of characters for conversational purposes. In UGC mode, content is produced by all users and circulated smoothly in the marketplace offered by the platform, which is attributed to decentralization, high diversification, low cost, and a high degree of freedom  \cite{kasapakis2017user}. In UGC mode, the users control the content creation process. In PGC mode, the professional content creators create the content (such as games) on the platform, and ordinary users can only participate and experience/view the content. Meanwhile, the PUGC mode merges the UGC and PGC modes and enables users and professionals to jointly produce  content. 

\begin{figure*}[!ht]
  \begin{center}
  \includegraphics[width=7.5in]{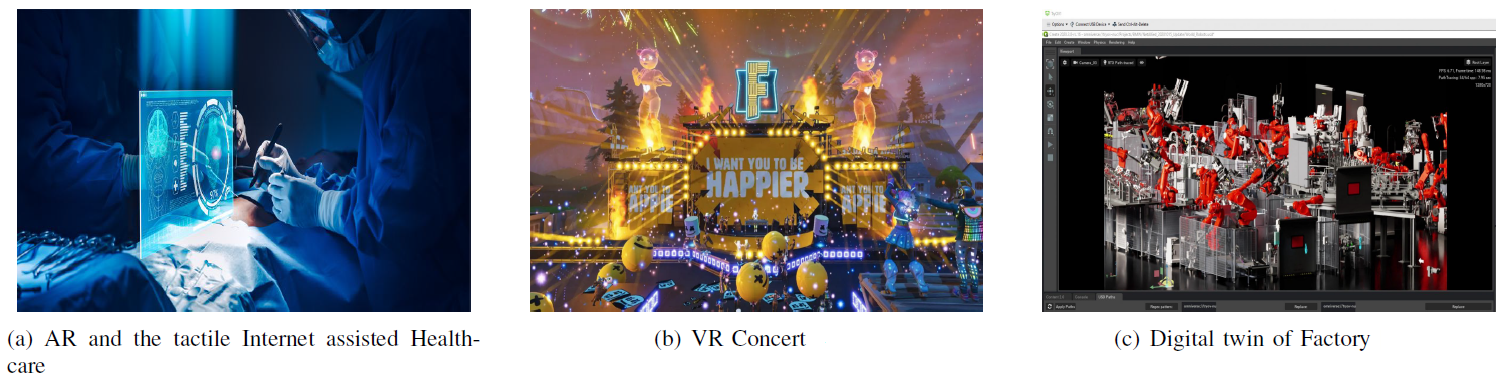}\\
  \caption{The Metaverse examples in communication for DT, the tactile Internet, and AR/VR \cite{xu2022full}}\label{mv-comm}
  \end{center}
 \end{figure*}

 \begin{table*}[]
\caption{Communication requirements for Metaverse applications}
\centering
\begin{tabular}{|l|l|l|l|}
\hline
\textbf{Applications}                     & \textbf{Latency (ms)} & \textbf{Reliability (\%)}           & \textbf{Data Rate (Mbit/s)}                                   \\ \hline
Hologram education (light field) & 20           & $1 - 10^{-5}$ & $1 \times 10^5 - 1 \times 10^6$ \\ \hline
AR smart healthcare              & 5            & $1 - 10^{-5}$ & 10000                                                \\ \hline
Digital twin (smart city)        & 5-10         & $1 - 10^{-5}$& 10                                                   \\ \hline
VR entertainment                 & 7-15         & $1 - 10^{-5}$ & 250                                                  \\ \hline
Tactile internet                 & 1            & $1 - 10^{-6}$& 1                                                    \\ \hline
\end{tabular}
\end{table*}
\section {Communication and Networking}
The Metaverse projects presented by early researchers were highly demanding in term that they might need the involvement of sophisticated edge devices for its successful deployment in real world. Henceforth, making the Metaverse accessibility as an inevitable challenge. It is anticipated that the mobile edge networks will offer effectual networking and communication coverage to the users with wide area, low delay, and high-speed wireless access for a real-time, flawless, and immersive Metaverse experience \cite{duan2021metaverse}.This is why the role of communication and networking and their importance for efficient Metaverse deployment are discussed in this section. 

Strict communication requirements must be met for users to have a flawless Metaverse experience as shown in Table 2 . Contrary to conventional 2D image transmission, 3D virtual scenarios and items need to be transmitted in the Metaverse to provide users with an immersive experience. However, this puts a substantial load on the existing network infrastructure. On the whole, the following Metaverse attributes pose notable and novel challenges for mobile edge networks when offering Metaverse services:

\begin{enumerate}
    \item Embodied 3D Worlds: 
    Metaverse contains several 3D worlds in which users can use embodied cognition to immerse themselves as avatars. Users must be ubiquitously connected with the Metaverse for getting access to the 3D multimedia services, which includes tactile Internet and VR/AR , anytime and anywhere.
    \item Reliable Immersion: The Metaverse provides substantially reliable and persistent 3D multimedia services \cite{lee2021all}. The users may surf the Metaverse for notably long time periods and can complete several tasks ranging from work to recreation. Mobile networks must thus provide Metaverse users with, highly persistent and reliable connectivity.
    \item Friction of Interaction: High-level interactions, which include machine-to-machine (M2M), human-to-machine (H2M), and human-to-human (H2H), interactions create the basis for users interactions and enables synchronization between the physical and virtual worlds in the Metaverse. However, some strict requirements for must be met when delivering multimedia services in the Metaverse, such as a low, like haptic perception rate, low interaction latency, and motion-to-photon latency \cite{zhao2017estimating} , to ensure low interaction friction. This is to ensure that users in the Metaverse enjoys the same experience as humans in the physical world. 
    \item Multi-dimensional Cooperation: The Metaverse maintenance involves multi-dimensional cooperation between the physical service providers (PSPs) and VSPs in the virtual as well as physical worlds \cite{pan2021network}.
\end{enumerate}

To address the aforementioned issues and facilitate an edge-driven Metaverse, an analysis of modern networking and communication solutions is provided in this section to. Firstly, Metaverse users must be able to experience telepresence through the seamless delivery of 3D multimedia services that help devices and users establish a smooth connection with the Metaverse, which includes the support from AR adaptation and VR broadcasting systems. Unlike classic content delivery networks, the networking and communication supporting Metaverse should prioritize user-focused considerations while delivering the content. The reason for this is that the limited bandwidth and rapid data growth require an evident shift in paradigm away from the traditional focal points of conventional information theory. We thus review goal-oriented/semantic communication solutions that may help in alleviating the spectrum limitation for future multimedia applications. Last but not least, the real-time bidirectional synchronization of the physical and virtual worlds, i.e. digital twinning is an essential requirements for constructing the Metaverse. Smart communication infrastructures, such as UAVs and  Reconfigurable Intelligent Surface (RIS) must be leveraged for the actuation and sensing interactions between the virtual and physical worlds. 
 
\begin{figure}
  \begin{center}
  \includegraphics[width=3.5in]{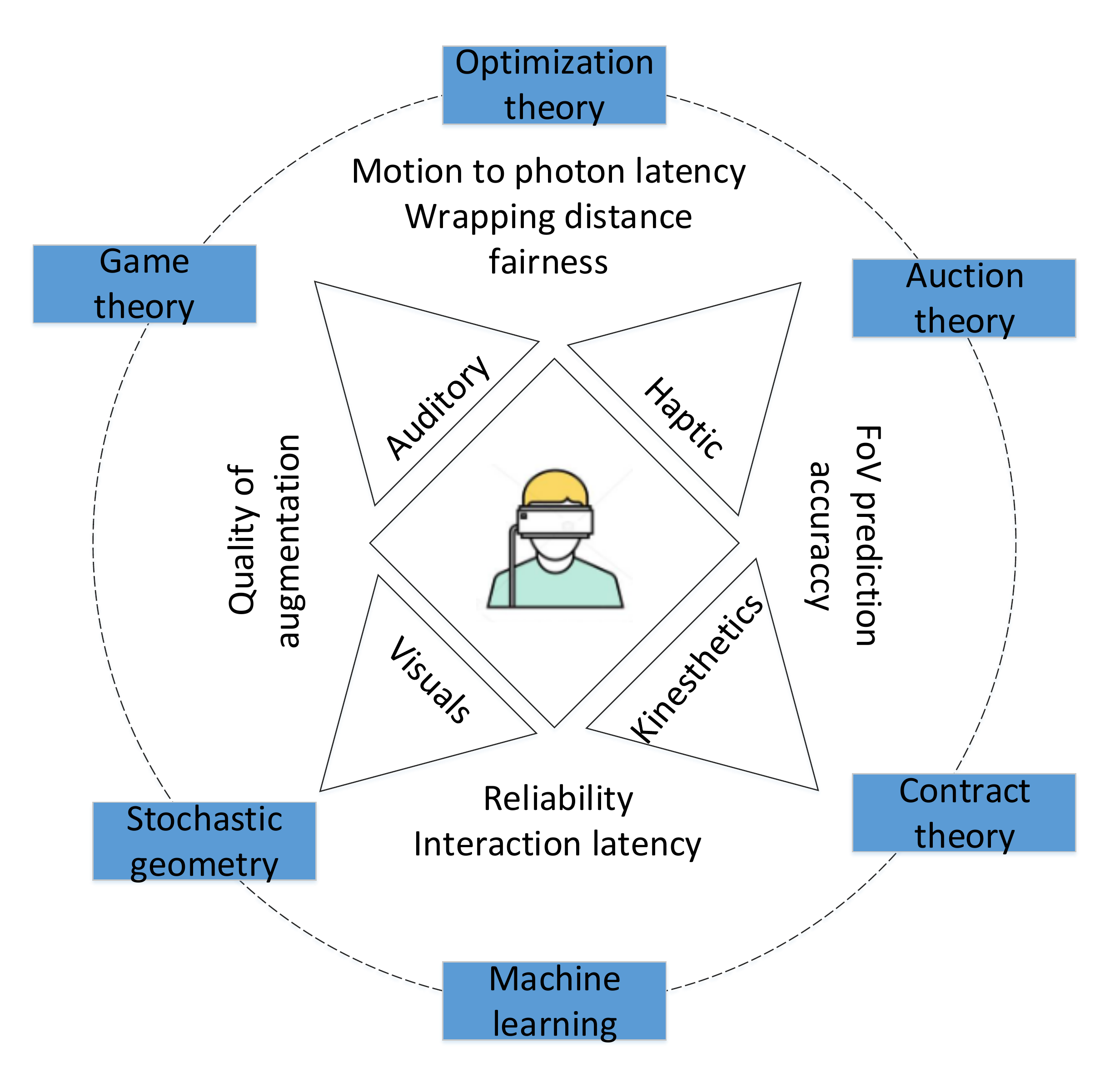}\\
  \caption{Mathematical tools for multi-sensory multimedia
network design 
}\label{tools}
  \end{center}
\end{figure}

Fig.~\ref{tools} provides the fundamental human-oriented metrics and mathematical tools considered. Moreover, Table III summarizes the studies reviewed, in terms of their mathematical tools, performance metrics, problems, and scenarios. 

\begin{table*}\centering
\caption{Edge Communications for VR/AR, semantic communication, and tactile Internet}
 \begin{tabular}{||p{1.5cm}|p{3cm}|p{4cm}|p{3cm}|p{3cm}||}  \hline
    Reference & Scenarios & Problems & Mathematical Tools & Performance Metrics \\ [0.5ex] 
     \hline
     \cite{guo2020adaptive} & Wireless mmWave-enabled VR system & Task offloading to render VR in real-time & DRL, Game theory & Convergence time, QoE \\ \hline 
     \cite{feng2020smart} & VR broadcasting in 5G Het-Nets & Selection of mode for broadcasting & RL, unsupervised learning & VR broadcasting throughput, convergence of algorithm \\ \hline 
     \cite{gimenez20195g} & MBMS over 5G & Designing and assessment of physical layer & Flexible configuration, numerology & Coverage of VR service,flexibility, capacity \\ \hline
     \cite{elbamby2018edge} &  MmWave-powered interactive VR gaming & Resource allocation & Matching theory & Interaction latency\\ \hline
     \cite{militano2015single} & D2D-based MBMS with single frequency & D2D radio resource management & Greedy algorithm, iterative search algorithm & Aggregate data rate, short-time fairness\\ \hline
     \cite{zhang2021buffer} & Buffer-aware VR video streaming & Personalized and private viewport prediction & DRL, FL & User-based streaming utility and accuracy \\ \hline
     \cite{wang2022meta} & Indoor VR over THz/VLC wireless networks & VAP selection and User association & Meta RL & Bit-rate, persistence of VR services \\ \hline
     \cite{xu2021wireless} & Non-panoramic wireless VR & Economic design  & Auction theory,DRL &  Social welfare, cost of auction information exchange \\ \hline
     \cite{liu2018dare} & MAR with dynamic edge resource & MAR service configuration & Optimization methods & Quality of Augmentation \\ \hline
     \cite{liu2018edge} & Edge-assisted MAR systems & Designing and analysis of system & Optimization approaches & Service latency and analytics accuracy  \\ \hline
     \cite{ren2020edge} & Mobile Web AR over 5G networks & Adaptive motion-aware selection of key frame & Planning approaches & Communication efficiency, feature extraction-tracking accuracy \\ \hline 
     \cite{ran2019sharear} & Shared multi-user AR & Implementation and evaluation of the system & Fine-grained control approach & Accuracy of Virtual Object Pose, jitter, drift, end-to-end latency \\ \hline
     \cite{mahzari2018fov} & Adaptive $360^{\circ}$ VR streaming & FoV-aware caching replacement & Probabilistic learning &  High quality for FoV, number and duration of rebuffering events\\\hline
     \cite{zhang2019exploiting} & Cinematic VR services in real-time & VR caching replacement on HMD & Linear regression & Warping distance and consistency of VR experience \\ \hline
     \cite{seo2021novel} & MEC-enabled MAR services & Joint power and mobile cache management & Optimization approach & Service latency, energy consumption \\ \hline
     \cite{boabang2021machine} & The 5G-based tactile Internet remote robotic surgery & Prediction of lost or/and delayed content & Gaussian process regression & Relative error, prediction time  \\ \hline
     \cite{hou2018burstiness} &  URLLC's UL transmission in the Tactile Internet & Burstiness-aware reservation of bandwidth & Unsupervised learning & Bandwidth efficiency \\ \hline
     \cite{yan2022resource} & Semantic-aware wireless networks & Semantic-aware resource allocation & Optimization method & Semantic spectral efficiency \\ \hline
     \cite{liew2022economics} & Semantic communication system in wireless-assisted IoT & Energy allocation problem & DL-based auction & Sentence similarity score and BLEU score of the hybrid access point and IC, IoT, or IR devices\\ \hline
 \end{tabular}
\end{table*}

\subsection {Data Rate, Reliability, and Latency of 3D Multimedia Networks}
Networking is provided via communication infrastructure based on heterogeneous wired and wireless networks. However, the TCP/IP architecture needs a novel design to forward messages about the virtual world via immersion. Table II shows that traditional TCP/IP communication may not satisfy the stringent QoS requirements for Metaverse applications.
Moreover, the immersive and flawless telepresence experience provided by the Metaverse through VR/AR puts high expectations on the network and communication infrastructures of mobile edge networks in terms of latency, reliability, and data rate \cite{saad2019vision}. In the following subsection we discuss the literature related to resource allocation techniques for different Metaverse applications.

\subsubsection {Resource Allocation for VR Broadcasting} VR broadcasting is the most direct method that aids wireless users in getting access to Metaverse. Those who use VR to gain access to the 3D worlds require data-oriented edge networks to ensure low latency, and high reliability and data rate \cite{bastug2017toward}. However, mobile edge and wireless network devices experience resource allocation issues that limit the operation of VR broadcasting services. Trade-offs must be made between various multi-objective functions that addresses the distribution and quality of service (QoS) issues to improve performance and optimize VR operations.

The VR broadcast is transmitted in heterogeneous manner in Metaverse due to the heterogeneous nature of wireless and edge devices \cite{feng2020smart}. This allows the users to access the Metaverse using  macrocell broadcasting, millimeter wave (mmWave) small-cell unicasting, and device-to-device (D2D) multicasting. In macrocell broadcasting, the base station offers broadcast multimedia services (BMS) to users experiencing bad network quality \cite{gimenez20195g}. The main reason for this is that in macrocell multimedia broadcasting, all users within a given coverage area receive the same QoS. In case of poor network connection with macrocell the user can establish the good quality connection with a neighbour mmWave cell for better quality of VR services \cite{elbamby2018edge}. 

Given the above mentioned fact, the authors of \cite{feng2020smart} proposed an intelligent framework for transmission mode selection, that improves the throughput of VR transmission by employing reinforcement algorithm (RL) algorithms for online transmission mode selection and unsupervised learning for D2D clustering, respectively. The performance results showed that their proposed hybrid transmission mode selection framework achieved a much higher data rate than the baseline approaches they considered that relied only on macrocell broadcasting or small-cell unicasting. 
 
Apart from dynamic transmission mode selection for VR services, some other promising solutions for enhancing the experience of users in $360^{\circ}$ VR video streaming includes prefetching and tiling \cite{hu2021virtual}. For instance, the latency of 3D multimedia services might be reduced through the prefetching of VR content. Furthermore, VR content tiling can play a crucial role in saving bandwidth by transmitting only those pixels that have been watched.  However, the performance of both these techniques relies on their field-of-view prediction accuracy and allocated resources. Hence, the balance of the fetched VR scenes and the allocation of communication resources must be considered when streaming to several users. For addressing this complicated issue, Zhang et al.  \cite{zhang2021buffer} proposed an federated learning (FL)-based scheme to train VR viewpoint prediction models in a privacy-preserving and distributed manner. Using their scheme and having efficient VR viewpoint prediction models may reduce the number of unrendered VR tiles that are downloaded, and thus improve the resource allocation efficiency of edge networks significantly reducing the amount of data transferred \cite{chung2018hand}. 

The self-adaptability and scalability of the VR services providing systems at mobile edge networks is a critical consideration for the Metaverse. A joint optimization algorithm is therefore proposed in \cite{guo2020adaptive} for maximizing the quality of experience (QoE) of VR users. In another study \cite{wang2022meta}, the authors developed a VR services provider model that uses visible light communication (VLC) / terahertz(THz) wireless networks for providing indoor services to users with high accuracy. The VLC/THz networks can provide massive bandwidth for the transmission of VR content that requires remarkably higher data rates  \cite{abbas2010constructing}. The proposed architecture enabled users to choose the base station to which they were linked and in opening an appropriate visible light access point. The authors proposed a meta RL algorithm to quickly adapt to new user movement patterns and maximize the average number of users that are served successfully. 

Since edge resources are leveraged in the above-mentioned studies for supporting high-quality content delivery, therefore, it is important to keep in mind the price charged by scarce edge resources to provide service. The economic strategies used to price the services of different entities involved in the Metaverse, including data owners, PSPs, and VSPs, must be considered to facilitate the combined efforts to construct Metaverse. In a recent study, Xu et al. \cite{xu2021wireless} designed a double Dutch auction approach based on learning for pricing and matching VR service providers (for example, base stations) with users in the Metaverse. Under this approach, the service providers and VR users can quickly complete transactions inside the call market to meet short-term VR services demand. The output showed that the proposed approach is capable of achieving near-optimal social welfare by reducing the cost of exchanging auction information. 

\subsubsection {Resource Allocation for Adapting AR} Besides VR broadcasting, one more method for accessing the Metaverse involves mobile AR (MAR) or AR to adapt the real-world environment  \cite{siriwardhana2021survey}. The users in the mobile edge networks use edge-supported AR devices to upload and analyze the environment. After the customization of AR, the suitable AR content is downloaded from the edge server and access to the Metaverse is gained. Nevertheless, edge network resources are limited, while there is an enormous amount of communication and customization resources employed for AR adaptation \cite{qiu2018avr}, \cite{liu2019edge}. In particular, users have to make resource allocation problem must therefore be addressed for effective AR adaptation. In particular, users have to make a trade-off between interaction latency and the adaptation accuracy, i.e., the longer the traversal time, the greater the AR adaptation accuracy.

The mobile edge network can provide two kinds of MAR services, namely approximate AR and precise AR,  to the Metaverse users \cite{liu2018dare}. In approximate AR, only those targets are detected where the AR user's visual attention is focused on, which reduces the edge servers' communication and computation overheads. In precise AR, on the other hand, all the elements in a given image are detected using computer vision algorithms to enhance the user's AR interaction experiences. Precise AR typically requires a greater service delay and more computing capacity than approximate AR does. 

\subsection {Human-in-the-Loop Communication}

For MAR services, service latency and analysis accuracy are reflected in the quality of augmentation (QoA). Liu et al. \cite{liu2018dare} proposed a framework to dynamically configure adaptive MAR services in order to maximize the QoA and service latency experienced by users with different computing capabilities and networking conditions. Under their proposed framework, the environmental content can be uploaded by MAR users based on the network conditions and service requirements. An 
object detection algorithm \cite{redmon2017yolo9000}, \cite{liu2016ssd} is selected by the service provider on the basis on the computation resources that are available and then the resources are allocated for processing the environmental details uploaded by MAR users following MAR request analysis. The AR configuration information is sent back by the AR service provider to the MAR user as per its computation capacity. The video frame rate and frame size are adjusted by the MAR user in accordance with its tolerable latency and requirements and the AR service provider's description of its computation resources and processing model and the modified AR configuration information is uploaded back to the AR service provider. Eventually, the AR user's sent content is processed by the AR service provider in accordance with its preferred configuration and the resultant AR content is sent back to the MAR user. This way, the AR service provider can provide the MAR user with the computing resources and processing model it needs by dynamically adjusting the MAR user's video frame rate and frame size. 

Generally, the greater the AR service latency, the greater the service analysis accuracy. For achieving performance improvement over edge MAR systems, Liu et al. \cite{liu2018edge} developed a network coordinator for balancing the service latency and analysis accuracy trade-off. The authors relied on the block coordinate descent approach \cite{grippo2000convergence} for developing a multi-objective analysis algorithm, which permits accurate and quick AR content target analysis. Their algorithm makes it possible to enhance the service provider's the resource allocation efficiency and in turn better serve the MAR users having different frame rate and frame size requirements. Moreover, in popular regions, multiple users simultaneously request AR services from AR service providers. To extend MAR to multi-user scenarios, Edge AR X5, a multi-user cooperative MAR architecture for edge networks, is proposed in \cite{ren2020edge}. The authors of \cite{ren2020edge} presented a multiple user prediction and communication protocol that takes economic computation, efficient communication, and cross-platform interaction requirements of heterogeneous AR users in edge networks into consideration. Furthermore, to improve sharing and interactivity in MAR, Ran et al. \cite{ran2019sharear} proposed a shared AR architecture, through which user's social acceptance and immersion in the Metaverse can be significantly enhanced by the MAR services providers \cite{duan2021metaverse}.

\subsubsection {Edge Caching for VR/AR Content} Low-friction VR/AR interactions are needed to immerse users in the shared 3D worlds of the Metaverse. In mobile edge networks, the VR/AR content caches might be placed on edge devices and edge servers to reduce transmission overheads  \cite{hennessy2011computer},\cite{paschos2018role},\cite{li2018survey},\cite{gao2020design}. Nevertheless, these shared worlds pose a novel challenge for distributed edge caching. The issue of effectively using edge cache while ensuring Metaverse content uniformity across the geographically-distributed edge servers needs to be addressed. Caching resources can be leveraged by the base station to improve the QoE of wireless users \cite{wang2017survey}. The majority of the provided services in the Metaverse are closely linked with the context in which the content has been provided, like DT and VR/AR. Such services might benefit from caching popular content on the edge servers for ensuring efficient resource allocation and low latency. In the Metaverse, massive amount of heterogeneous edge data is received by the caching devices on the edge. It is therefore still a formidable challenge to build an integrated Metaverse with distributed caching for mobile edge networks. 

Panoramic VR videos can be cached by the edge servers in mobile edge networks to reduce the consumption of bandwidth through a user's  field-of-view (FoV)-based adaptive $360^{\circ}$ VR streaming solution \cite{mahzari2018fov}. In \cite{mahzari2018fov}, the authors permitted the edge servers to utilize probabilistic learning for developing caching policies based on FoV for VR video delivery through current user preferences and historic VR video viewing data. Their results demonstrated that the bandwidth-savings/cache-hit ratio can be improved by a factor of two by using their proposed FoV aware caching approach, thereby reducing the load on the networks bandwidth and minimizing the networks latency. 

A VR user's behavior (such as rapid head movements or changes in position) might also affect the QoS of VR. Particularly, the VR content being streamed may experience playback quality degradation and unwanted jitter if the predictions of these movements is not considered. Zhang et al. \cite{zhang2019exploiting} recently proposed a novel approach for VR content caching, in which wireless Head Mounted Displays (HMD) caching is used to address the above-mentioned problem. In there, the authors used linear regression model for motion prediction to reduce the latency of VR services. The HMD was used to render the image based on values predicted for the user in question and then used it for VR services. Seo et al.\cite{seo2021novel} provided an architecture for AR content cache management on mobile devices to optimize the transmission power and mobile cache size, and introduced an energy consumption and service latency optimization problem for acceptable MAR service provision. 

\subsubsection {Holography} 
Another way users can gain access to Metaverse is to use wireless holographic communication for projecting all the details about themselves into the physical world. Holography \cite{huo2017wireless} is a complicated approach that is reliant on the diffraction and interference of the visible light with the objects for recording and reproducing the phase and amplitude of optical wavefronts, which include digital holography (DH), computer graphic holography (CGH), and optical holography. In \cite{huo2017wireless}, Huo et al. presented an unequal error protection-based coding scheme for forward error correction for optimizing the QoS of wireless holographic communication at varied coding levels. Moreover, they also introduced an immersive wireless holographic type communication mode for rendering light field \cite{karafin2017support} interaction,  (for instance, point clouds \cite{mekuria2016design} of remote holographic data over wireless networks).

The Metaverse is considered to be a massive machine-type communication supported human-centric 3D virtual environment. Human-centric applications, for instance, healthcare over the tactile Internet, immersive VR gaming, and holographic AR social media, are anticipated to drive the next-generation wireless systems \cite{fettweis2014tactile}. The development of simulated XR systems requires the interdisciplinary examination of human body physiology, cognition, and perception for haptic communications \cite{simsek20165g},\cite{hirche2012human}, \cite{antonakoglou2018toward} along with the incorporation of engineering in service provision (storage, computation, communication). In haptic communication, haptic refers to both tactile perception (understanding of the friction and surface texture sensed by varied kinds of mechanoreceptors present in the human's skin) and kinesthetic perception (having knowledge of velocity, position, torque, and the forces that are felt by tendons, joints, and muscles in the human body). Moreover, a few new metrics, e.g., quality of physical experience (QoPE) \cite{saad2019vision} are presented for assessing the communication quality and evaluate service provider performance in the Metaverse. A major attribute of these metrics involves blending conventional QoE inputs (average opinion score) and QoS inputs, such as reliability, data rate, and latency, with the human users' physical factors, i.e., psychological and physiological perceptions. Furthermore, several human-related activities, including gestures, physiological experience, and cognition, influence the factors that affect QoPE. 

\subsubsection {The Tactile Internet with Haptic Feedback} Contrary to the classic multimedia applications (video and audio) of the Internet, the services within Metaverse facilitate users in immersing themselves in the 3D environment in a multi-sensory, holistic manner. Luckily, the materialization of the Metaverse, wherein users are linked through responsive and reliable networks that enable real-time interactions, has become possible through the advanced tactile Internet \cite{sharma2020toward}. The 3D Metaverse world needs kinematic and haptic interaction along with the $360^{\circ}$ auditory and visual content for a completely immersive experience. In the teleoperation of the Metaverse, the tactile information transmission is more sensitive to the system's latency and stability and has a stringent limits about 1ms end-to-end latency \cite{lawrence1993stability} as a supplement for audio and visual information. Moreover, apart from the gaming sector, such advancement in haptic-visual feedback control might alter the human approach for communication in today's world, i.e. human-in-the-loop communication \cite{duan2017human}.

\textbf {Edge Testbeds for Tactile Internet:}
Even though the tactile Internet is capable of revolutionizing the wireless networking future, however, its large-scale deployment is still not expected in the near future. Research in this domain is in the testing phase and is facing two main challenges: 1) different deployment levels among different tactile Internet disciplines, and 2) a lack of consensus on the tactile Internet performance. To address these concerns, Gokhale et al. \cite{gokhale2020tixt} designed a scalable, modular, and universal testbed. The authors gave two examples of tactile Internet in the virtual and real spaces for providing a proof-of-concept for their presented testbed. Another study \cite{polachan2019towards} developed an open tactile cyber-physical systems (TCPS) testbed that is focused on the swift prototyping and assessment of the TCPS applications. Contrary to the previously mentioned testbed that lacked evaluation tools, the authors equipped this testbed with means for assessing two major attributes of TCPS, i.e. management performance and latency. 

\textbf {Resource Allocation Schemes for Tactile Internet:} The tactile Internets, highly demanding good communication latency poses a novel challenges for the allocation of resources within the mobile edge networks. As a medium for the real-time transmission of motion and touch, the tactile Internet in the Metaverse, alongside VR/AR, might offer multi-sensory multimedia services for the users. Unlike VR/AR, which require significant computational and bandwidth resources, the tactile Internet demands low and very reliable latency for the human-computer and human-human interactions and is highly sensitive towards the jitters in the network \cite{sarathchandra2021enabling}. Thus, Boabang et al. \cite{boabang2021machine} proposed a prediction algorithm for lost or delayed data, that uses scalable Gaussian process regression (GPR) to predict content. However, the resource allocation in the edge network becomes more complicated because of the excessive and rapid packets arrival over the tactile Internet. In light of this, Hou et al. \cite{hou2018burstiness} formulated an optimization problem to minimize the amount of bandwidth reserved with under reliability and service constraints. They proposed a technical data-based strategy and an unsupervised-learning-based model to cluster the fractional arrival procedure of users in order to improve the longitudinal spectrum efficiency and reduce classification error in wireless networks. 

\subsubsection {Goal-Aware/Semantic Communication} Metaverse is a complex system, which includes several semantically-oriented applications. The content dissemination services of the Metaverse can put a huge load on existing data-oriented communications \cite{strinati20216g}, wherein a channel having massive capacity is required by the real-time traffic. Thus, Metaverse applications need service-level optimization and service diversity to reduce the burden on wireless channels in edge networks. In  human-in-the-loop communication, the original data is not simply presented to the users in bits, however, it involves semantic structures for presenting information to users. In semantic communication,  the reasoning tools and knowledge representation are combined with ML algorithms to pave the path for semantic coordination and knowledge modeling. 

\begin{figure}
  \begin{center}
  \includegraphics[width=8.5cm, height=8cm]{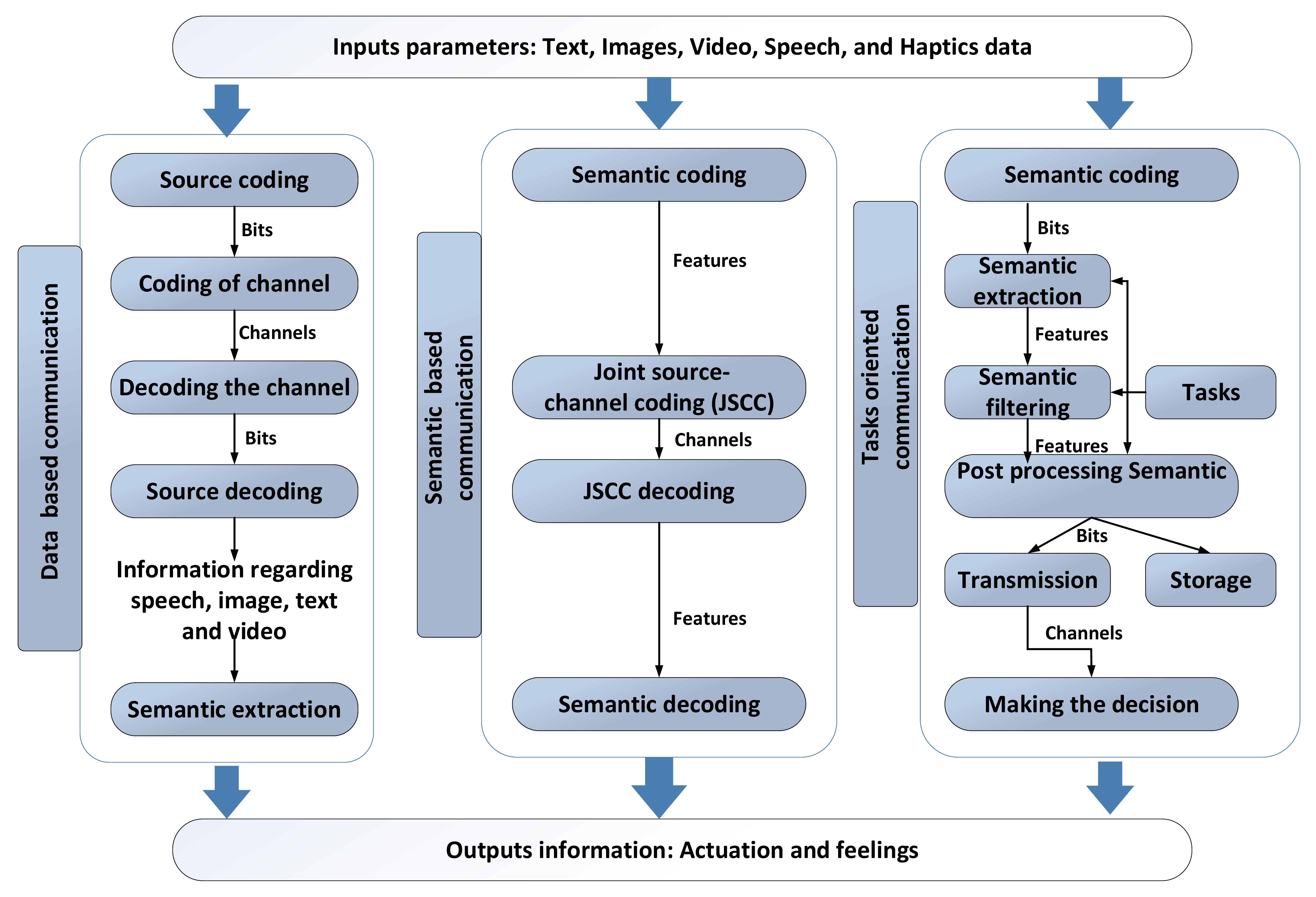}\\
  \caption{Comparison among goal-oriented,semantic-oriented, and data-oriented communication systems 
  }\label{comparison}
  \end{center}
\end{figure}

Fig.~\ref{comparison}, shows the three components of semantic communication in a semantic-aware communication system \cite{qin2021semantic}, \cite{yang2022semantic}:

\begin{itemize}
    \item The semantic information is detected and identified by the semantic encoder and the unnecessary information from the ongoing conversation is compressed or removed. 
    \item  The received semantic information is decoded by the semantic decoder and is restored in such a  way that it can easily be comprehended by the target user.
    \item The semantic information in transmissions is interrupted by semantic noise, which causes misperception or misunderstanding at the receiver. This is quite common in semantic decoding, transmission, and encoding.
\end{itemize}

Semantic communication has the potential to overcome the Shannon capacity limit \cite{qin2021semantic}. It is an intelligent solution at the physical layer, where only the information needed for a particular task is transmitted. For instance, deep neural networks (DNNs) have been widely used to detect and extract needed information from signals in wireless communications. A recent study \cite{qin2019deep} took that one step further and proposed a robust end-to-end communication system to facilitate intelligent communications at the physical layer. Moreover, for the goal-oriented communication  \cite{kalfa2021towards} illustrated in Fig.~\ref{comparison}, Bourtsoulatze et al. \cite{bourtsoulatze2019deep} proposed a joint source-channel coding (JSSC) scheme to extract and transmit semantic attributes in messages for performing particular tasks. 

Contrary to the classic bit error rate in data-oriented communications, in semantic communication, the important factor of human perception needs to be considered as one of the performance metrics. A new metric, named sentence similarity, is presented in \cite{xie2021deep} for measuring the semantic error of transmitted sentences in semantic communication systems. The authors applied  natural language processing to the decoding and encoding process of physical layer communication based on this metric. More specifically, they used deep learning-based machine translation schemes to extract and interpret the messages being transmitted. Their results showed that their proposed system transmits information more efficiently than the benchmark system, particularly at a low signal-to-noise ratio.
 
The deployment of these communication systems in edge networks requires effective resource allocation to reduce the computational resources consumption in semantic communications \cite{yang2022semantic}. Yan et al.\cite{yan2022resource} noted that the resource allocation problem as it pertains to semantic networks had not been explored much and thus presented an optimization method for semantic spectral efficiency (S-SE) maximization. Liew et al. \cite{liew2022economics}, for their part, proposed a deep learning (DL)-based auction to allocate energy to IoT devices. They proposed a bilingual evaluation score (BLEU)  \cite{papineni2002bleu} and sentence similarity score-based evaluation scheme for semantic communication systems. Their results showed that the incentive compatibility (IC) and individual rationality (IR) results were sufficient to maximize the revenue of semantic service providers. 

\begin{figure}
  \begin{center}
  \includegraphics[width=9cm, height=8cm]{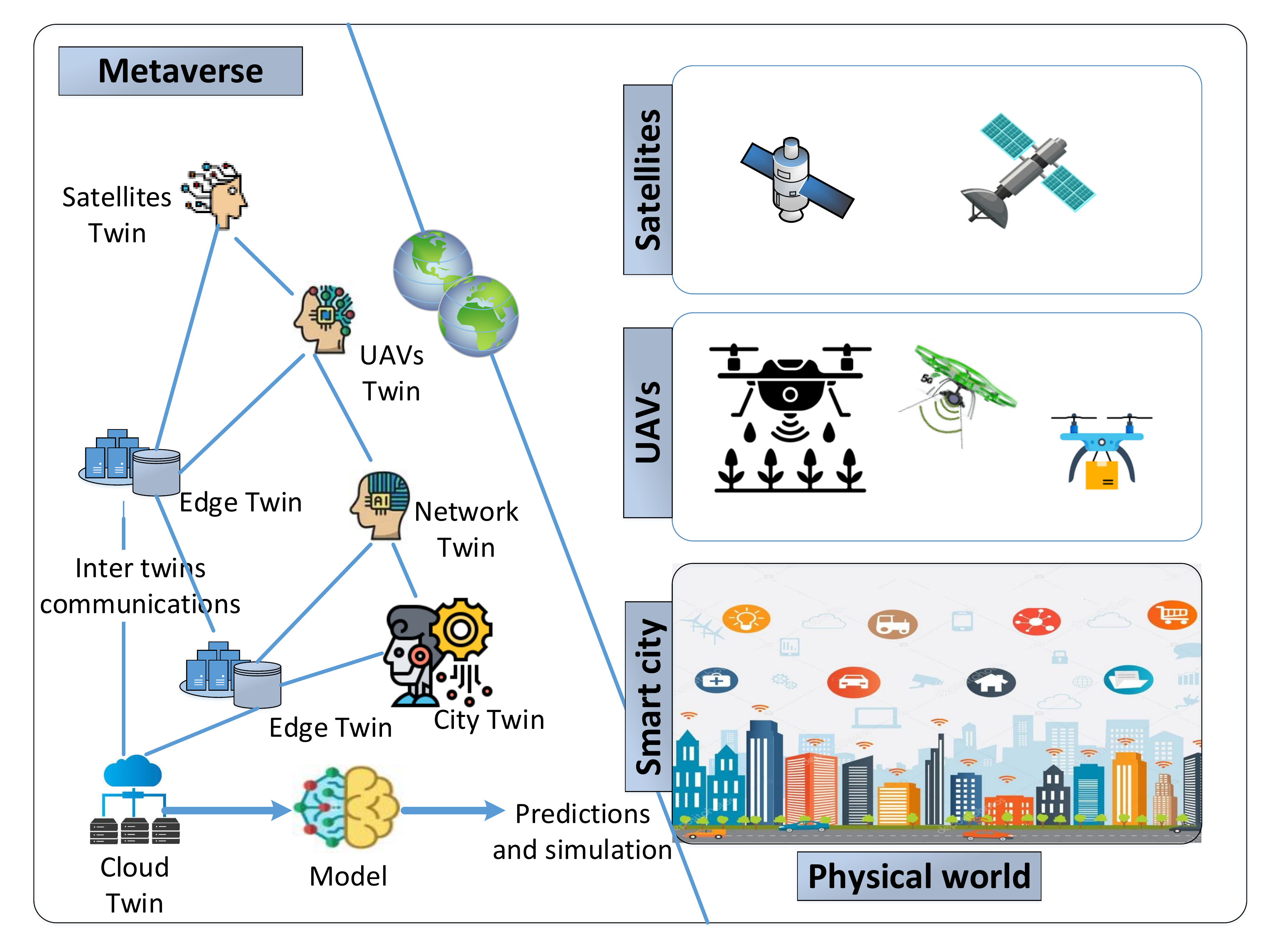}\\
  \caption{Real-time virtual/physical synchronization
between the intelligent edge networks and the Metaverse 
}\label{sync}
  \end{center}
\end{figure}

\subsection { Real-Time Bidirectional Synchronization Between the Virtual and Physical Worlds} 

The Metaverse includes several virtual spaces that are based on fractions on the real/physical world  \cite{duan2021metaverse}. For instance, digital copies of physical offices can be built by the employees in the Metaverse using DT technology to facilitate working from home. For maintaining bidirectional synchronization between the physical world and Metaverse in real-time, extensively distributed edge servers and edge devices are needed for actuation and sensing. Fig.~\ref{sync} shows that DT, such as copies of smart factories and city twins, is one promising solution for such virtual services \cite{han2021dynamic}. Moreover, the Metaverse could be employed for improving the efficiency of edge networks through the virtualization of real-world edge networks. More specifically, users could directly calibrate and control their physical units in the real world through Metaverse by monitoring the DT of edge devices and infrastructures, which include space-air-ground integrated network (SAGIN), UAVs, and IRS \cite{cheng2019space}, \cite{cheng2018air}.

\subsubsection {Resource Allocation for Synchronizing the Virtual and Physical Worlds} 
Real-time smooth immersion in the Metaverse is made possible mainly by effectively allocating resources in internal communications between the virtual and physical entities in mobile edge networks while synchronizing the virtual and physical worlds \cite{shen2021holistic}. Considering this, Han et al. \cite{han2021dynamic} designed a data collection incentive mechanism for edge devices. In their proposed approach, the IoT devices functioned as PSPs for collecting data in the physical world. A dynamic ground-breaking strategy is used in this approach that helps device owners in updating their policies by entering into several contracts. Sun et al. \cite{sun2021dynamic} extended \cite{han2021dynamic} by taking dynamic internet of vehicles (IoV) scenarios into consideration. They proposed a distributed incentive method for incentivizing vehicles that contribute their resources to the synchronization of virtual and physical worlds in the Metaverse. This method can maximize the overall energy efficiency of vehicular twin networks in dynamic scenarios.

\subsubsection {Physical/ Virtual Synchronization Through Intelligent Edge Networks} DT technology offers the classic edge networks a potential smart solution for enhancing their performance via physical/virtual synchronization \cite{khan2022digital},\cite{li2021slicing}. For instance, DT coordination and communication in the Metaverse can improve the energy efficiency of edge networks in the real world. In \cite{dong2019deep}, the authors designed a DT that models the network topology for mirroring the mobile edge computing systems. The system's optimization target
is to support its delay-tolerant and ultra reliable low latency communications (URLLC) services as energy efficiently as possible. The authors used DT data to train a DL framework for jointly determining offloading probability,  resource allocation, and user grouping so as to reduce the consumption of energy in the system. Furthermore, with the increase in the diversity of mobile applications and the complexity of edge networks, fine-grained services can be provided to mobile users by slicing edge networks. Wang et al.  \cite{wang2020graph} proposed a DT-based smart network slicing framework, that uses graph neural networks to capture the interconnection relationship within slices in varied physical network scenarios, and employed DT for monitoring the end-to-end scalability metrics of the slices.

Reconfigurable intelligent surfaces (RIS) is an important technology for 6G communications, that has the potential of transforming the wireless environments by phasing RIS elements \cite{renzo2019smart}. However, restricting the RIS phase and the beamforming signal's power allocation remains an open challenge. In light of this an environmental twin (Env-Twin) scheme is presented in  \cite{sheen2020digital}, that virtualizes wireless network infrastructure into a DT network. The optimization of the wireless edge environment can be done as an intelligent edge environment by controlling and monitoring different spectrum ranges (such as Thz, mm waves, and VLS), cell layouts (i.e, non-terrestrial networks, terrestrial networks, small cells, macro cells), and RIS in complicated edge networks. In the proposed Env-Twin scheme, the most suitable RIS configuration of each new receiver site in a given wireless network is predicted using DL techniques to maximize spectral efficiency in the edge networks. 

To provide Metaverse users with broad coverage, the RIS, UAVs, and IoV in edge networks have to create a SAGIN with satellites \cite{zhou2020deep}. Recently, Yin et al. \cite{yin2021physical} implemented a DT network for SAGIN in a virtual environments to enable reliable vehicular communication in a SAGIN. The real physical entities may ensures the secure and efficient transmission of information within vehicles through communication with virtual entities in the virtual environment. Moreover, even though spectrum sharing can improve spectrum utilization in a SAGIN, it still remains at risk of malicious threats and eavesdropping. To address this issue, a mathematical model is proposed by the authors for the secrecy rate of the base station-vehicle link and the satellite-vehicle link. They used planning and relaxation approaches to solve the secure communication rate issues in SAGINs. Aside from that another issue that is faced by VSP in the Metaverse is DTs synchronization. To address that authors in \cite{han2022dynamic} evaluate the temporal value decay dynamics for DTs and observes how they affect the VSP synchronization. Furthermore, they uses UAVs to collect the most updated state of the physical body for DTs. Then optimal optimal controls for VSPs were determined using Nash algorithms and adopted differential game to provide solution to upper level problem.

\subsection {Summary and Lessons Learned}
\subsubsection {High-Performance Networks for Real-Time User Interaction and Multimodal Content Delivery} The combination of tactile Internet, AR, and VR drives the Metaverse, which includes the multimodal delivery of content that incorporates touch, audio, and images. Furthermore, contrary to the classic AR/VR that is focused on scenarios involving a single user, the Metaverse involves multiple users that co-exist and interact with one another in virtual worlds. There is a critical need for effective resource allocation to optimize the delivery of services to a large number of users on the edge for these purposes. Moreover, the rendering of 3D FoVs and high quality user interactions in real-time will requires extensive calculations. 

\subsubsection {AI for Smart Edge Communication }: AI techniques are required for efficient communication resource management in dynamic and complex network environments. Some of the work reviewed used AI for improving the efficiency of resource allocation at the edge, for instance, efficient bandwidth allocation and task offloading. AI can be used to supplement classic optimization tools for pricing PSPs' services, for example, by increasing auctions convergence and reducing the cost of communication. However, the issue remains that training and storing AI models is quite computationally costly on resource-constrained devices. 

\subsubsection {Semantic-Aware Content Popularity} The Metaverse is a self-sustainable and scalable 3D embodied Internet having huge semantic-aware details about the personalized avatars of users. In the Metaverse, virtual services are deployed online, which means that the precise forecasting of semantic-aware content popularity may enhance the efficiency of communication in mobile edge networks. For instance, different kinds of semantic-aware content are requested by the users from VSPs while attending different events. Moreover, different physical environment contexts, like the degree of mobility and location; networking contexts, like network topology and connection congestion; and social network contexts, like social moods and neighborhood affect the popularity. 

\subsubsection {Human-Centric Service Provisioning} The approach for evaluating the user experience has gradually evolved from being a supply-side approach to a demand-side approach in human-in-the-loop communication. For instance, the service quality in human-centric applications is evaluated by the users based on their perspectives. However, human experience modeling is a complex interdisciplinary issue that is subject to several sources of uncertainty, e.g, users' emotions. Thus, future VSP and PSPs should considers all the factors while accurately modeling the users’ perceptual utility function with human-in-the-loop variate.

\subsubsection {Break-in-Presence Feeling} Unlike current URLLC service, which need redundant resources to prevent a break-in-presence feeling. In the Metaverse, the URLLC should be supported by PSPs in emergencies or extreme cases with temporally and spatially charging traffic patterns, device densities, and infrastructure and spectrum availability.

\section {Privacy and Security Threats to Metaverse}

\begin{figure*}[!ht]
  \begin{center}
  \includegraphics[width=7.5in]{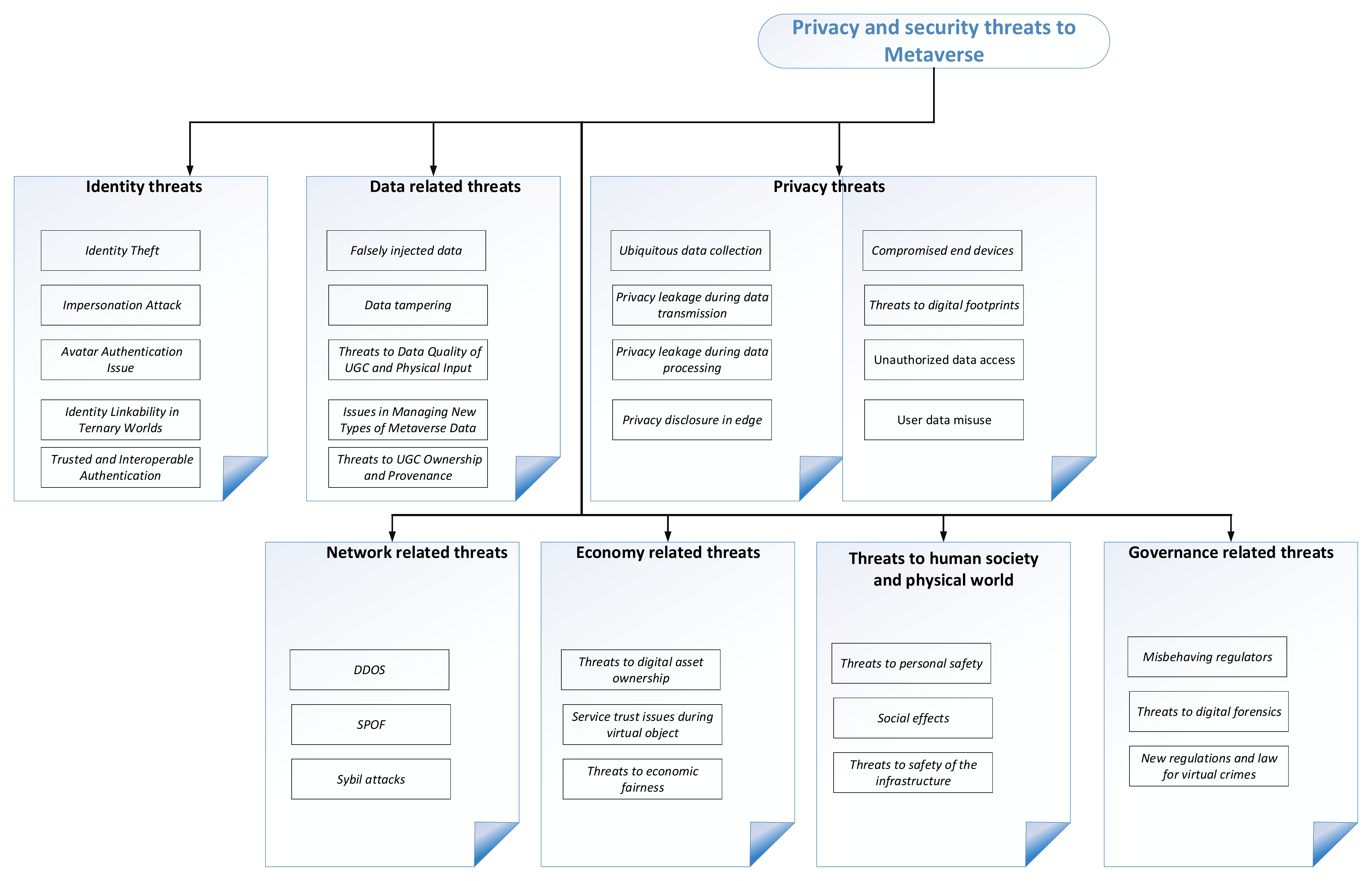}\\
  \caption{ Security threats in Metaverse: A taxonomy 
  }\label{security-threats}
  \end{center}
 \end{figure*}

So far, we have discussed the building blocks of Metaverse and the terminologies that will help in developing an efficient Metaverse architecture. This section now looks into the challenges that Metaverse will face from another perspective that are security and privacy threats. In this section we discuss Metaverse security and privacy threats related to governance, social/physical effects, the economy, privacy, data, and identity. The taxonomy of these threats is shown in Fig.~\ref{security-threats}. 

\subsection {Identity-Related Threats}
Identity management plays an important role in the Metaverse because, the identity of avatars/users can be impersonated and stolen. Moreover, there could be interoperability problems during the authentication process across virtual worlds.

\subsubsection {Issues Related to Avatar Authentication} In comparison to identity authentication in the real world, the avatars' authentication for Metaverse users might be trickier because it may involve the verification of video footage, and one's voice, and facial features, among other details. Moreover, adversaries can create AI bots in the virtual world (for instance, Roblox) that behave, hear, and look the same as the real avatar of the user by copying the behaviors, voice, and appearance of the user \cite{falchuk2018social}. Consequently, supplementary personal details may be needed as proof of identity to ensure the secure authentication of avatars, which might also lead to new privacy breach problems.

\subsubsection {Interoperable and Trusted Authentication}  Metaverse avatar/user identity authentication must be trusted, efficient, and fast across virtual worlds and service domains, constructed on unique platforms like blockchains \cite{dionisio20133d}. 

\subsubsection {Impersonation Attacks} The impersonation attack can be launch by an attacker by pretending to be an authorized body to gain access to a system or service in the Metaverse  \cite{greitzer2008combating}. For example, an adversary can create fake digital copies of users and used them for malicious activity, like fraud and or even committing crimes against other Metaverse users.

\subsubsection {Identity Theft} In the case of stolen user identity in the Metaverse, the user's avatar, digital life, social relationships, and digital assets can be lost and leaked. Identity theft could be more critical in the Metaverse than in conventional information systems. For instance, users' personal information of the users (such as their banking details, secret keys for digital assets, and full names) can be stolen by hackers in Roblox through authentication loopholes, phishing email scams, and hacked personal VR glasses for committing crimes (such as stealing victims, digital assets and avatar) and fraud \cite{wang2022survey}. 

\subsubsection {Identity Linkability in the Ternary Worlds} As the reality is assimilated by the Metaverse into itself, the virtual, physical, and human worlds are smoothly incorporated into the Metaverse, resulting in identity linkability issues in the ternary worlds \cite{hendaoui20083d}. For instance, in Roblox, a malicious user $A$ can identify the position of user B's real body searching for their name in the Metaverse and obtaining their personal details.

\subsection {Data-Related Threats}
The data generated or gathered by avatars/users and wearable devices in the Metaverse is vulnerable to several risks including intellectual property violation, provenance/ownership tracing, low-quality UGC, false data injection, and data tampering. 

\subsubsection {Falsely Injected Data} The adversaries can inject false details, e.g. wrong instructions and false messages for misleading the Metaverse systems \cite{liang20162015}. For instance, AI-assisted content creation may facilitate in improving the immersiveness of users in the early-stage Metaverse, and attackers may inject poisoned gradients or malicious training samples during distributed or centralized AI training, respectively, for generating biased AI models. The false instructions or feedback that are returned might threaten the safety of people or physical equipment. For example, by manifesting the level pain being felt by the users in Metaverse game for being shot, the user might expose to severe consequences or, in the worst case scenario, might lead to the human users' death.

\subsubsection {Data Tampering} Identity features ensure the efficient monitoring and identification of alterations in data during communication between sub-Metaverses and ternary worlds. The attackers may remove, replace, forge, or modify raw data during the lifespan of Metaverse data services for interfering with the usual activities of the physical entities, avatars, or users \cite{su2020lvbs}. Furthermore, the attackers might go unnoticed by distorting corresponding message-digest outcomes or log files to hide the traces of their illicit activity in the virtual environment. 

\subsubsection {Threats to Physical Input and UGC Data Quality} The selfish avatars/users may contribute poor-quality content to the Metaverse in UGC mode, like unreal experiences in synthetic environments, to reduce their expenses  \cite{guo2017availability}. This can be achieved by using uncalibrated sensory devices, which often generates low-quality data, and leads to the creation of poor DT model in Metaverse, and could result in undermining users' experience.

\subsubsection {Problems in Managing New Forms of Metaverse Data} Unlike today's Internet, new devices and hardware is required by the Metaverse for gathering several new forms of data (for instance, head movement, facial expression, and eye movements) that has never before been collected to give users a fully immersive experience  \cite{kumar2008second}. Moreover, in the Metaverse, fingerprints, iris biometrics, and other user-sensitive information might be captured by the end devices (such as haptic gloves and VR glasses). The collection, management, and storage of these massive user-sensitive details that may be collected could pose new challenges for the physical/cyber security of Metaverse devices.

\subsubsection {Threats to UGC Provenance and Ownership} Contrary to the real-world has a government-supervised procedure of asset registration, the Metaverse is a completely autonomous and open space with no centralization control. This lack of control makes it difficult to trace the provenance and ownership of several UGCs generated by a massive number of avatars in different sub-Metaverses and to turn UGC into secured assets \cite{liang2017provchain}.

\subsection {Privacy Threats}
While having digital live experiences in the Metaverse, users' privacy, including private details about their lifestyle, habits, and location could become compromised during the life-cycle of different data-related application services, such as the storage, processing, collection, governance, and transmission of data. 

\subsubsection {Ubiquitous Data Collection} Ubiquitous user profiling at an irrationally coarse level, which includes brain wave patterns, biometric and speech features, hand/eye movements, and facial expressions, is required for immersively interaction with avatars \cite{falchuk2018social}. Moreover, it can be helpful when tracking users and analyzing users' attributes and physical movements using advanced HCI and XR technologies \cite{shang2020arspy}. For instance, the built-in cameras and motion sensors in the Oculus helmet make it possible to track our head movement and direction, draw our rooms, and monitor our environment and position in real time with high accuracy, when browsing and interacting with other avatars in Roblox. The hacking of this device by attackers may lead to serious crimes based on the massive amount of sensitive information they collect. 

\subsubsection {Privacy Leakage during Data Transmission} The Metaverse systems involve the transmission of the massive amount of sensitive personal information that is collected via wearable devices be transmitted through wireless and wired communications. Unauthorized services and individuals should not be allowed to access that confidential information \cite{ometov2016facilitating}. Even though the information is transferred confidentially and communication is encrypted, attackers might eavesdrop on the particular channel to access raw information or even use advanced inference attacks \cite{wasserkrug2008inference} and differential attacks \cite{wei2020ldp} to track users' location. 

\subsubsection {Privacy Leakage During Data Processing} It is essential to aggregate and collect the massive amount of data that is gathered from humans and their surroundings by Metaverse services to create and render virtual environments and avatars, wherein sensitive data of the users might be leaked in the process \cite{li2021verifiable}. Moreover, attackers might be able to infer private details about users by analyzing and linking of the published processing outcomes (for instance, synthetic avatars) in different virtual environments like Fortnite and Roblox.

\subsubsection {Privacy Leakage in Edge/Cloud Storage} Storing sensitive and private information about a massive number of users on edge devices or cloud servers may result in privacy leakage problems. For instance, adversaries may carry out distributed denial-of-service (DDoS) attacks to compromise the edge/cloud storage \cite{bertino2017botnets} or differential attacks for deducting private details about users through frequent queries \cite{wei2020ldp}. One such attack happened in 2006 and the hacker was able to obtain personal information regarding the users such as addresses and username as well as more sensitive information like their payment details \cite{kirkpatrick2006metaverse}.

\subsubsection {Compromised or Rogue End Devices} The Metaverse will involve the placement of a lot of wearable sensors on humans' bodies and surroundings to enable avatars in reflecting facial expressions, capturing hand gestures, and making natural eye contact in real-time. A major threat is that the wearable devices may have a completely authenticated sense of who you are and how you express, feel, behave, and talk. In the Metaverse, compromised or rogue wearable end devices are turning out to be an entryway for malware invasions and data breaches. This issue may become more critical as the popularity of such devices for entry into the Metaverse grows \cite{shang2020arspy}. Under the control of compromised or rogue end devices, the avatars might become a data collection source in the Metaverse, which would lead to user privacy infringement. 

\subsubsection {Threats to Digital Footprints} In the Metaverse, avatars' activities, habits, preferences, and behavior patterns may reflect the actual status of their physical counterparts. Their digital footprints could therefore be collected by attackers and  exploited for precise user profiling and even illicit activities because they are so similar to that of real users \cite{ning2021survey}. Furthermore, in the Metaverse, a third-person view is typically offered with a broader viewing angle of avatars' surroundings than is seen in the real-world \cite{leenes2007privacy}. This might infringe other users' behavior and privacy without their knowledge. 

\subsubsection {Unauthorized Access to Data}
New types of personal profiling data (such as user habits, daily routine, and biometric information) will be generated by the complex Metaverse services. Therefore, various VSPs in different sub-Metaverses are required to access avatar/user profiling activities in real time to deliver flawless customized services (e.g, personalized avatar appearance) in Metaverse \cite{xu2021wireless}. Moreover, attacks could be carried out by malicious VSPs to obtain (and benefit from having) unauthorized access to the data. For instance, malicious VSPs could illicitly increase their data access rights by tampering with access control lists and buffer overflow attacks \cite{yu2018leveraging}. Besides that for smooth operation of Metaverse, excessive amount of personal data is generated and transmitted, which makes it highly complex as to what type of information must be shared and with whom and for what purpose.

\subsubsection {Avatar/User Data Misuse} In Metaverse, data related to avatars/users could be disclosed unintentionally by VSPs or intentionally by attackers in the life-cycle of data services for facilitating targeted advertising activities and user profiling. Furthermore, in the large-scale Metaverse, data misuse may be difficult to trace because some sub-Metaverses might not be interoperable. 

\subsection {Network-Related Threats} Since the Metaverse is evolved from the existing Internet and incorporates existing wireless communication technologies, the conventional threats that communications networks are exposed to might apply to the Metaverse as well. A few typical threats are:

\subsubsection {DDoS} Since the Metaverse will require that numerous small wearable devices be worn, adversaries might take interest in compromising them and incorporating them in a botnet (e.g, Mirai) \cite{bertino2017botnets} to carry out DDoS attacks. These attacks overwhelm the centralized server with a massive amount of traffic in a short period of time, to make service unavailable and cause a network outage. 
\subsubsection {SPoF} The centralized framework (for instance, cloud-based architecture) simplifies the avatar/user management and reduces operational costs during the construction of Metaverse systems  \cite{ali2018applications}. However, it is vulnerable to SPoF caused by DDoS attacks and could damage server operation. Moreover, this raises the issue of transparency and trust concerns in the trust-free exchange of digital assets, virtual currencies, virtual goods, etc. between sub-Metaverses.  
\subsubsection {Sybil Attacks} Sybil adversaries could manipulate multiple stolen/fake identities to gain too much influence \cite{zhang2014sybil} over Metaverse services (e.g., blockchain consensus, reputation service, etc.) or even take over the Metaverse network and thus compromise the system's, the effectiveness. 

\subsection {Economy-Related Threats} In the Metaverse, different types of attacks may pose threats to the creator economy from different aspects including economic fairness, digital asset ownership, and service trust. 

\subsubsection {Threats to Digital Asset Ownership} The complicated ownership (e.g shared ownership and collective ownership), circulation forms that exists and the absence of central authority in the distributed Metaverse bring massive challenges for the ownership traceability, trusted trading, pricing, and generation of digital assets in the creator economy \cite{ritzdorf2018toward}). Backed by blockchain technology, the irreplaceable, tamper-proof, and indivisible NFT has the potential to solve the ownership provenance and asset identification issues in the Metaverse \cite{wang2021non}. Nevertheless, NFTs are also at risk of several threats including phishing attacks, scams, and ransomware. For instance, a single NFT can be simultaneously minted by the attackers on multiple blockchains. Moreover, shares can be cashed out or sold by the evil actors after inflating the NFTs' values to gain benefits without mining anything.

\subsubsection {Service Trust Issues During Virtual Object Trading \& UGC} The avatars in the public Metaverse marketplace can be unreliable elements with no authentic interactions. There is an inherent risk of fraud (e.g, refusal-to-pay and repudiation) during virtual object trading and UGC among different shareholders in Metaverse \cite{de2018swarm}. Moreover, while using DT technology for constructing virtual objects, the reliability and authenticity of the generated and deployed digital copies must be ensured by the Metaverse  \cite{suhail2021trustworthy}. For instance, the adversary can buy the UGCs and other accessories in Roblox and may sell their digital imprints to other users to earn profit. A practical use case of this type of attack can be found in \cite{wang2022survey}, where the Metaverse project initiated by Binance suffer the loss of 1.7 million US dollars due to flaws in smart contracts. 

\subsubsection {Threats to Economic Fairness} In the creator economy, well-planned incentives \cite{zhang2021privacy}, \cite{xu2014collusion} are favorable impetuses to promote user participation and creativity in the trading of digital resources and the sharing of resources.  Following are the three main threats to economic fairness; 
\begin{itemize}
    \item In Metaverse, ] avatars/users may collude with the VSP or with one another for manipulating the market and gaining economic advantages \cite{xu2014collusion}.
    \item Free-riding avatars/users may enjoy the services of the Metaverse and unfairly gain revenue without making any contributions to the Metaverse market  \cite{li2008free}. This could eventually compromise the viability of the creator economy.
    \item In the Metaverse, the digital market might be manipulated by the strategic avatars/users to make huge profits by breaking the demand and supply statuses \cite{zhang2021privacy}. For instance, may inflate bids strategic avatars in Metaverse auctions rather than use true valuation to manipulate the auction market and win auctions. 
\end{itemize}

\subsection {Threats to Human Society and the Physical World} Cyber systems, human society, and physical systems are interlinked in the Metaverse through complicated interactions. This configuration makes Metaverse as an extended version of the cyber-physical social system (CPSS) \cite{zhou2019cyber}. Human society, personal safety, and physical infrastructures are badly affected by the threats in the virtual world. 

\subsubsection {Threats to Personal Safety} Indoor sensors (such as cameras) and wearable devices, e.g, XR helmets, can be attacked by hackers in the Metaverse for tracking the users' position in real time and obtain their daily routine to facilitate bulgary and threaten their safety \cite{casey2019immersive}. Moreover, novel chances for crime and misconduct can also be opened up by the Metaverse. The risk of physical traumas is limited in the Metaverse; however, users may suffer mental stress. In addition, abusive or criminal actions may further increase due to the lack of regulations and laws. 
 
\subsubsection {Social Effects} Even though an exciting digital space is offered by the Metaverse, , it may also have critical side effects in human society, e.g, simulated terrorist camps \cite{d2022terrorist}, cyberstalking \cite{leenes2007privacy}, cyber-bullying, extortion, biased outcomes, child pornography, rumor fueling \cite{zhu2020activity}, and user addiction \cite{valluripally2021modeling}. For instance, the immersive Metaverse could offer terrorists and extremists new opportunities by reducing the cost of identifying new targets, providing novel training methods, and simplifying the meet-up and recruitment process.

\subsubsection {Threats to the Safety of Infrastructure} In a highly incorporated Metaverse, the hackers can detect the vulnerabilities in the system or software, and the compromised devices can be exploited by them as an entry points
\cite{vellaithurai2014cpindex} for invading crucial national infrastructures (such as high-speed rail systems and power grid systems) through advanced persistent threats (APT)  attacks \cite{hu2015dynamic}.

\subsection {Governance-Related Threats}
Like in the real world, where regulations and social norms are in effect, in the Metaverse, interactions between avatars (e.g, virtual economy, during social activities, and creation of content ) must be aligned with the regulations and norms of the digital world for ensuring compliance  \cite{almeida2021ecosystem}. In the Metaverse's governance and supervision process, the security and efficiency of the system could be deteriorated by the following threats. 

\subsubsection {New Regulations \& Laws for Virtual Crimes} It is tricky to determine whether virtual crimes are the same as real ones. Thus, the direct application of rules and laws used in real life are hard for enforcing penalties for criminal acts  \cite{hendaoui20083d}, such as virtual spying/stalking, virtual harassment, abusive language, and so on. 

\subsubsection {Misbehaving Regulators} Regulators could misbehave and paralyze the system as a result. Thus, effective and dynamic reward/punishment mechanisms must be enforced to encourage honesty and discourage misbehavior among regulators. Moreover, reward and punishment rules must be upheld by the majority of the avatars in a democratic and decentralized manner to ensure sustainability \cite{bai2021public}. One promising solution involves the automated implementation of regulations through smart contracts without needing to depend on reliable intermediaries. However, this may lead to new problems, such as vulnerability to re-entry attacks and short address attacks, mishandled exceptions, and information disclosure \cite{sayeed2020smart}.

\subsubsection {Threats to Digital Forensics} In the Metaverse, digital forensics stands for the virtual recreation of cybercrimes through the identification, extraction, fusion, and analysis of evidence acquired in the virtual and real worlds \cite{li2021towards}. However, because several virtual worlds have interoperability problems and are highly dynamic, the effective forensics analysis, which includes tracking anonymous avatars/users exhibiting different behaviors, and connecting entities and behaviors, is quite challenging. Moreover, the unclear boundary between the virtual world and the real world makes it difficult to differentiate between false and true. For instance, the malicious avatars can manipulate their identities using deepfakes to confused law enforcement agencies in Metaverse. 

\subsection {Summary and Lessons Learned}
In this section, we discuss Metaverse security and privacy threats related to seven areas, including identity, governance, social/physical, economy, network, privacy, and data. 

\begin{itemize}
\item Since the Metaverse integrates numerous evolving technologies, it might inherit their complex vulnerabilities and flaws as well. Besides that, the impact of the current threats may be intensified in the Metaverse.
\item The personal details (such as biometrics data and user profile details) gathered and processed in the Metaverse might be quite raw and remarkably pervasive to ensure a completely immersive experience, so the devices used to acquire, transmit, process, and store data must be well-secured. 
\item With the expansion and flourishing of next-generation Metaverse systems will emerge new threats that will require novel defense mechanisms. 
\item The threats that arise in the virtual world could also have an adverse impact in the physical world and threaten human society. 
\item The aforementioned threats could pose massive challenges for Metaverse regulators and lawmakers to address regulation challenges. 
\item Because of the innate Metaverse's attributes (e.g., heterogeneity, scalability, decentralization, and interoperability), a sequence of crucial challenges may arise when applying existing security countermeasures in the Metaverse. Thus, advanced privacy and security solutions that are customized to the Metaverse context are required. 
\end{itemize}

\begin{table*}\centering
\caption{Summary of current/future security countermeasures in the Metaverse}
 \begin{tabular}{||p{1cm}|p{2cm}|p{3cm}|p{3cm}|p{3cm}|p{3cm}||}  \hline
    Ref. & Type of Threat & Security Threat & Employed technologies & Advantages & Limitations \\ [0.5ex] 
     \hline
     \cite{li2017secret} & Identity-related threat & Eavesdropping, RSS trajectory predictions & Secret key establishment based on RSS trajectory for wearables & Protection against eavesdropping and efficient performance in outdoor/indoor scenarios  & Workable only for wearable with short-range communication \\ \hline
     \cite{liu2018cooperative} & Identity-related threat, Privacy threats & Leakage of privacy & MinHash, CP-Abe, bloom filter, edge computing & Offered privacy protection with less system overheads & Did not conduct detailed  real-world evaluations \\ \hline
    \cite{shen2020blockchain} & Identity-related threat, Privacy threats & Eavesdropping, impersonation, and man-in-the-middle & Blockchain & Anonymous identity authentication and low overhead & Slow response speed because of less blockchain throughput\\ \hline 
    \cite{chen2021xauth} & Identity-related threat, Privacy threats & Data tampering, impersonation & Multiple Merkle tree, Blockchain & Fast response, anonymous authentication, and reduced overhead & Lack real-world testing at a large-scale \\ 
    \hline
   \cite{gehrmann2019digital} & Data-related threat & Threats to DT & DT, Cloud computing & Low  synchronization latency and computational cost & Does not ensure the reliability of data collected from different data silos \\ \hline
     \cite{ruth2019secure} & Data-related threats &  Insecure sharing of AR content & Multi-user AR & Feasibility through the validation of prototype on Microsoft HoloLens & Does not protect location privacy in AR applications \\ \hline
     \cite{lv2020industrial} & Network-related threats & Intrusion of VR control system & SVM & High detection and classification accuracy & Incapable of resisting new/unknown types of attacks \\ \hline
     \cite{shahsavari2019situational} & Network-related threats & Harmful activities in distribution grid & Multi-class SVM & Highly accurate labeling of harmful activities & Depends on supplementary expert knowledge for expensive labeling of activities \\ \hline
     \cite{li2019toward} & Economy-related threat &  Economic fairness, free-riding attack & Smart contracts & Enable conditional likability and anonymity & Cannot support group verification of combined dissemination proofs \\ \hline
     \cite{jiang2021cooperative} & Economy-related threat & Fraud in DT construction & DT, Transparent DT, FL & DAG model training and resource trading & Missing scalability and efficiency analysis of the DAG blockchain \\ \hline
     \cite{zhu2020activity} & Social/Physical effects &  Butterfly effect in information spreading & Heuristic greedy & Reduced misinformation interactions and misinformation spreading value & Difficult to be applied to the  time-varying and dynamic Metaverse \\ \hline
     \cite{lau2021coalitional} & Social/Physical effects & High premium stipulation & Cyber-insurance & High protection level with lasting low premiums & Lack dependence analysis of cyber threats and dynamic insurance design \\ \hline
     \cite{he2021datingsec} & Governance-related threat & Abnormal social accounts & Attention-based LSTM & High AUC and F1-score on a real-world dataset collected from Momo & Hard to apply on blockchain-based dating applications \\ \hline
     \cite{zou2018multigranularity} & Governance-related threat & Violation of privacy & Cloud forensics & High detection efficiency of privacy leakage paths on real malware samples & Only limited privacy leakage paths and detection attributes are considered \\ \hline
     \cite{bai2021public} & Governance-related threat &  Centralized governance risks & Stackelberg game, Blockchain & High time efficiency and user utility in decentralized governance & Security and scalability problems in the practical deployment of the system  \\ \hline
 \end{tabular}
\end{table*}

\section {Metaverse Security Enhancement}

This section summarizes some existing and potential approaches to protect against the above-mentioned security and privacy threats in the Metaverse.

\subsection{ Security Threat Countermeasures}
Wearable devices, e.g, HoloLens headsets and Oculus helmets are the expected to be the key terminals for entering the Metaverse. The management of Key, which includes generation, recovery, revocation, update, distribution, and negotiation, is essential for wearable devices for receiving immersive services, delivering sensory information, and establishing secure communication. In light of this, Li et al. \cite{li2017secret} devised a novel scheme for contactless secret key establishment among small wearable devices. They utilized unique wireless channel attributes depending on a given wearable device positioning. In particular, the received signal strength (RSS) trajectories of two moving wearables is leveraged for constructing the secret key by shaking or moving the devices. Their meticulous security assessment proved their scheme protected against eavesdropping, and their results validated their proposed feasibility approach for frequent movements and short-range communications. The channel impulse response (CIR), besides RSS, is also an important physical-layer attribute among the communication parties. In \cite{liu2018cooperative}, the authors designed a secure mechanism for authenticating the identity of wearable devices given the spatio-temporal context, through the combination of ciphertext-policy attribute-based encryption (CP-ABE), bloom filter, and MinHash in the edge computing environments. Their security analysis demonstrated that their proposed scheme preserves cooperative privacy. 

Generally, the Metaverse includes several administrative security domains designed by different standards/operators. The authentication of identity across different administrative areas (e.g., AR/VR services managed by different VSPs) in the Metaverse is crucial for delivering flawless Metaverse services for avatars/users. Conventional cross-domain authentication approaches are heavily reliant on a reliable intermediary and massively increase key management overheads. For addressing this concern, blockchain technology is employed in \cite{shen2020blockchain} for designing a transparent and decentralized cross-domain authentication approach for industrial IoT devices in distinct areas. Shen et al.~\cite{shen2020blockchain} employed a consortium blockchain for establishing trust between different domains and used identity-based encryption (IBE) for authenticating devices. Chen et al. \cite{chen2021xauth} recently, proposed XAuth, an effective cross-domain authentication approach, by exploiting the distributed blockchain consensus. They designed privacy preservation functions and an optimal blockchain scheme that exploit the low throughput of blockchain to protect user privacy and increase response speed. They implement a proof-of-concept (PoC) prototype to prove the feasibility and functionality of their proposed approach. 

 Gehrmann et al. \cite{gehrmann2019digital} discussed the data reliability of DTs in the Metaverse and proposed a reliable state replication approach for synchronizing DTs in industrial applications and identifying major requirements when designing security architecture. A programmable logic controller-assisted PoC implementation was used to validate the proposed approach's efficiency. Nevertheless, the authors did not study the reliability of data gathered through distinct data repositories.

Since most Metaverse applications involve multiple users, for instance, multiple players in remote collaborations and gaming. Keeping that in view. Ruth et al. \cite{ruth2019secure} attempted to devise a secure approach for sharing content in multi-user AR applications. They studied a novel mechanism for controlled content sharing and implemented a prototype on HoloLens to enable co-located or remote users to share AR content with outbound and inbound control. 

Recently, an intelligent intrusion detection model is presented by Lv et al. \cite{lv2020industrial} for detecting attack behaviors on 3D support vector machine (SVM)-based industrial control systems using VR. Their acquired results showed that mean accuracy can be maintained at over 90\%. Nevertheless, new/unknown types of attacks cannot be resisted by their proposed model. In another paper \cite{shahsavari2019situational}, a multi-class SVM classifier is proposed to extract malicious activities from raw metering information. Nevertheless, this scheme is reliant upon the supplementary expert understanding of expensive event labeling. Jiang et al. \cite{jiang2021cooperative} recently introduced  FL-based DT edge networks, wherein the access points (APs) function as edge nodes to help end devices build DT models. They used a directed acyclic graph (DAG) blockchain to safely record global and local model between users and APs. 

Existing approaches for preventing free-riders (those who enjoy the perks of services/goods without making any contributions) are majorly focused on contribution certification, cryptographic schemes, and node behavior modeling. For mitigating free-riding attacks, Li et al. \cite{li2019toward} used Zero-Knowledge Proof (ZKP) and smart contracts in blockchain systems for generating proof-of-ad-receiving commitments with a guarantee of conditional linkability and anonymity. Moreover, the quick dissemination of information in the Metaverse increases the challenges for the known “butterfly effect” in public safety and social governance in the real world. For addressing these concerns,  Zhu et al.  \cite{zhu2020activity}  attempted to reduce the influence of misinformation in online social networks (OSNs) through a dynamic selection of a series of nodes that needs to be blocked from the OSNs. This is workable only in classic static OSNs, however, it is quite difficult to apply in the completely interactive Metaverse with a time-varying and massive social graph structure.

 AI is expected to play a vital part in enabling the fully-automated digital governance of society and individuals in the Metaverse by extensively fusing actuation, computing, and perception. AI techniques could be used to detect Sybil or abnormal accounts and misbehaving entities in the Metaverse. In a recent study, He et al. \cite{he2021datingsec} exploited a multi-factor attention-based long short term memory (LSTM) model for dynamically identify malicious signals from illicit accounts in online applications. The detection accuracy of their proposed approach was found effective through the obtained results on real-world datasets. Moreover, some promising decentralized solutions are offered by blockchain technologies for collaborative governance in the Metaverse, including a simple strategy for automated decentralized governance by smart contracts. Recently a decentralized framework based on blockchain technology was presented in \cite{bai2021public} to encourage active user engagement and witness all administrative procedures. The approach involved dynamically selecting of a verifier group from digital citizens for verifying transactions in hybrid blockchains. The authors devised a private-prior peer prediction method to prevent collusion between the verifiers and designed a Stackelberg game theoretical approach to motivate people to participate. 

In the Metaverse, digital forensics is believed to be an accountability enabler during disputes. It has been scrutinized widely in videos and images. Moreover, it can also be used for privacy violations' accountability. A privacy disclosure forensics scheme with RAM mirroring and taint analysis is proposed in \cite{zou2018multigranularity} to obtain digital proofs without touching the private data of the users in a simulated virtual environment. Nevertheless, there is room for more research with reference to QoS enhancement, collaboration, resilience, and privacy preservation in digital forensics implementation specific to Metaverse applications. 

Table III summarizes the potential Metaverse security countermeasures. 

\subsection{Blockchain for Securing Metaverse}
\begin{figure*}[!ht]
  \begin{center}
  \includegraphics[width=6.5in]{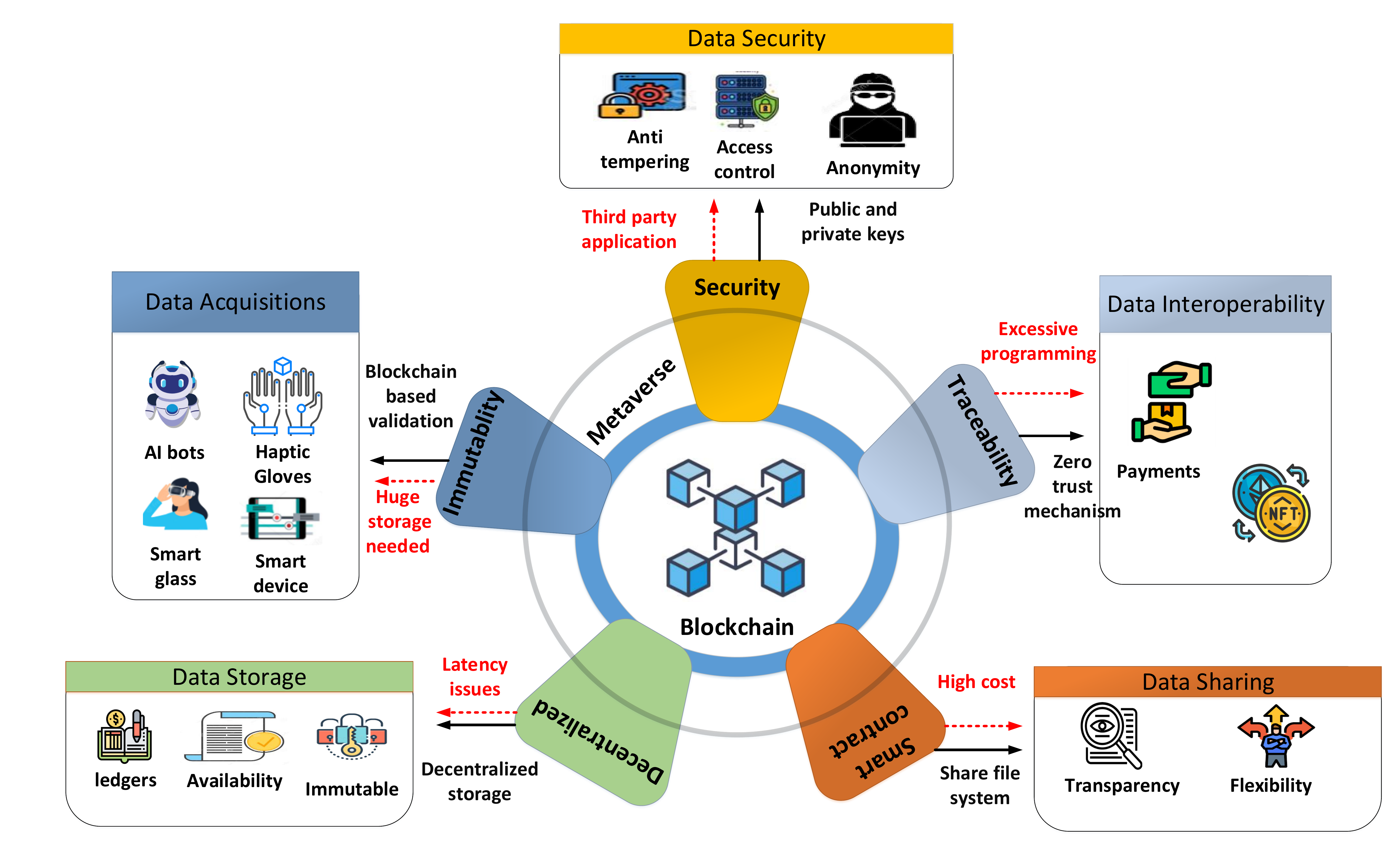}
  \caption{Blockchain for technical aspects in the Metaverse 
  }\label{blockchain-Metaverse}
  \end{center}
 \end{figure*}

Contemporary blockchain-based approaches for securing Metaverse are investigated in this section with reference to data acquisition, storage, sharing, interoperability, and privacy preservation. Fig.~\ref{blockchain-Metaverse} depicts blockchain technology from the aforementioned technical perspectives in the Metaverse. 

\subsubsection {\textbf{Data Acquisition}} 
This is a critical stage in the Metaverse ecosystem. It facilitates the training of machine learning algorithms that may help in marketing, recommendation systems, and digital products development, and decision-making in the Metaverse \cite{wang2022metasocieties}. Data acquisition could help applications to adapt to novel conditions and generate improved insights. But decentralized applications, such as Ampliative Art, 4G Capital, Etheria, and WeiFund will produce an abundance of unstructured real-time data in the Metaverse. The acquisition of such massive data can be a huge challenge. Thus, data integrity will become more crucial in developing certain applications in the Metaverse, such as recommender systems, which will be affected by the acquisition of data from unreliable/unknown sources since such data might impact systems reliability \cite{tao2018secured}.

Blockchain technology can simplify the authentic data acquisition for social networking and other applications in the Metaverse. Moreover, data tracing and transactions verification is made possible in Metaverse by the distributed ledger in blockchain \cite{islam2019buav}, \cite{deepa2022survey}. Consequently, the acquisition of data is immune to attacks since any modifications in Metaverse data need to be approved by a majority of the nodes in the ledger \cite{xu2021light}. In the Metaverse, all acquired data must undergo a blockchain-based validation process that is assisted by consensus methods \cite{bouraga2021taxonomy}, \cite{lashkari2021comprehensive}. Moreover, the data acquired through any block cannot be tampered with \cite{zhang2021research}, and there is almost zero possibility of creating duplicate blocks, which guarantees zero duplication in the data acquisition process. All blockchain blocks are authorized, so the data obtained in the Metaverse using blockchain-powered systems will be highly reliable \cite{guo2021reliable}.

Data acquisition in Metaverse poses challenges in terms of ensuring that data is authentic and of high quality. Even though blockchain techology does address these issues, however, it can be slow because of its distributed and complex nature  \cite{xu2021latency}. Blockchain transactions can be costly and time-consuming \cite{alrubei2020latency}. Moreover, the data collected using blockchain needs to be copied along the chain, which increases storage demand \cite{chen2022blockchain}. More research is thus needed to address these issues and devising matured blockchain systems for the Metaverse.

\subsubsection {\textbf{Data Storage}} Metaverse involves the parallel existence of the physical world and a digital reality. The Metaverse will generate an astonishing amount of data as more and more people are added to the digital spaces. Once the Metaverse will start operating at its full capacity then the capacity of data storage in the physical world will quickly become inadequate. Thus, data storage might be the main concern when deploying healthcare, real estate, entertainment, gaming, and other similar types of applications in Metaverse \cite{bian2021demystifying}. 

Using blockchain technology might lead to several blocks that contribute to the distribution of data, and thus increases the availability of data life support alerts, vital monitoring and other critical health applications in the Metaverse. Also, blockchain's decentralized nature means less time is needed to identify and label data. Fig.~\ref{blockchain-Metaverse} shows how blockchain ensures data availability, transparency, and reliability are provided in the Metaverse. Each block in the chain will back up the data, and a consensus-based distributed ledger resists data duplication and tampering \cite{xie2019survey}. Nevertheless, there is room for more research for addressing the latency issues, since all the acquired data needs to be mirrored throughout the entire chain. 

\subsubsection {\textbf{Sharing of Acquired Data}} A wide range of  Metaverse stakeholders can benefit from data sharing in a variety of ways. As the same platform is shared by the applications and people, collaboration among them can more more effectively than when they use separate platforms, as shown in  Fig.~\ref{blockchain-Metaverse}. The sharing of data may benefit people from different walks of life ranging from the general public to scientists  \cite{kraus2022facebook}. However, users' private and sensitive data might be exposed to huge risks if it is shared in centralized platforms \cite{liu2020blockchain}, \cite{egliston2021critical}. In a conventional sharing environment, the data is very mutable, which leads to decreased data availability and increased data latency. It is quite challenging to scale mutable data than the immutable one \cite{yu2021blockchain}.  A massive amount of data will be generated by many Metaverse applications including entertainment, media, traffic optimization, and healthcare applications in real time. Thus, data flexibility becomes the main concern when there is an increase in the demand for real-time data in a conventional data sharing scenario.

The precision and transparency of transactions in education, crypto exchange, and other applications can be increased by using blockchain technology \cite{egliston2021critical}. Finance, governance and other similar applications will create immutable, decentralized records of all transactions, that can be accessed by stakeholders. Thus, more data transparency will benefit Metaverse shareholders \cite{rashid2021blockchain} and will boost the confidence of the users \cite{vashistha2021echain}. Moreover, the information can be completely controlled by the data owner. Distributed ledger technology will also facilitate data audits. Consequently, blockchain reduces the amount of money and time spent on data validation \cite{min2022portrait}. Furthermore, data sharing can be made more flexible through smart contracts, which are generally employed to automate the agreement execution process so that all parties are aware of the immediate outcomes without needing to involve any intermediary or losing time. Moreover, blockchain supports the heterogeneous programming of smart contracts, so applications such as Applicature, UBS, Tracr, Ascribe, and  Nmusik can benefit from it \cite{ali2021comparative}. Blockchain will improve the adaptability and flexibility of Metaverse data, because copies of data must be replicated across a chain, which delays the information transfer \cite{luo2019blockchain}. The number of blocks must increase as the number of Metaverse users does, which necessitates the utilization of huge computing resources  \cite{gao2021b} and will make the verification of shared transactions very costly for users. Next-generation blockchain needs to cater to this problem for efficient data sharing in the Metaverse.

\subsubsection {\textbf{Data Interoperability}} Interoperability is expected to be one of the primary driving forces behind the Metaverse. The creation of the Metaverse will involve the amalgamation of several digital realms. However, existing conventional centralized digital platforms are unorganized and disjointed, and users need to create their payment infrastructure, hardware, avatars, and accounts to participate in those various realms \cite{bian2021demystifying}. A cross-chain protocol is expected to be an ideal solution for ensuring interoperability among the numerous  sub-Metaverses \cite{belchior2021survey}, \cite{madine2021appxchain}. Fig.~\ref{blockchain-Metaverse} shows that interoperability enables the exchange of possessions such as payment, NFTs, and avatars between virtual spaces, and cross-blockchain technology can enable interoperability between virtual worlds \cite{jabbar2020blockchain}, which eliminates the need for intermediaries in the Metaverse. Thus, blockchain can simplify the connection of applications and people in the Metaverse. 

\subsubsection {\textbf{Data Privacy Preservation}} 
Cutting-edge HCI technologies will make it possible for users to take part in social interactions in the Metaverse and interact with their virtual environments \cite{siyaev2021towards}. There are significant data privacy concerns in Web 2.0, which is centralized. With the increase in the complexity and scope of Web 3.0 (the Internet of Metaverse), the boundaries between the virtual and real worlds will be reduced  \cite{arvas2022gutenberg}. Some concerns regarding the impact of Web 2.0 on the protection of personal rights remain unaddressed. The data privacy problems facing the upcoming Web 3.0 will be more complex \cite{kostenko2022electronic}. 

Blockchain technology enables Metaverse users to manage their data via public and private keys and therefore have the ownership of their information/data. The third-party intermediaries in the blockchain-powered Metaverse are not allowed to gain access to or misuse other parties' data \cite{kumar2021ppsf}. Moreover, the blockchain ledgers come standard with an audit trail, which ensures that Metaverse transactions are consistent and complete, as shown in Fig.~\ref{blockchain-Metaverse}.

Even though blockchain technology can help users in preserving their data privacy, however, blockchain security and data privacy can be compromised  by even a small human error, like losing a private key. Moreover, third-party applications in the Metaverse can easily be targeted by attackers, who exploit insufficient security approaches that are adopted by the third parties and compromise personal data  \cite{hassan2019privacy}. Thus, the use of blockchain technology to ensure the privacy of user data must be further investigated. 

\subsection {Blockchain for DT in the Metaverse}
\subsubsection {Introduction} Everything in the Metaverse, from simple to complex surroundings and products, is digitally represented using DT technology  \cite{ramu2022federated}. Thus, everything related to the users' needs can be made an element of the ecosystem utilizing DTs. Bidirectional IoT links facilitate users in bringing their chosen models to life while staying synced with the real world. The Metaverse applications cannot function properly unless a an adequate link is made between the digital and physical worlds in the beginning \cite{chen2021research}. DT can also help to predict the users' needs in the Metaverse or the right time to service hardware \cite{yoon2021interfacing}.

\subsubsection {Challenges} In the Metaverse, the information acquired through numerous remote sensors is used to develop DT models. The data quality used to create the models affects the accuracy of the DT models. In other words, the data from the source must be of high quality and genuine \cite{zhuang2021digital}. Moreover,  DTs from different virtual worlds should cooperate for improve Metaverse outcomes, and DTs from different markets, for example, finance and to healthcare need to interact and connect. In addition, DTs need to be able to identify and respond to rapidly occurring changes in the virtual worlds and should be able to detect and fix problems for more consistent and accurate communication. However, when various sensors and devices are combined to develop DT models utilizing real-time data, it becomes quite difficult to keep data safe from malware and botnets  \cite{khan2022digital}.

\subsubsection {Significance of blockchain} 
\begin{figure*}[!ht]
  \begin{center}
  \includegraphics[width=7.5in]{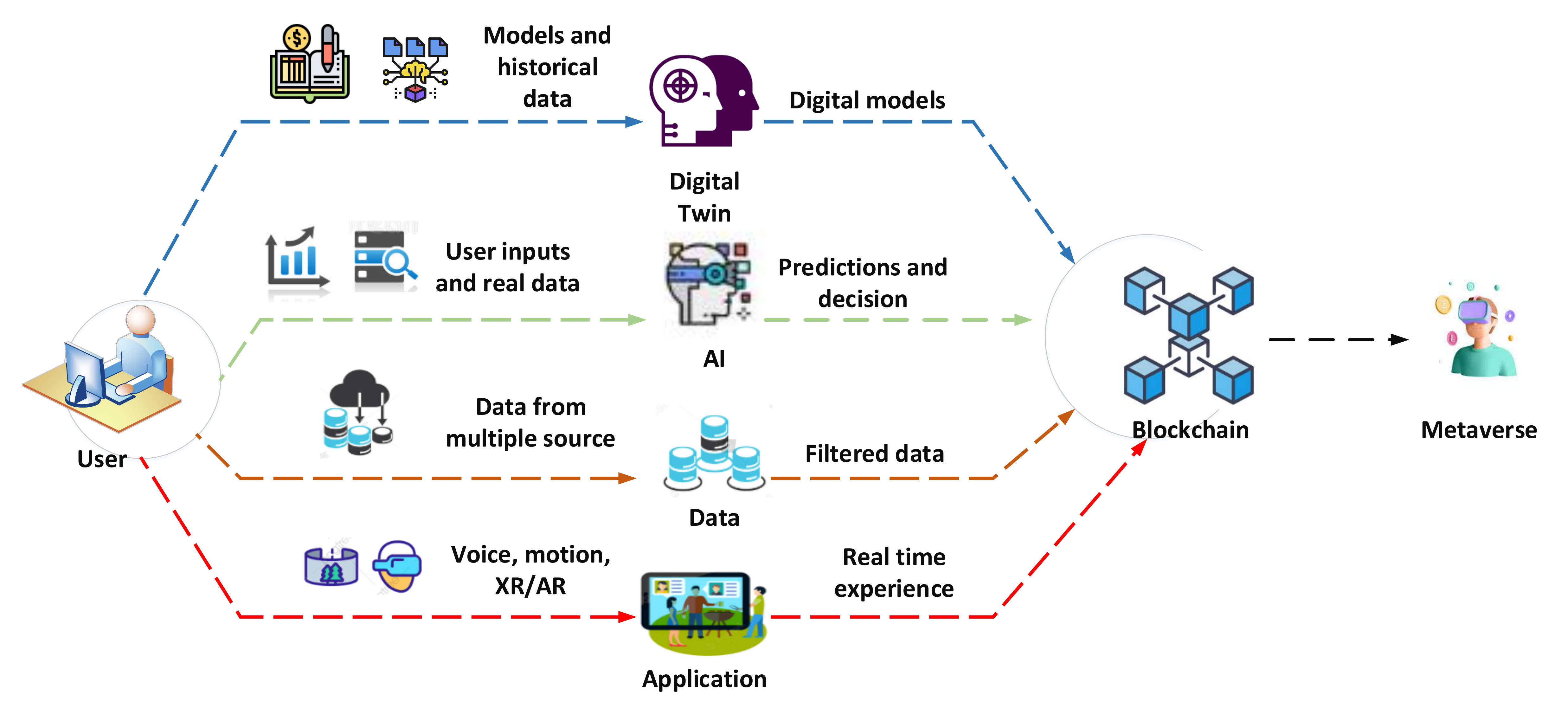}\\
  \caption{Blockchain for key enabling technologies of the Metaverse 
  }\label{block-tech}
  \end{center}
 \end{figure*}

Blockchain's classic data transparency and encryption capabilities enable DTs to resist attacks and securely share data between different virtual environments \cite{lee2021integrated}. An intelligent distributed ledger can be used for data sharing between DTs in virtual spaces. Fig.~\ref{block-tech} shows that an intelligent distributed ledger will be used for storing real-world objects on the blockchain and synchronizing to DTs. Moreover, DT deployment on the blockchain would help to resolve data privacy and security issues \cite{shen2021secure}. Furthermore, the tracking of sensor data and production of high-quality DTs can be made more viable by incorporating AI and blockchain. In Metaverse, all the DT actions are recorded as immutable transactions on blockchain that require consensus to be altered \cite{lee2021integrated}.

\subsubsection {Summary} The integration of blockchain technology with DTs helps Metaverse stakeholders effectively manage their data in a shared distributed ledger in addition to addressing data safety, integrity, and trust concerns. Scalability, privacy, and standardization are the major issues that need to be effectively handled for the successful implementation of blockchain in DT applications in the Metaverse. Moreover, the quality of DTs in Metaverse can be improved by combining FL approaches, AI, and blockchain \cite{wang2021explainable}.

\subsection{Summary and Lessons Learned}

From a macro perspective, Metaverse blurs the boundary between the virtual and real worlds by blending human and digital worlds. On the other hand, from a micro perspective, Metaverse is the interconnection of the different virtual worlds whose services are personalized by the avatars. 
\begin{enumerate}
    \item We have learned that for effective and authentic identity management of the users, a fusion of sensory signals from IoTs and wearable devices with biometrics can be very beneficial. Besides, using blockchain, highly secured digital identities for the users can be made.
    \item It has been observed that incorporating different technologies leads to more attack areas on inputs, UGC, and outputs from the Metaverse. In this regard, data reliability for DTs creation can be ensured using blockchain.
    \item We have learned that privacy issues can be amplified in Metaverse, as new types of threats will emerge in the Metaverse environment that will mostly affect digital footprints. 
    \item Incorporating AR, AI, and SDN technologies, situation awareness can be built in the Metaverse. Besides, threats that target multiple sub-Metaverse can be monitored through situational awareness.
\end{enumerate}

\section {Metaverse for Smart Cities}
\subsection {The Role of MetaVerse in Power Systems}
The Metaverse is expected to have huge potential for asset-heavy sectors, such as utilities and power. The Metaverse enablers include cryptography, blockchain, 3D reconstruction, AI, the IoT, MR, VR, and AR, and soon, XR. Surprisingly, most of the evolving technologies are not new to the power sector. Grassroots groups/initiatives and utility companies are moving toward microgrids in which blockchain-based smart contract management is leveraged to facilitate the efficient and transparent measurement of energy at all levels and enables peer-to-peer trading between electricity prosumers. It is a cumbersome job to maintain and monitor far-reaching generation assets because of their geographic expansion. This issue is directly addressed by MR/AR/VR technologies through cost-efficient and operationally-effective maintenance, remote assistance for workers in the field, and virtual site visits. These technologies can considerably enhance manpower productivity and operational efficiency through advanced analytics \cite{song2021build}.

MR/VR/AR technology and IoT devices can be used to create virtual power plants to streamline remote operations; when one can view the complete power asset in 3D and carry out virtual operations, it may be possible. These technologies can convert energy companies into distribution service operators (DSO) while offering varied services to the consumers and assisting in developing smart societies. Since these companies are focused on energy efficiency, thus there would be a more critical technology role in sharing and analyzing information in real-time and taking corrective measures. The system as a whole might function as a virtual world with VR/AR. It could be used not only to implement of shared operation centers and manpower optimization but also to effectively monitor assets in real time \cite{kumar2022virtual}.

Moreover, automated signals and alerts may help to quickly detect and resolve bottlenecks and problems and in turn reduce failures and downtime. Metaverse may help practitioners in learning the dynamics of the power sector in a more practical and accessible way. The early adopters all around the world are inquisitive about finding approaches for efficient shaping of their future strategies and reap the dividends early. Each bitcoin transaction currently requires 2264.93 kWh of energy, which is more than the per-capita consumption of several countries. Though it can boost the generation of renewable energy, it raises concerns regarding the sustainability and inevitability of this level of consumption. Countries with enormous renewable energy potential and a massive skilled workforce seem to have a bright future in a Metaverse-driven world \cite{kumar2022virtual}.

\subsection {Metaverse for Internet of Health Things}
Over the past few years, telemedicine applications and online health communities have emerged and played a vital part in boosting the digital economy in the medical/health sector. Digital health providers offer stakeholders, including pharmaceutical companies, online platforms, patients,and doctors substantial benefits.  Like many other numerous technical buzzwords of their modern era, such as blockchain, AI, big data, and cloud computing, the Metaverse is driving industry innovations and technological advancements in several domains despite its unresolved challenges and issues \cite{chen2022exploring}. 

Owens et al. \cite{owens2011empirical} indicated that the Metaverse is characterized by that technological and social exchange and social existence. In the medical health sector, the attribute of Metaverse's attributes currently meet the stringent requirements of shareholders, patients, and doctors, and thus open up applications such as surgery assistance,  health management, virtual care, and telemedicine. The medical health Metaverse is designed to elevate virtual care from a 2D experience to a 3D experience and reform health and medical informatics.

The health Metaverse has various promising applications, such as improving capabilities of tracking physical fitness, tracking heart rate, and blood glucose monitoring, analyzing clinical patient details, and keeping an eye on seriously ill patients in immersive 3D. Presently, peer-to-peer support is turning out to be very important for interactions between patients and medical professionals in online medical communities \cite{furstenau2019process}. Conventionally, the efficiency of interaction between patients and doctors arises from the support of medical personnel and institutions, while the role of the intermediary is played by the online platform. This whole scenario is changed with the emergence of the health Metaverse. Professional medical treatment and advice are yet regarded as the gold standard in the health Metaverse. It is therefore crucial to develop effective immersive health and medical services in a medical-knowledge-based 3D virtual space. The majority of preliminary Metaverse applications have focused on more entertainment and games. Thus, a new attraction for online users is to pay to consult medical health care providers in the health Metaverse \cite{chen2022exploring}.

Nevertheless, some opportunities and challenges related to the health Metaverse have yet to be addressed. Firstly, the guarding of a patient's life and privacy in the health Metaverse raises many concerns and questions \cite{cheung2020disambiguating}. The medical practice is expected to undergo significant changes on various fronts in the health Metaverse \cite{leenes2007privacy}. A patient's health conditions might be endangered by the inappropriate handling of the mental and physical health-related activities pertaining to them. Thus, relevant regulations and constraints are required considering that the healthcare applications might be gamified in the Metaverse. The entertainment of health-related applications must be carefully regulated by health Metaverse builders, otherwise patients might face serious consequences. 

The stakeholders that initially started building the health Metaverse were mostly tech giants in the health and non-medical sector. The gamification of medical expertise and turning it into entertainment has since led to the moral crisis that poses huge challenges \cite{chen2022exploring}. The challenges and problems that are anticipated for the health Metaverse are:
\begin{itemize}
    \item Gamifying health-related services and making them entertainment has a detrimental impact.
    \item Existing online platforms must be further upgraded to meet virtual user requirements.
    \item Regulation and censorship contradict the freedom and equality that are integral to  Metaverse. 
    \item Concerns remain regarding users'  personalization, safety, and privacy.
    \item Forcing people to escape reality leads to numerous social consequences.
\end{itemize}

In the future, further consolidation of necessary applications and technologies in the health Metaverse is needed from different health care perspectives:
\begin{itemize}
    \item Transforming health and medical-related services providers.
    \item Encouraging inevitable technological research and  development in the health Metaverse.
    \item Updating health and social governance concepts in the health Metaverse \cite{chen2022exploring}.
\end{itemize} 

\subsection {Distributed Metaverse}

Seeking authority and the advantages that come with having authority is the major driving force behind blockchain fragmentation. Even though, cloud-based Metaverse is not favorable with respect to visualization quality and fails to serve the universal Metaverse vision. Nevertheless, it provides companies, that have deployed their own personal metaverse with authority to control the metaverse data of the users of their metaverse on the servers and thereby results in numerous separate metaverses instead of one universal one. For instance in Second Life,, there are several virtual worlds that use centralized architecture, and are each is partitioned into smaller sectors, each of which is controlled by a dedicated server. Each sector's centralized server carries out most of the computations needed to run the simulations in the virtual world, e.g, collision detection and 3D animation. Thus, the communication and computational capacity of the server limits the number of users accessing each sector. Thus, the target assumption that the virtual world can accommodate as many avatars as the physical world can accommodate people is disproven by these limitations. 

\begin{figure}
  \begin{center}
  \includegraphics[width=3.5in]{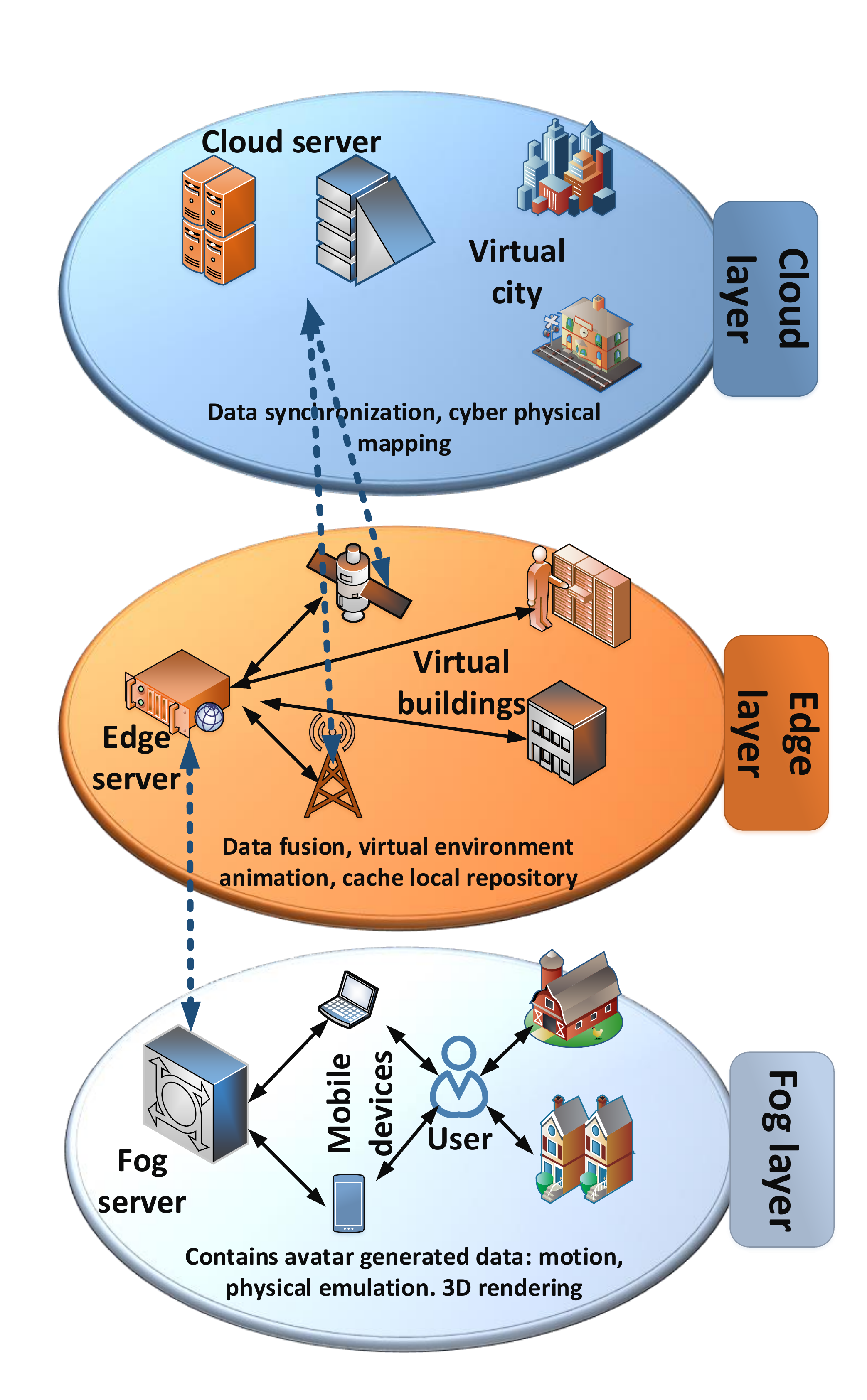}\\
  \caption{Hybrid Edge-Fog Framework for
Metaverse applications 
}\label{dmv}
  \end{center}
\end{figure}

The computational bottleneck and fragmentation problem of the Metaverse can be solved by adopting a universal Metaverse approach, wherein every entity in the physical world is responsible for the computation requirement and data control of its corresponding entity in the virtual world. In light of this,  a distributed architecture is proposed in \cite{dhelim2022edge} to solve the computational bottleneck by achieving a universal Metaverse. Each users' data can access general Metaverse data, including environmental context information and virtual world spatial data, whereas the entities in the physical world manage all the personal details about their counterparts in virtual space. Moreover, the physical entity's end device is responsible for the portion of computational cost of every entity, including physics emulation and collision detection. For instance, for all avatars the Metaverse, the edge devices of each avatar's respective user is responsible for of performing the computations related to the physical movements of the avatar, such as its mass, its momentum, and the physical forces of surroundings bodies. Such a distributed framework requires a dependable independent setting that manages the virtual space and updates and synchronizes events to all the end devices of related entities.  In Fig.~\ref{dmv}, there are dedicated cloud servers for simulating the virtual world, whereas the edge servers are used to perform computations in a particular area of the virtual world grid. The medium-level institutes can be represented by these areas. Eventually, computational tasks near end users are performed by fog servers in a smart home environments. The layered framework has two advantages: 1) data is controlled by users which makes it easier for organizations to access one universal Metaverse instead of several separate metaverses. 2) the computational costs of heavy jobs are distributed which resolves the computational bottleneck issues \cite{dhelim2022edge}. 

\subsection {Use Case: Distributed Social Metaverse}
Dhelim et al. \cite{dhelim2022edge} devised a distributed social Metaverse application to demonstrate the efficiency of the mentioned distributed Metaverse architecture. In the application they developed, computational tasks are carried out at the edge devices. Moreover, users rely on a virtual map for navigation and text messaging for communicating with the neighboring users. In addition, users can also place orders to buy some digital assets. The authors used the iFogSim simulator \cite{gupta2017ifogsim} for simulating the network architecture and extended it to simulate their proposed Metaverse application. Contextual spatial information-based user classes were created to simulate users and their assets in the Metaverse. The Transaction/Block/Blockchain classes were employed for trading digit assets in the Metaverse, and some other classes were used for managing events. The history of digital assets trading was kept in the public blockchain, while the personal details of all the users including navigation history and messages were stored at the end devices. Every user was linked to a computational task to simulate the computational costs of 3D simulations. These tasks were performed at the edge server and local fog server for social interaction management and spatial navigation, respectively in the proposed architecture. The cloud-based framework is set as the baseline where all the data management and computation is done at the cloud server. The difference in latency of the two approaches was calculated with different Metaverse users count, which showed that latency increases substantially in the -based Metaverse when the user count is increased. The fog-edge Metaverse , on the other hand, has the shortest latency even when the user count increases. 

\subsection {Metaverse for Social Good}

Even though Metaverse, referred to as human-centered computing, is a virtual world, however, it has a considerable positive influence on the real world, particularly with reference to humanity, equality, diversity, and accessibility. This section lists a few representative applications that reflect Metaverse being used for for social good.

\subsubsection {\textbf{Humanity}}
A humanistic spirit is a form of universal human self-care that is manifested in concern for, and the pursuit and maintenance of human destiny, value, and dignity. Humanity, in society, enjoys several cultural and spiritual phenomena that the older generations have passed on. The Metaverse is expected to be an excellent mean of cultural protection and communication. For instance, the protection of cultural relics is possible in the Metaverse. In the real world, cultural relics are exposed to years of weathering, become  are delicate and can easily be destroyed by natural disasters or man-made damages. For instance, in 2019, a fire broke out in Notre Dame de Paris that caused serious damage to the wooden sections of the cathedral. Luckily, the Notre Dame de Paris is reconstructed by Ubisoft as a digital 3D model in Assassin’s Creed Unity that might be used for supporting its reconstruction. The digital reconstruction of cultural relics in the Metaverse can not only be carried out anywhere in the world but also provides evidence for restoring relics  \cite{duan2021metaverse}.

\subsubsection {\textbf{Diversity}}
Due to certain physical restrictions (such as language, geography, etc.), several elements cannot be integrated into the real world in one place to satisfy the needs of many people. Nevertheless, the Metaverse supports smooth scene transformation and infinite extension space, which might efficiently attain diversity. In the Metaverse, numerous interesting scenarios can take place in different fields scenes, ranging from haunted houses to pets, artwork, political campaigns, shopping, and education. Thus, the Metaverse can greatly satisfy the varied needs of the physical world  \cite{duan2021metaverse}.

\subsubsection {\textbf{Equality}}
Equality is a divine quest for humans; however, it is actually influenced by several factors including property ownership, disability, gender, and race. In the Metaverse, everyone has the power to control their personalized avatars and exercise their authority to build a sustainable and fair society. For instance, as an automated ecosystem, a feature of democracy is included in the Metaverse that helps users maintain order and regular operations. For example, there is a decentralized autonomous organization in Decentraland that enables users to propose and vote on the policies to implement to regulate the behavior of the virtual world (e.g., the types of wearable items that are allowed).  \cite{duan2021metaverse}.

\subsubsection {\textbf{Accessibility}}
Increasing globalization has led to frequent cooperation and communication between countries all across the globe. However,  geographical distance remains an objective barrier that might cause an increase in cost during the process. Recently, several events were suspended due to the COVID-19 pandemic for preventing its spread, however, Metaverse supported converting several physical events into virtual ones. In 2020, UC Berkeley held its graduation ceremony on Minecraft. Furthermore, numerous virtual events are held daily on Fortnite. These examples show that the Metaverse has already become a part of everyday activities in the physical world and can fulfill our social needs with greater security and lower costs \cite{duan2021metaverse}.

\subsection{Virtual Metaverse Campus}
To show the importance of Metaverse in social good, the university prototype was introduce by the authors of \cite{duan2021metaverse}. They designed and demonstrated the functionality of an early-stage prototype of the Chinese University of Hong Kong was designed and demonstrated. The proposed Metaverse model provides environment for student on campus, wherein any activity of students in the real world will affects the virtual world, and vice versa. 

The proposed demo model of university Metaverse was developed using Unity, which is environment development cross platform engine, hence the application designed can easily be used on smartphones and other devices. Furthermore, Blender was used to design the 3D models that are incorporated in the Metaverse university campus. Moreover, for secure and reliable transactions among users blockchain-enabled smart contracts were used. In the university Metaverse prototype, the UGC creation is encouraged, which allows user to create almost anything the user wants. For instance, the users can trade the UGCs, or display it in their personal profile room on a billboard. Moreover, they can get personal rooms in hostel and can redesigned them to their liking.

\subsection{Entertainment}
Metaverse integrates numerous XR-based applications and the virtual world and provides limitless creative functions for users. In Metaverse, the users only require a single account to access different virtual worlds while maintaining their assets, such as digital wallets and appearance. After logging in to their respective account, the avatar may start from their virtual homes, which may be in the form of a flat, house, or even an island that users own. Furthermore, the user can select the type of application he likes and even invites friends to join that application. Moreover, to mimic real-world entertainment, Metaverse users may ask to buy a ticket to join the application. In addition, limitless creativity options are available to Metaverse users, allowing them to create assets and virtual objects using different tools provided by the platform. Based on these functionalities, Metaverse allows its users to develop their virtual application and share it with other users. Henceforth, Metaverse entertainment applications do not come solely from companies but can also come from Metaverse users, leading to many Metaverse applications. 

The one entertainment application that will be significantly transformed and affected by Metaverse is a concert. As of now, we either attend a concert physically or virtually. However, in Metaverse, we can attend the concert both physically and virtually at the same time \cite{24}. More specifically, a concert will be arranged in the physical world and can be attended physically, but at the same time a digital version of the concert will also open in Metaverse, and people who chose not to attend physically can participate in a concert virtually. Thanks to the effort of XR technologies, both types of the audience can interact with each other as if they are attending the same events. As Metaverse is still in its early stage of development, many features of online entertainment might be introduced in the near future.

\subsection{Gaming}
For years, the revenue collected through gaming has been massive compared to other industries, such as music and films \cite{6}. However, the users in Metaverse can enjoy an entirely new gaming experience through XR technologies. XR technologies completely change how users interact with the gaming environment. For instance, Hololens, an XR gaming device, replaces the mouse and keypad as it recognizes user gestures. Similarly, another gaming device, the Omni One VR platform, takes advantage of user movements like jumping and running to control the gaming character movements. These experience not only help us to burn calories due to the continuous movement of the body but also reduces chances for health diseases. Besides, Metaverse provides an environment that uses both the physical and virtual worlds, allowing users to interact with virtual objects placed in the physical world. One such example of the usage of XR technologies for gaming in Metaverse is Pokemon Go, a massive phenomenon in mobile gaming that can be seen as a primitive form of this genre.

\subsection{Autonomous Vehicles }
In past years, the automotive industry has significantly improved, especially in autonomous vehicles. Due to advanced technologies like IoT and AI, significant features are added to vehicles, which assist humans in driving and ensure safety. Moreover, due to improved AI algorithms and advances IoT sensing and control devices like image sensing and processing, vehicles operate more precisely with less human involvement. In addition, drivers can continuously do their work without worrying about driving. However, automotive industries still face some difficulties in equipping vehicles with such features. Moreover, before deploying the autonomous vehicle in the real world, it must be tested in both simulated and physical worlds. However, both the simulated and physical worlds are different, and there is a possibility that effective results may not be achieved. In this regard, the vehicles' testing environment can be created in Metaverse \cite{33}, and almost real-world driving experience can be achieved. Furthermore, in Metaverse, the vehicle will be tested and trained on much richer data gathered from different autonomous vehicles and drivers running worldwide.

\subsection{Summary and Lessons Learned}
In this section, we have learned about the role of Metaverse in smart cities. We have observed that Metaverse can play a significant role in designing, operating, and maintaining power grids. With the help of MR and XR, the operator can determine which section of the line is compromised to resolve it effectively. Moreover, through advanced security and privacy algorithms, the users' sensitive data will be protected and can not be used by malicious activists. From a health perspective, we have learned how Metaverse's involvement can help improve the health system. Unlike traditional health systems, Metaverse doctors can interact with the patients' 3D models and provide detailed medical diagnoses \cite{m44}. Moreover, with the help of AR, XR, and DTs, doctors can work on the simulated patient model and observe the effect of the environment and medicine on the patient's health, thus enabling doctors to provide better treatment.

As observed from the applications mentioned above, Metaverse will play an important role in entertainment industries and rectify users' experiences to the next level. However, to achieve this, some challenges need to be handled effectively for the efficient and reliable operation of Metaverse. For example, entertainment activities include millions of attendees. Although current online gaming platforms did support that amount of users'. However, it is worth noting that these platforms do not use XR technologies. Henceforth, if XR technologies are to be used, it will put tremendous stress on existing infrastructure, as it requires huge computation power and low latency connections \cite{42}.

Metaverse is a step toward the future of E-Learning. We have learned that Metaverse can be used effectively in the education sector as it involves using different advanced technologies, such as XR, AR, MR, and DTs. In addition, the Metaverse environment can resolve many issues, such as limited spaces, disease, or pandemics, and provide interactive sessions for the students.

\section {Metaverse Projects}
\begin{figure*}[!ht]
  \begin{center}
  \includegraphics[width=7.5in]{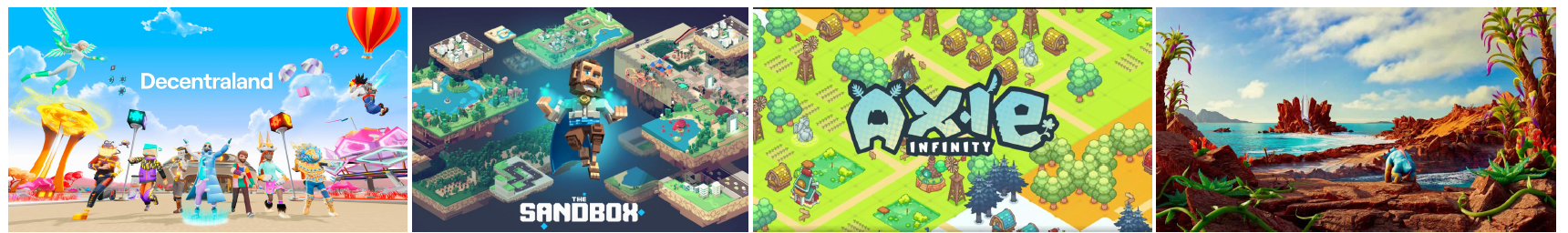}\\
  \caption{Metaverse projects \cite{gadekallu2022blockchain}}\label{Metaverseprojects}
  \end{center}
 \end{figure*}

Some of the famous Metaverse projects, including Illuvium, Axie Infinity, Sandbox, and Decentraland, are briefly introduced in this section. These projects have utilized blockchain as the main technology for creating and developing Metaverse, and for delivering diversified blockchain-based applications and services in the virtual world, ranging from E-commerce to real estate. Fig.~\ref{Metaverseprojects} shows the landscapes inside the virtual spaces of these projects.

\subsubsection {Sandbox}
Sandbox is an Ethereum blockchain-based decentralized Metaverse, wherein users can use SAND to build, own and monetize the immersive gaming experience. Sandbox takes inspiration from Roblox and Minecraft and raises the gaming experience from a 2D environment to a complete 3D space through the voxel gaming platform. In this Metaverse project, 3D objects (including tools, buildings, animals, and people) can be freely created and animated by the users using VoxEdit- a built-in voxel gaming package. In the Sandbox marketplace, creations can be traded as game assets, and the creator may get their incentives/rewards in SAND tokens.
Moreover, InterPlanetary File System is adopted by Sandbox for storing the real digital assets and ensuring modifications to assets without the permission of the owners. However, scalability remains a major concern in Ethereum-based projects. In light of this, the Sandbox team was motivated to find out layer-2 solutions (i.e, scaling an application by managing transactions outside of the Ethereum Main net and exploiting decentralized security). 

\subsubsection{Decentraland}
Decentraland is an Ethereum blockchain-powered VR platform that enables users to experience, create and monetize applications, hyper-real content, and economic assets. Community members have complete control and oversight over their creative actions and are the permanent owners of the land in Decentraland. In the virtual Decentraland world, a virtual land is uniquely determined to be a scarce, transferable, and non-fungible digital asset that is stored in an Ethereum smart contract. Contrary to the other conventional virtual social networks and spaces, in Decentraland, there is no centralized organization managing and supervising the virtual world, which means that no single agent has the right to change the economic mechanism, content, and software rules, or preventing others from gaining access to the world, offering services, and trading digital products. 

With reference to user cases, Decentraland supports applications (dynamic 3D scenes with scripting language toolsets, gambling, games), social settings (multi-player games, chat groups, forums), advertisements, content curation, and other functions (like virtual tourism, education, 3D design, and therapy). In terms of architecture, there are three layers in the Decentraland protocol, i.e, real-time layer, the land content layer, and the consensus layer, along with two support systems, which include an identity system and payment channel framework. To maintain an ownership ledger for an area of land using an Ethereum smart contract, non-fungible digital assets are remarked by Decentraland by burning MANA tokens. Decentraland uses the Ethereum Name Service for ownership identification. 

\subsubsection {Illuvium} Illuvium is usually lauded as the pioneer Ethereum blockchain-based open-world fantasy game that provides entertainment to decentralized finance users and regular gamers. Illuvium's virtual space includes fantasy creatures, named Illuvials, that can be caught by players who defeat them in usual battles. Following that, these fantasy creatures become a devoted member of that player's collection and are called into combating in a random PvP game played against other players. In simpler terms, the Illuvium game combines a PvP battle game and an open-world exploration game, in which varied gameplays can be immersed by the players, i.e. the players can freely explore the virtual spaces and plan tactics for battle. There is a unique NFT linked to each Illuvial that can be traded with zero gas fees on an external exchange platform or in-game marketplace. Immutable X, a layer-2 Ethereum scaling solution, is leveraged by Illuvium for obtaining application scalability with NFT functionality. This solution enables users to trade called zero-knowledge rollup. 

\subsubsection {Axie Infinity} Axie Infinity is regarded as a ground-breaking play-to-earn project in the Metaverse. A crypto-meet-Pokemon game universe is built in Axie Infinity with fantasy creatures, named Axies. Players can collect, raise, or breed Axies, or pit them against one another in battle to build their Axies' kingdoms. Axie Infinity owns a user-based economy system, similar to Sandbox and Decentraland, that enables users to truly own, sell, buy and trade in-game resources and thus contributing to the ecosystem. The main difference between conventional games and Axie Infinity is that Axie's economic scheme helps players increase their assets by improving their in-game skills to reach specific levels. The the Axie Metaverse's ERC-20 governance token, which is called Axie Infinity Shards (AXS), may be claimed as a reward when players put their AXS tokens at stake, play games, and take part in governance jobs. It is possible to buy sell and trade virtual real estate and Axies' creatures as NFTs in the in-game marketplace. Strangely, the majority of transactions are processed on Rohin, an Ethereum-linked sidechain, which is designed specifically to have fewer charges compared to the standard Ethereum blockchain. 

\section {Future directions and Challenges}

A new era of the Internet of Everything that is opened up by 6G-based edge intelligence that has made the interconnection of the cloud, devices, and people a reality irrespective of place and time.  The evolving smart service applications in the wireless networks are altering our lifestyles and substantially upgrading our life quality. Metaverse is considered to be the most trading next-generation Internet application, aimed at connecting millions of individuals and creating a shared space where reality and virtuality can merge. Nevertheless, the complete achievement of the Metaverse vision of interoperability, materialization, and immersion is restricted by sensory devices, computational power, and resources, and is still quite far off \cite{chang20226g}.
 
\subsection {Security} A major attribute of the Metaverse is providing users an immersive experience. This experience is majorly reliant on the intelligence of the wearable devices worn by the user. In users' everyday use of the Metaverse, their, personal details, voice, habits, iris, and fingerprints can be read and stored by the wearable devices. After that, the data is uploaded to the Metaverse servers for meeting the demands of users' immersive experiences. Nevertheless, when the Metaverse server is attacked or collapsed, this may result in the leakage of users' identity data. It might lead to some identity crisis problems, such as body swapping, identity theft, and false avatar scandals. Thus, the Metaverse security needs to be strengthened to improve the ability to protect users' information. The encryption algorithm and consensus method adopted by the blockchain ensure data transparency, immutability, and security. Thus, the blockchain-based Metaverse system is anticipated to be an effective solution for security enhancement \cite{chang20226g}.
 
\subsection {Code of Ethics} There are ethical concerns in the Metaverse, due to conflicts between avatars' behaviors in the virtual worlds and the moral values of the physical world. More specifically, people are given another identity, 
(i.e, an avatar) in the Metaverse and then offered free virtual space. Nevertheless, with the inclusion of more users in the Metaverse, the unfettered avatars' behavior causes more complicated social relations in the virtual world than in the real world. Such behaviors, which are in contradiction with the moral norms of the real world, might harm minors that , who are still maturing \cite{slater2020ethics}. Therefore, in the future, users' behaviors must be constrained and controlled by the Metaverse, and clear moral and ethical norms must be established to maintain an orderly and fine ecological Metaverse environment \cite{ning2021survey}.

\subsection {Perceptual Realization}
One of the Metaverse's main aims is to provide a multi-sensory immersive experience to the users. This experience depends heavily on the efficacy of wearable devices, like mobile and VR headsets. Nevertheless, the user's tactile, auditory, and visual perceptions when interacting with the virtual world are utilized by the wearable devices. Furthermore, feedback coming from the users' brains is also important. Brain computer interface (BCI) is the communication link between external devices and users' brain that exchanges information between the device and brain and then operates the devices according to the ideas produced by the brain. This can substantially enhance the quality of users’ immersive experiences. There are two forms of BCI technology, i.e, non-invasive and invasive. The case of non-invasive BCI is easier to wear, but a particular error occurs during brain waves recognition. The invasive BCI involves the implantation of deep electrodes in the brain to obtain precise signals through the implanted signal acquisition equipment. Thus, obtaining a balance between convenience and precision is a promising direction for the application of BCI technology in Metaverse \cite{chang20226g}.

\subsection {Balance Between Real and Virtual}
Metaverse is a real-world-based virtual space that constructs platforms for entertainment, life, and social interactions in the virtual world and incorporates the virtual and real worlds. For instance, in the movie Ready Player One, wearable devices and VR can be used as an entry point to access the virtual space for getting an immersive experience. In particular, the somatosensory feeling of flying and falling can be experienced by a user in the virtual world via the completely automated haptic chair. In addition, somatosensory clothing enables the user to feel the pain of getting attacked on their body. But the users that are too much involved in the Metaverse might not differentiate between the real and virtual worlds, which will badly impact their studies, life, and work in real life. Therefore, achieving a balance between reality and virtual is an important future research direction so that a high-quality virtuality experience can be enjoyed by the users without getting overly caught up in it and harming normal life. There could be three potential research directions: 1) limiting usage time and access age of the users, 2) setting up a regulatory body for scrutinizing if any unnecessary negative effects have occurred due to virtual functions, 3) reducing the virtual world's realism to minimize the chances of user addiction \cite{chang20226g}.

\subsection {Energy-Efficient and Green Metaverse}
Metaverse is believed to have a huge potential in the industry. Sean et al. \cite{Sean2021} anticipated that in the next 10-15  years Metaverse might be a \$10 trillion to \$30 trillion opportunity. However, a notable possible drawback of this technology is its impact on the environment. The data centers needed to run the persistent Metaverse world are quite costly in terms of energy consumption. Intel, one known vendor, has determined that a thousand-fold increase in power is required over its existing combined computational capacity to power the Metaverse, which would further increase its carbon footprint \cite{Kyle2022}.  

Recently, numerous organizations have promised to take action to reduce the adverse environmental impact associated with Metaverse-powered data centers. With reference to the blockchain end, energy-efficient alternates of proof-of-work have been drawing considerable attention. The future of Ethereum will shift towards proof-of-stake that would facilitate cryptocurrency owners in the collateral staking of their assets for validating transactions in the network through consensus \cite{Kyle2022}. Nevertheless, the cost expected to be associated with Metaverse is quite clear. Thus, more fruitful research is required in the future on offsetting the costs by distributing its impact geographically to have an energy-efficient Metaverse. 

\subsection {Cloud-Edge-End Orchestrated Metaverse}
At present, massive multiplayer online (MMO) games can host hundreds of players in one game session, but doing so involves highly specific GPU requirements. However, there are a limited number of VR MMO games in the industry that might require devices like HMDs to be connected with with powerful computers for rendering interactions with hundreds of other players and the immersive virtual worlds. Moreover, the cloud-edge-end computational paradigm is a promising solution for enabling ubiquitous Metaverse access \cite{lim2022realizing}. In particular, tasks that consume little resources might be able to be computed locally on end devices. To reduce the load on the cloud and in turn minimize end-to-end latency and enhance scalability, edge servers could be used to perform the expensive foreground rendering that requires very little graphical information and lower latency \cite{guo2020adaptive}. The less delay-sensitive and more computationally-intensive tasks, such as background rendering may then be performed on cloud servers. Furthermore, for reduced computational overheads and effective retrieval, popular content should be cached at the network edge. Cooperation and orchestration between the edge and the cloud may offer a critical computing framework for the Metaverse applications, and therefore requires more productive research in the future. 

\subsection {Digital Twin Edge Network}
Even though the DT technology refers particularly to the simulation of future experiences, context calculation, and the development of a chain of results, it may facilitate edge computing in terms of the real-time optimization of its algorithm and development of a substantially safe environment. For example, if an autonomous vehicle experiences a drastic change in weather, there might be a jam or change in course in the newly generated traffic. Having a DT of the vehicle and the terrain may help computing technologies to predict the changes that should be made to ensure the journey completed safely. Therefore, incorporating DT technology is expected to be a promising approach to radically enhancing edge computing and requires more investigation. 

\subsection{Communications and Protocol Design}

{The metaverse communication requirements will feature novel and differentiated service provisioning. A paradigm shift in communication protocol will be required that should be goal oriented and semantic aware. Moreover,  communication protocol design will need to consider the vision of a seamless Metaverse experience. Finally, a unified model design will bee required to standardize the communication protocols for the Metaverse that can be flexibly accessed from heterogeneous communication systems in different virtual worlds. }

\subsection {Intelligent Blockchain}
Lately, blockchain has gained considerable research attention from both industry and academia in fields ranging from transportation to entertainment, health-care, and finance. Nevertheless, there are still several critical issues in the existing blockchain technology, such as limited interoperability, its lack of finance and standardization, and the risk that malicious activities obstruct the deployment of blockchain applications at a large scale. When integrated with effective AI schemes, intelligent blockchain is believed to be able to play a huge part in future networking and communication, as it could be leveraged to provide novel solutions to numerous issues arising in the next-generation networking and communication, such as smart resource allocation, intelligent access permission control, malicious behaviors monitoring, and secure data transmission. Thus, innovative intelligent blockchain-based ideas, technologies, frameworks, mechanisms, and designs are needed for providing better support future networking and communications, and thus ensure a better Metaverse experience. 
 
\section{Conclusion}
We have provided a comprehensive review of the Metaverse's basic principles, applications, and associated privacy and security considerations in enabling novel technologies in smart cities. In particular, we have discussed the architecture, major attributes, enabling technologies, and state-of-the-art prototypes of the Metaverse in detail. We have also investigated the privacy and security risks along with the crucial challenges associated with privacy preservation and security defenses in the context of a distributed Metaverse architecture, and identified the possible countermeasures. Furthermore, the current/future procedures for networking and communication in the Metaverse have been comprehensively reviewed. Most importantly, the role of the modern Metaverse concept in power systems, the health care sector, and overall society has been explored. We expect that this survey paper will be a key reference and play an important role in the development of enabling technologies for the Metaverse, and inspire more pioneering work in this promising field. 


\ifCLASSOPTIONcaptionsoff
  \newpage
\fi




\bibliographystyle{IEEEtran}
\bibliography{IEEEabrv,Bibliography}

\begin{thebibliography}{100}
\providecommand{\url}[1]{#1}
\csname url@samestyle\endcsname
\providecommand{\newblock}{\relax}
\providecommand{\bibinfo}[2]{#2}
\providecommand{\BIBentrySTDinterwordspacing}{\spaceskip=0pt\relax}
\providecommand{\BIBentryALTinterwordstretchfactor}{4}
\providecommand{\BIBentryALTinterwordspacing}{\spaceskip=\fontdimen2\font plus
\BIBentryALTinterwordstretchfactor\fontdimen3\font minus
  \fontdimen4\font\relax}
\providecommand{\BIBforeignlanguage}[2]{{%
\expandafter\ifx\csname l@#1\endcsname\relax
\typeout{** WARNING: IEEEtran.bst: No hyphenation pattern has been}%
\typeout{** loaded for the language `#1'. Using the pattern for}%
\typeout{** the default language instead.}%
\else
\language=\csname l@#1\endcsname
\fi
#2}}
\providecommand{\BIBdecl}{\relax}
\BIBdecl
\renewcommand{\BIBentryALTinterwordstretchfactor}{4}

\bibitem{joshua2017information}
J.~Joshua, ``Information bodies: Computational anxiety in neal stephenson's
  snow crash,'' \emph{Interdisciplinary Literary Studies}, vol.~19, no.~1, pp.
  17--47, 2017.

\bibitem{cheng2022will}
R.~Cheng, N.~Wu, S.~Chen, and B.~Han, ``Will metaverse be nextg internet?
  vision, hype, and reality,'' \emph{arXiv preprint arXiv:2201.12894}, 2022.

\bibitem{burns2018everything}
W.~Burns~III, ``Everything you know about the metaverse is wrong,'' 2018.

\bibitem{bruun2019lifelogging}
A.~Bruun and M.~L. Stentoft, ``Lifelogging in the wild: Participant experiences
  of using lifelogging as a research tool,'' in \emph{IFIP Conference on
  Human-Computer Interaction}.\hskip 1em plus 0.5em minus 0.4em\relax Springer,
  2019, pp. 431--451.

\bibitem{chayka2021facebook}
K.~Chayka, ``Facebook wants us to live in the metaverse,'' 2021.

\bibitem{milgram1995augmented}
P.~Milgram, H.~Takemura, A.~Utsumi, and F.~Kishino, ``Augmented reality: A
  class of displays on the reality-virtuality continuum,'' in
  \emph{Telemanipulator and telepresence technologies}, vol. 2351.\hskip 1em
  plus 0.5em minus 0.4em\relax International Society for Optics and Photonics,
  1995, pp. 282--292.

\bibitem{lee2021all}
L.-H. Lee, T.~Braud, P.~Zhou, L.~Wang, D.~Xu, Z.~Lin, A.~Kumar, C.~Bermejo, and
  P.~Hui, ``All one needs to know about metaverse: A complete survey on
  technological singularity, virtual ecosystem, and research agenda,''
  \emph{arXiv preprint arXiv:2110.05352}, 2021.

\bibitem{9}
M.~Xu, W.~C. Ng, W.~Y.~B. Lim, J.~Kang, Z.~Xiong, D.~Niyato, Q.~Yang, X.~S.
  Shen, and C.~Miao, ``A full dive into realizing the edge-enabled metaverse:
  Visions, enabling technologies, and challenges,'' \emph{arXiv preprint
  arXiv:2203.05471}, 2022.

\bibitem{8}
H.~Duan, J.~Li, S.~Fan, Z.~Lin, X.~Wu, and W.~Cai, ``Metaverse for social good:
  A university campus prototype,'' in \emph{Proceedings of the 29th ACM
  International Conference on Multimedia}, 2021, pp. 153--161.

\bibitem{wang2022survey}
Y.~Wang, Z.~Su, N.~Zhang, D.~Liu, R.~Xing, T.~H. Luan, and X.~Shen, ``A survey
  on metaverse: Fundamentals, security, and privacy,'' \emph{arXiv preprint
  arXiv:2203.02662}, 2022.

\bibitem{statista_2020}
\BIBentryALTinterwordspacing
NA, ``Leading applications of immersive technologies in smart cities according
  to xr/ar/vr/mr industry experts in the united states in 2020,'' April 2020.
  [Online]. Available:
  \url{https://www.statista.com/statistics/1185244/applications-immersive-technologies-xr-ar-vr-mr-smart-cities/}
\BIBentrySTDinterwordspacing

\bibitem{kemecreality}
A.~Kemec, ``From reality to virtuality: Re-discussing cities with the concept
  of the metaverse,'' \emph{International Journal of Management and
  Accounting}, vol.~4, no.~1, pp. 12--20, 2022.

\bibitem{11}
A.~Treadgold and J.~Reynolds, \emph{Navigating the new retail landscape: A
  guide for business leaders}.\hskip 1em plus 0.5em minus 0.4em\relax Oxford
  university press, 2020.

\bibitem{gadalla2013metaverse}
E.~Gadalla, K.~Keeling, and I.~Abosag, ``Metaverse-retail service quality: A
  future framework for retail service quality in the 3d internet,''
  \emph{Journal of Marketing Management}, vol.~29, no. 13-14, pp. 1493--1517,
  2013.

\bibitem{nguyen2021metachain}
C.~T. Nguyen, D.~T. Hoang, D.~N. Nguyen, and E.~Dutkiewicz, ``Metachain: A
  novel blockchain-based framework for metaverse applications,'' \emph{arXiv
  preprint arXiv:2201.00759}, 2021.

\bibitem{coinmarket_2022}
\BIBentryALTinterwordspacing
NA, ``Largest metaverse by market cap,'' June 2022. [Online]. Available:
  \url{https://coinmarketcap.com/tr/view/metaverse/}
\BIBentrySTDinterwordspacing

\bibitem{gelernter1993mirror}
D.~Gelernter, \emph{Mirror worlds: Or the day software puts the universe in a
  shoebox... How it will happen and what it will mean}.\hskip 1em plus 0.5em
  minus 0.4em\relax Oxford University Press, 1993.

\bibitem{benedikt1991cyberspace}
M.~Benedikt, \emph{Cyberspace: first steps}.\hskip 1em plus 0.5em minus
  0.4em\relax Mit Press, 1991.

\bibitem{cardullo2019right}
P.~Cardullo, C.~Di~Feliciantonio, and R.~Kitchin, \emph{The right to the smart
  city}.\hskip 1em plus 0.5em minus 0.4em\relax Emerald Group Publishing, 2019.

\bibitem{foth2015citizen}
M.~Foth, M.~Brynskov, and T.~Ojala, ``Citizen’s right to the digital city,''
  \emph{Berlin: Springer. doi}, vol.~10, pp. 978--981, 2015.

\bibitem{graham1997virtual}
S.~Graham and A.~Aurigi, ``Virtual cities, social polarization, and the crisis
  in urban public space,'' \emph{Journal of Urban Technology}, vol.~4, no.~1,
  pp. 19--52, 1997.

\bibitem{leenes2007privacy}
R.~Leenes, ``Privacy in the metaverse,'' in \emph{IFIP International Summer
  School on the Future of Identity in the Information Society}.\hskip 1em plus
  0.5em minus 0.4em\relax Springer, 2007, pp. 95--112.

\bibitem{falchuk2018social}
B.~Falchuk, S.~Loeb, and R.~Neff, ``The social metaverse: Battle for privacy,''
  \emph{IEEE Technology and Society Magazine}, vol.~37, no.~2, pp. 52--61,
  2018.

\bibitem{hu2015dynamic}
P.~Hu, H.~Li, H.~Fu, D.~Cansever, and P.~Mohapatra, ``Dynamic defense strategy
  against advanced persistent threat with insiders,'' in \emph{2015 IEEE
  Conference on Computer Communications (INFOCOM)}.\hskip 1em plus 0.5em minus
  0.4em\relax IEEE, 2015, pp. 747--755.

\bibitem{nevelsteen2018virtual}
K.~J. Nevelsteen, ``Virtual world, defined from a technological perspective and
  applied to video games, mixed reality, and the metaverse,'' \emph{Computer
  Animation and Virtual Worlds}, vol.~29, no.~1, p. e1752, 2018.

\bibitem{yoon2021interfacing}
K.~Yoon, S.-K. Kim, S.~P. Jeong, and J.-H. Choi, ``Interfacing cyber and
  physical worlds: Introduction to ieee 2888 standards,'' in \emph{2021 IEEE
  International Conference on Intelligent Reality (ICIR)}.\hskip 1em plus 0.5em
  minus 0.4em\relax IEEE, 2021, pp. 49--50.

\bibitem{dionisio20133d}
J.~D.~N. Dionisio, W.~G.~B. III, and R.~Gilbert, ``3d virtual worlds and the
  metaverse: Current status and future possibilities,'' \emph{ACM Computing
  Surveys (CSUR)}, vol.~45, no.~3, pp. 1--38, 2013.

\bibitem{ning2021survey}
H.~Ning, H.~Wang, Y.~Lin, W.~Wang, S.~Dhelim, F.~Farha, J.~Ding, and
  M.~Daneshmand, ``A survey on metaverse: the state-of-the-art, technologies,
  applications, and challenges,'' \emph{arXiv preprint arXiv:2111.09673}, 2021.

\bibitem{yang2022fusing}
Q.~Yang, Y.~Zhao, H.~Huang, and Z.~Zheng, ``Fusing blockchain and ai with
  metaverse: A survey,'' \emph{arXiv preprint arXiv:2201.03201}, 2022.

\bibitem{huynh2022artificial}
T.~Huynh-The, Q.-V. Pham, X.-Q. Pham, T.~T. Nguyen, Z.~Han, and D.-S. Kim,
  ``Artificial intelligence for the metaverse: A survey,'' \emph{arXiv preprint
  arXiv:2202.10336}, 2022.

\bibitem{park2022metaverse}
S.-M. Park and Y.-G. Kim, ``A metaverse: taxonomy, components, applications,
  and open challenges,'' \emph{IEEE Access}, 2022.

\bibitem{xu2022full}
M.~Xu, W.~C. Ng, W.~Y.~B. Lim, J.~Kang, Z.~Xiong, D.~Niyato, Q.~Yang, X.~S.
  Shen, and C.~Miao, ``A full dive into realizing the edge-enabled metaverse:
  Visions, enabling technologies, and challenges,'' \emph{arXiv preprint
  arXiv:2203.05471}, 2022.

\bibitem{duan2021metaverse}
H.~Duan, J.~Li, S.~Fan, Z.~Lin, X.~Wu, and W.~Cai, ``Metaverse for social good:
  A university campus prototype,'' in \emph{Proceedings of the 29th ACM
  International Conference on Multimedia}, 2021, pp. 153--161.

\bibitem{lee2021creators}
L.-H. Lee, Z.~Lin, R.~Hu, Z.~Gong, A.~Kumar, T.~Li, S.~Li, and P.~Hui, ``When
  creators meet the metaverse: A survey on computational arts,'' \emph{arXiv
  preprint arXiv:2111.13486}, 2021.

\bibitem{diaz2020virtual}
J.~D{\'\i}az, C.~Salda{\~n}a, and C.~Avila, ``Virtual world as a resource for
  hybrid education,'' \emph{International Journal of Emerging Technologies in
  Learning (iJET)}, vol.~15, no.~15, pp. 94--109, 2020.

\bibitem{bourlakis2009retail}
M.~Bourlakis, S.~Papagiannidis, and F.~Li, ``Retail spatial evolution: paving
  the way from traditional to metaverse retailing,'' \emph{Electronic Commerce
  Research}, vol.~9, no.~1, pp. 135--148, 2009.

\bibitem{saad2019vision}
W.~Saad, M.~Bennis, and M.~Chen, ``A vision of 6g wireless systems:
  Applications, trends, technologies, and open research problems,'' \emph{IEEE
  network}, vol.~34, no.~3, pp. 134--142, 2019.

\bibitem{heller2016avatars}
L.~Heller and L.~Goodman, ``What do avatars want now? posthuman embodiment and
  the technological sublime,'' in \emph{2016 22nd International Conference on
  Virtual System \& Multimedia (VSMM)}.\hskip 1em plus 0.5em minus 0.4em\relax
  IEEE, 2016, pp. 1--4.

\bibitem{genay2021being}
A.~C.~S. Genay, A.~L{\'e}cuyer, and M.~Hachet, ``Being an avatar" for real": a
  survey on virtual embodiment in augmented reality,'' \emph{IEEE Transactions
  on Visualization and Computer Graphics}, 2021.

\bibitem{kai2020collaborative}
C.~Kai, H.~Zhou, Y.~Yi, and W.~Huang, ``Collaborative cloud-edge-end task
  offloading in mobile-edge computing networks with limited communication
  capability,'' \emph{IEEE Transactions on Cognitive Communications and
  Networking}, vol.~7, no.~2, pp. 624--634, 2020.

\bibitem{2888}
\BIBentryALTinterwordspacing
A.~N, ``Ieee 2888 standards,'' NA NA. [Online]. Available:
  \url{https://sagroups.ieee.org/2888/}
\BIBentrySTDinterwordspacing

\bibitem{MPEG-V}
\BIBentryALTinterwordspacing
------, ``Iso/iec 23005 (mpeg-v) standards,'' NA NA. [Online]. Available:
  \url{https://mpeg.chiariglione.org/standards/mpeg-v}
\BIBentrySTDinterwordspacing

\bibitem{jayasinghe2018machine}
U.~Jayasinghe, G.~M. Lee, T.-W. Um, and Q.~Shi, ``Machine learning based trust
  computational model for iot services,'' \emph{IEEE Transactions on
  Sustainable Computing}, vol.~4, no.~1, pp. 39--52, 2018.

\bibitem{han2010user}
J.~Han, J.~Yun, J.~Jang, and K.-R. Park, ``User-friendly home automation based
  on 3d virtual world,'' \emph{IEEE Transactions on consumer electronics},
  vol.~56, no.~3, pp. 1843--1847, 2010.

\bibitem{sugimoto2021extended}
M.~Sugimoto, ``Extended reality (xr: Vr/ar/mr), 3d printing, holography, ai,
  radiomics, and online vr tele-medicine for precision surgery,'' in
  \emph{Surgery and Operating Room Innovation}.\hskip 1em plus 0.5em minus
  0.4em\relax Springer, 2021, pp. 65--70.

\bibitem{jaynes2003metaverse}
C.~Jaynes, W.~B. Seales, K.~Calvert, Z.~Fei, and J.~Griffioen, ``The metaverse:
  a networked collection of inexpensive, self-configuring, immersive
  environments,'' in \emph{Proceedings of the workshop on Virtual environments
  2003}, 2003, pp. 115--124.

\bibitem{wu2021digital}
Y.~Wu, K.~Zhang, and Y.~Zhang, ``Digital twin networks: a survey,'' \emph{IEEE
  Internet of Things Journal}, vol.~8, no.~18, pp. 13\,789--13\,804, 2021.

\bibitem{vural2012survey}
S.~Vural, D.~Wei, and K.~Moessner, ``Survey of experimental evaluation studies
  for wireless mesh network deployments in urban areas towards ubiquitous
  internet,'' \emph{IEEE Communications Surveys \& Tutorials}, vol.~15, no.~1,
  pp. 223--239, 2012.

\bibitem{du2021optimal}
H.~Du, D.~Niyato, J.~Kang, D.~I. Kim, and C.~Miao, ``Optimal targeted
  advertising strategy for secure wireless edge metaverse,'' \emph{arXiv
  preprint arXiv:2111.00511}, 2021.

\bibitem{wang2021blockchain}
Y.~Wang, Z.~Su, J.~Ni, N.~Zhang, and X.~Shen, ``Blockchain-empowered
  space-air-ground integrated networks: Opportunities, challenges, and
  solutions,'' \emph{IEEE Communications Surveys \& Tutorials}, 2021.

\bibitem{wu2020interactive}
E.~H.-K. Wu, C.-S. Chen, T.-K. Yeh, and S.-C. Yeh, ``Interactive medical vr
  streaming service based on software-defined network: Design and
  implementation,'' in \emph{2020 IEEE International Conference on Consumer
  Electronics-Taiwan (ICCE-Taiwan)}.\hskip 1em plus 0.5em minus 0.4em\relax
  IEEE, 2020, pp. 1--2.

\bibitem{wang2021non}
Q.~Wang, R.~Li, Q.~Wang, and S.~Chen, ``Non-fungible token (nft): Overview,
  evaluation, opportunities and challenges,'' \emph{arXiv preprint
  arXiv:2105.07447}, 2021.

\bibitem{59}
R.~Scheiding, ``Designing the future? the metaverse, nfts, \& the future as
  defined by unity users,'' \emph{Games and Culture}, p. 15554120221139218,
  2022.

\bibitem{60}
T.~Ward, A.~Bolt, N.~Hemmings, S.~Carter, M.~Sanchez, R.~Barreira, S.~Noury,
  K.~Anderson, J.~Lemmon, J.~Coe \emph{et~al.}, ``Using unity to help solve
  intelligence,'' \emph{arXiv preprint arXiv:2011.09294}, 2020.

\bibitem{62}
A.~Palumbo, ``Microsoft hololens 2 in medical and healthcare context: State of
  the art and future prospects,'' \emph{Sensors}, vol.~22, no.~20, p. 7709,
  2022.

\bibitem{63}
H.~Vo, ``Interior design studio in the new normal era: A virtual reality case
  study,'' in \emph{2022 International Symposium on Educational Technology
  (ISET)}.\hskip 1em plus 0.5em minus 0.4em\relax IEEE, 2022, pp. 168--172.

\bibitem{44}
A.~Chia, ``The metaverse, but not the way you think: game engines and
  automation beyond game development,'' \emph{Critical Studies in Media
  Communication}, pp. 1--10, 2022.

\bibitem{kasapakis2017user}
V.~Kasapakis and D.~Gavalas, ``User-generated content in pervasive games,''
  \emph{Computers in Entertainment (CIE)}, vol.~16, no.~1, pp. 1--23, 2017.

\bibitem{zhao2017estimating}
J.~Zhao, R.~S. Allison, M.~Vinnikov, and S.~Jennings, ``Estimating the
  motion-to-photon latency in head mounted displays,'' in \emph{2017 IEEE
  Virtual Reality (VR)}.\hskip 1em plus 0.5em minus 0.4em\relax IEEE, 2017, pp.
  313--314.

\bibitem{pan2021network}
J.~Pan, L.~Cai, S.~Yan, and X.~S. Shen, ``Network for ai and ai for network:
  Challenges and opportunities for learning-oriented networks,'' \emph{IEEE
  Network}, vol.~35, no.~6, pp. 270--277, 2021.

\bibitem{guo2020adaptive}
F.~Guo, F.~R. Yu, H.~Zhang, H.~Ji, V.~C. Leung, and X.~Li, ``An adaptive
  wireless virtual reality framework in future wireless networks: A distributed
  learning approach,'' \emph{IEEE Transactions on Vehicular Technology},
  vol.~69, no.~8, pp. 8514--8528, 2020.

\bibitem{feng2020smart}
L.~Feng, Z.~Yang, Y.~Yang, X.~Que, and K.~Zhang, ``Smart mode selection using
  online reinforcement learning for vr broadband broadcasting in d2d assisted
  5g hetnets,'' \emph{IEEE Transactions on Broadcasting}, vol.~66, no.~2, pp.
  600--611, 2020.

\bibitem{gimenez20195g}
J.~J. Gimenez, J.~L. Carcel, M.~Fuentes, E.~Garro, S.~Elliott, D.~Vargas,
  C.~Menzel, and D.~Gomez-Barquero, ``5g new radio for terrestrial broadcast: A
  forward-looking approach for nr-mbms,'' \emph{IEEE Transactions on
  Broadcasting}, vol.~65, no.~2, pp. 356--368, 2019.

\bibitem{elbamby2018edge}
M.~S. Elbamby, C.~Perfecto, M.~Bennis, and K.~Doppler, ``Edge computing meets
  millimeter-wave enabled vr: Paving the way to cutting the cord,'' in
  \emph{2018 IEEE Wireless Communications and Networking Conference
  (WCNC)}.\hskip 1em plus 0.5em minus 0.4em\relax IEEE, 2018, pp. 1--6.

\bibitem{militano2015single}
L.~Militano, M.~Condoluci, G.~Araniti, A.~Molinaro, A.~Iera, and G.-M. Muntean,
  ``Single frequency-based device-to-device-enhanced video delivery for evolved
  multimedia broadcast and multicast services,'' \emph{IEEE Transactions on
  Broadcasting}, vol.~61, no.~2, pp. 263--278, 2015.

\bibitem{zhang2021buffer}
R.~Zhang, J.~Liu, F.~Liu, T.~Huang, Q.~Tang, S.~Wang, and F.~R. Yu,
  ``Buffer-aware virtual reality video streaming with personalized and private
  viewport prediction,'' \emph{IEEE Journal on Selected Areas in
  Communications}, 2021.

\bibitem{wang2022meta}
Y.~Wang, M.~Chen, Z.~Yang, W.~Saad, T.~Luo, S.~Cui, and H.~V. Poor,
  ``Meta-reinforcement learning for reliable communication in thz/vlc wireless
  vr networks,'' \emph{IEEE Transactions on Wireless Communications}, 2022.

\bibitem{xu2021wireless}
M.~Xu, D.~Niyato, J.~Kang, Z.~Xiong, C.~Miao, and D.~I. Kim, ``Wireless
  edge-empowered metaverse: A learning-based incentive mechanism for virtual
  reality,'' \emph{arXiv preprint arXiv:2111.03776}, 2021.

\bibitem{liu2018dare}
Q.~Liu and T.~Han, ``Dare: Dynamic adaptive mobile augmented reality with edge
  computing,'' in \emph{2018 IEEE 26th International Conference on Network
  Protocols (ICNP)}.\hskip 1em plus 0.5em minus 0.4em\relax IEEE, 2018, pp.
  1--11.

\bibitem{liu2018edge}
Q.~Liu, S.~Huang, J.~Opadere, and T.~Han, ``An edge network orchestrator for
  mobile augmented reality,'' in \emph{IEEE INFOCOM 2018-IEEE Conference on
  Computer Communications}.\hskip 1em plus 0.5em minus 0.4em\relax IEEE, 2018,
  pp. 756--764.

\bibitem{ren2020edge}
P.~Ren, X.~Qiao, Y.~Huang, L.~Liu, C.~Pu, S.~Dustdar, and J.-L. Chen, ``Edge ar
  x5: An edge-assisted multi-user collaborative framework for mobile web
  augmented reality in 5g and beyond,'' \emph{IEEE Transactions on Cloud
  Computing}, 2020.

\bibitem{ran2019sharear}
X.~Ran, C.~Slocum, M.~Gorlatova, and J.~Chen, ``Sharear:
  Communication-efficient multi-user mobile augmented reality,'' in
  \emph{Proceedings of the 18th ACM Workshop on Hot Topics in Networks}, 2019,
  pp. 109--116.

\bibitem{mahzari2018fov}
A.~Mahzari, A.~Taghavi~Nasrabadi, A.~Samiei, and R.~Prakash, ``Fov-aware edge
  caching for adaptive 360 video streaming,'' in \emph{Proceedings of the 26th
  ACM international conference on Multimedia}, 2018, pp. 173--181.

\bibitem{zhang2019exploiting}
S.~Zhang, M.~Tao, and Z.~Chen, ``Exploiting caching and prediction to promote
  user experience for a real-time wireless vr service,'' in \emph{2019 IEEE
  Global Communications Conference (GLOBECOM)}.\hskip 1em plus 0.5em minus
  0.4em\relax IEEE, 2019, pp. 1--6.

\bibitem{seo2021novel}
Y.-J. Seo, J.~Lee, J.~Hwang, D.~Niyato, H.-S. Park, and J.~K. Choi, ``A novel
  joint mobile cache and power management scheme for energy-efficient mobile
  augmented reality service in mobile edge computing,'' \emph{IEEE Wireless
  Communications Letters}, vol.~10, no.~5, pp. 1061--1065, 2021.

\bibitem{boabang2021machine}
F.~Boabang, A.~Ebrahimzadeh, R.~H. Glitho, H.~Elbiaze, M.~Maier, and
  F.~Belqasmi, ``A machine learning framework for handling delayed/lost packets
  in tactile internet remote robotic surgery,'' \emph{IEEE Transactions on
  Network and Service Management}, vol.~18, no.~4, pp. 4829--4845, 2021.

\bibitem{hou2018burstiness}
Z.~Hou, C.~She, Y.~Li, T.~Q. Quek, and B.~Vucetic, ``Burstiness-aware bandwidth
  reservation for ultra-reliable and low-latency communications in tactile
  internet,'' \emph{IEEE Journal on Selected Areas in Communications}, vol.~36,
  no.~11, pp. 2401--2410, 2018.

\bibitem{yan2022resource}
L.~Yan, Z.~Qin, R.~Zhang, Y.~Li, and G.~Y. Li, ``Resource allocation for
  semantic-aware networks,'' \emph{arXiv preprint arXiv:2201.06023}, 2022.

\bibitem{liew2022economics}
Z.~Q. Liew, Y.~Cheng, W.~Y.~B. Lim, D.~Niyato, C.~Miao, and S.~Sun, ``Economics
  of semantic communication system in wireless powered internet of things,'' in
  \emph{ICASSP 2022-2022 IEEE International Conference on Acoustics, Speech and
  Signal Processing (ICASSP)}.\hskip 1em plus 0.5em minus 0.4em\relax IEEE,
  2022, pp. 8637--8641.

\bibitem{bastug2017toward}
E.~Bastug, M.~Bennis, M.~M{\'e}dard, and M.~Debbah, ``Toward interconnected
  virtual reality: Opportunities, challenges, and enablers,'' \emph{IEEE
  Communications Magazine}, vol.~55, no.~6, pp. 110--117, 2017.

\bibitem{hu2021virtual}
M.~Hu, X.~Luo, J.~Chen, Y.~C. Lee, Y.~Zhou, and D.~Wu, ``Virtual reality: A
  survey of enabling technologies and its applications in iot,'' \emph{Journal
  of Network and Computer Applications}, vol. 178, p. 102970, 2021.

\bibitem{chung2018hand}
S.-J. Chung, ``Hand pose estimation and prediction for virtual reality
  applications,'' Ph.D. dissertation, Carnegie Mellon University, 2018.

\bibitem{abbas2010constructing}
A.~E. Abbas, ``Constructing multiattribute utility functions for decision
  analysis,'' \emph{Risk and Optimization in an Uncertain World}, pp. 62--98,
  2010.

\bibitem{siriwardhana2021survey}
Y.~Siriwardhana, P.~Porambage, M.~Liyanage, and M.~Ylianttila, ``A survey on
  mobile augmented reality with 5g mobile edge computing: architectures,
  applications, and technical aspects,'' \emph{IEEE Communications Surveys \&
  Tutorials}, vol.~23, no.~2, pp. 1160--1192, 2021.

\bibitem{qiu2018avr}
H.~Qiu, F.~Ahmad, F.~Bai, M.~Gruteser, and R.~Govindan, ``Avr: Augmented
  vehicular reality,'' in \emph{Proceedings of the 16th Annual International
  Conference on Mobile Systems, Applications, and Services}, 2018, pp. 81--95.

\bibitem{liu2019edge}
L.~Liu, H.~Li, and M.~Gruteser, ``Edge assisted real-time object detection for
  mobile augmented reality,'' in \emph{The 25th Annual International Conference
  on Mobile Computing and Networking}, 2019, pp. 1--16.

\bibitem{redmon2017yolo9000}
J.~Redmon and A.~Farhadi, ``Yolo9000: better, faster, stronger,'' in
  \emph{Proceedings of the IEEE conference on computer vision and pattern
  recognition}, 2017, pp. 7263--7271.

\bibitem{liu2016ssd}
W.~Liu, D.~Anguelov, D.~Erhan, C.~Szegedy, S.~Reed, C.-Y. Fu, and A.~C. Berg,
  ``Ssd: Single shot multibox detector,'' in \emph{European conference on
  computer vision}.\hskip 1em plus 0.5em minus 0.4em\relax Springer, 2016, pp.
  21--37.

\bibitem{grippo2000convergence}
L.~Grippo and M.~Sciandrone, ``On the convergence of the block nonlinear
  gauss--seidel method under convex constraints,'' \emph{Operations research
  letters}, vol.~26, no.~3, pp. 127--136, 2000.

\bibitem{hennessy2011computer}
J.~L. Hennessy and D.~A. Patterson, \emph{Computer architecture: a quantitative
  approach}.\hskip 1em plus 0.5em minus 0.4em\relax Elsevier, 2011.

\bibitem{paschos2018role}
G.~S. Paschos, G.~Iosifidis, M.~Tao, D.~Towsley, and G.~Caire, ``The role of
  caching in future communication systems and networks,'' \emph{IEEE Journal on
  Selected Areas in Communications}, vol.~36, no.~6, pp. 1111--1125, 2018.

\bibitem{li2018survey}
L.~Li, G.~Zhao, and R.~S. Blum, ``A survey of caching techniques in cellular
  networks: Research issues and challenges in content placement and delivery
  strategies,'' \emph{IEEE Communications Surveys \& Tutorials}, vol.~20,
  no.~3, pp. 1710--1732, 2018.

\bibitem{gao2020design}
J.~Gao, S.~Zhang, L.~Zhao, and X.~Shen, ``The design of dynamic probabilistic
  caching with time-varying content popularity,'' \emph{IEEE Transactions on
  Mobile Computing}, vol.~20, no.~4, pp. 1672--1684, 2020.

\bibitem{wang2017survey}
S.~Wang, X.~Zhang, Y.~Zhang, L.~Wang, J.~Yang, and W.~Wang, ``A survey on
  mobile edge networks: Convergence of computing, caching and communications,''
  \emph{Ieee Access}, vol.~5, pp. 6757--6779, 2017.

\bibitem{huo2017wireless}
Y.~Huo, P.~T. Kov{\'a}cs, T.~J. Naughton, and L.~Hanzo, ``Wireless holographic
  image communications relying on unequal error protected bitplanes,''
  \emph{IEEE Transactions on Vehicular Technology}, vol.~66, no.~8, pp.
  7136--7148, 2017.

\bibitem{karafin2017support}
J.~Karafin, ``On the support of light field and holographic video display
  technologies,'' \emph{ISO/IEC JTC 1/SC 29/WG 11 Macau, CN}, 2017.

\bibitem{mekuria2016design}
R.~Mekuria, K.~Blom, and P.~Cesar, ``Design, implementation, and evaluation of
  a point cloud codec for tele-immersive video,'' \emph{IEEE Transactions on
  Circuits and Systems for Video Technology}, vol.~27, no.~4, pp. 828--842,
  2016.

\bibitem{fettweis2014tactile}
G.~P. Fettweis, ``The tactile internet: Applications and challenges,''
  \emph{IEEE Vehicular Technology Magazine}, vol.~9, no.~1, pp. 64--70, 2014.

\bibitem{simsek20165g}
M.~Simsek, A.~Aijaz, M.~Dohler, J.~Sachs, and G.~Fettweis, ``5g-enabled tactile
  internet,'' \emph{IEEE Journal on Selected Areas in Communications}, vol.~34,
  no.~3, pp. 460--473, 2016.

\bibitem{hirche2012human}
S.~Hirche and M.~Buss, ``Human-oriented control for haptic teleoperation,''
  \emph{Proceedings of the IEEE}, vol. 100, no.~3, pp. 623--647, 2012.

\bibitem{antonakoglou2018toward}
K.~Antonakoglou, X.~Xu, E.~Steinbach, T.~Mahmoodi, and M.~Dohler, ``Toward
  haptic communications over the 5g tactile internet,'' \emph{IEEE
  Communications Surveys \& Tutorials}, vol.~20, no.~4, pp. 3034--3059, 2018.

\bibitem{sharma2020toward}
S.~K. Sharma, I.~Woungang, A.~Anpalagan, and S.~Chatzinotas, ``Toward tactile
  internet in beyond 5g era: Recent advances, current issues, and future
  directions,'' \emph{IEEE Access}, vol.~8, pp. 56\,948--56\,991, 2020.

\bibitem{lawrence1993stability}
D.~A. Lawrence, ``Stability and transparency in bilateral teleoperation,''
  \emph{IEEE transactions on robotics and automation}, vol.~9, no.~5, pp.
  624--637, 1993.

\bibitem{duan2017human}
L.~Duan, L.~Huang, C.~Langbort, A.~Pozdnukhov, J.~Walrand, and L.~Zhang,
  ``Human-in-the-loop mobile networks: A survey of recent advancements,''
  \emph{IEEE Journal on Selected Areas in Communications}, vol.~35, no.~4, pp.
  813--831, 2017.

\bibitem{gokhale2020tixt}
V.~Gokhale, K.~Kroep, V.~S. Rao, J.~Verburg, and R.~Yechangunja, ``Tixt: An
  extensible testbed for tactile internet communication,'' \emph{IEEE Internet
  of Things Magazine}, vol.~3, no.~1, pp. 32--37, 2020.

\bibitem{polachan2019towards}
K.~Polachan, T.~Prabhakar, C.~Singh, and F.~A. Kuipers, ``Towards an open
  testbed for tactile cyber physical systems,'' in \emph{2019 11th
  International Conference on Communication Systems \& Networks
  (COMSNETS)}.\hskip 1em plus 0.5em minus 0.4em\relax IEEE, 2019, pp. 375--382.

\bibitem{sarathchandra2021enabling}
C.~Sarathchandra, K.~Haensge, S.~Robitzsch, M.~Ghassemian, and
  U.~Olvera-Hernandez, ``Enabling bi-directional haptic control in next
  generation communication systems: Research, standards, and vision,''
  \emph{arXiv preprint arXiv:2104.04297}, 2021.

\bibitem{strinati20216g}
E.~C. Strinati and S.~Barbarossa, ``6g networks: Beyond shannon towards
  semantic and goal-oriented communications,'' \emph{Computer Networks}, vol.
  190, p. 107930, 2021.

\bibitem{qin2021semantic}
Z.~Qin, X.~Tao, J.~Lu, and G.~Y. Li, ``Semantic communications: Principles and
  challenges,'' \emph{arXiv preprint arXiv:2201.01389}, 2021.

\bibitem{yang2022semantic}
W.~Yang, Z.~Q. Liew, W.~Y.~B. Lim, Z.~Xiong, D.~Niyato, X.~Chi, X.~Cao, and
  K.~B. Letaief, ``Semantic communication meets edge intelligence,''
  \emph{arXiv preprint arXiv:2202.06471}, 2022.

\bibitem{qin2019deep}
Z.~Qin, H.~Ye, G.~Y. Li, and B.-H.~F. Juang, ``Deep learning in physical layer
  communications,'' \emph{IEEE Wireless Communications}, vol.~26, no.~2, pp.
  93--99, 2019.

\bibitem{kalfa2021towards}
M.~Kalfa, M.~Gok, A.~Atalik, B.~Tegin, T.~M. Duman, and O.~Arikan, ``Towards
  goal-oriented semantic signal processing: Applications and future
  challenges,'' \emph{Digital Signal Processing}, vol. 119, p. 103134, 2021.

\bibitem{bourtsoulatze2019deep}
E.~Bourtsoulatze, D.~B. Kurka, and D.~G{\"u}nd{\"u}z, ``Deep joint
  source-channel coding for wireless image transmission,'' \emph{IEEE
  Transactions on Cognitive Communications and Networking}, vol.~5, no.~3, pp.
  567--579, 2019.

\bibitem{xie2021deep}
H.~Xie, Z.~Qin, G.~Y. Li, and B.-H. Juang, ``Deep learning enabled semantic
  communication systems,'' \emph{IEEE Transactions on Signal Processing},
  vol.~69, pp. 2663--2675, 2021.

\bibitem{papineni2002bleu}
K.~Papineni, S.~Roukos, T.~Ward, and W.-J. Zhu, ``Bleu: a method for automatic
  evaluation of machine translation,'' in \emph{Proceedings of the 40th annual
  meeting of the Association for Computational Linguistics}, 2002, pp.
  311--318.

\bibitem{han2021dynamic}
Y.~Han, D.~Niyato, C.~Leung, C.~Miao, and D.~I. Kim, ``A dynamic resource
  allocation framework for synchronizing metaverse with iot service and data,''
  \emph{arXiv preprint arXiv:2111.00431}, 2021.

\bibitem{cheng2019space}
N.~Cheng, F.~Lyu, W.~Quan, C.~Zhou, H.~He, W.~Shi, and X.~Shen,
  ``Space/aerial-assisted computing offloading for iot applications: A
  learning-based approach,'' \emph{IEEE Journal on Selected Areas in
  Communications}, vol.~37, no.~5, pp. 1117--1129, 2019.

\bibitem{cheng2018air}
N.~Cheng, W.~Xu, W.~Shi, Y.~Zhou, N.~Lu, H.~Zhou, and X.~Shen, ``Air-ground
  integrated mobile edge networks: Architecture, challenges, and
  opportunities,'' \emph{IEEE Communications Magazine}, vol.~56, no.~8, pp.
  26--32, 2018.

\bibitem{shen2021holistic}
X.~Shen, J.~Gao, W.~Wu, M.~Li, C.~Zhou, and W.~Zhuang, ``Holistic network
  virtualization and pervasive network intelligence for 6g,'' \emph{IEEE
  Communications Surveys \& Tutorials}, 2021.

\bibitem{sun2021dynamic}
W.~Sun, P.~Wang, N.~Xu, G.~Wang, and Y.~Zhang, ``Dynamic digital twin and
  distributed incentives for resource allocation in aerial-assisted internet of
  vehicles,'' \emph{IEEE Internet of Things Journal}, 2021.

\bibitem{khan2022digital}
L.~U. Khan, W.~Saad, D.~Niyato, Z.~Han, and C.~S. Hong, ``Digital-twin-enabled
  6g: Vision, architectural trends, and future directions,'' \emph{IEEE
  Communications Magazine}, vol.~60, no.~1, pp. 74--80, 2022.

\bibitem{li2021slicing}
M.~Li, J.~Gao, C.~Zhou, X.~S. Shen, and W.~Zhuang, ``Slicing-based artificial
  intelligence service provisioning on the network edge: Balancing ai service
  performance and resource consumption of data management,'' \emph{IEEE
  Vehicular Technology Magazine}, vol.~16, no.~4, pp. 16--26, 2021.

\bibitem{dong2019deep}
R.~Dong, C.~She, W.~Hardjawana, Y.~Li, and B.~Vucetic, ``Deep learning for
  hybrid 5g services in mobile edge computing systems: Learn from a digital
  twin,'' \emph{IEEE Transactions on Wireless Communications}, vol.~18, no.~10,
  pp. 4692--4707, 2019.

\bibitem{wang2020graph}
H.~Wang, Y.~Wu, G.~Min, and W.~Miao, ``A graph neural network-based digital
  twin for network slicing management,'' \emph{IEEE Transactions on Industrial
  Informatics}, vol.~18, no.~2, pp. 1367--1376, 2020.

\bibitem{renzo2019smart}
M.~D. Renzo, M.~Debbah, D.-T. Phan-Huy, A.~Zappone, M.-S. Alouini, C.~Yuen,
  V.~Sciancalepore, G.~C. Alexandropoulos, J.~Hoydis, H.~Gacanin \emph{et~al.},
  ``Smart radio environments empowered by reconfigurable ai meta-surfaces: An
  idea whose time has come,'' \emph{EURASIP Journal on Wireless Communications
  and Networking}, vol. 2019, no.~1, pp. 1--20, 2019.

\bibitem{sheen2020digital}
B.~Sheen, J.~Yang, X.~Feng, and M.~M.~U. Chowdhury, ``A digital twin for
  reconfigurable intelligent surface assisted wireless communication,''
  \emph{arXiv preprint arXiv:2009.00454}, 2020.

\bibitem{zhou2020deep}
C.~Zhou, W.~Wu, H.~He, P.~Yang, F.~Lyu, N.~Cheng, and X.~Shen, ``Deep
  reinforcement learning for delay-oriented iot task scheduling in sagin,''
  \emph{IEEE Transactions on Wireless Communications}, vol.~20, no.~2, pp.
  911--925, 2020.

\bibitem{yin2021physical}
Z.~Yin, N.~Cheng, T.~H. Luan, and P.~Wang, ``Physical layer security in
  cybertwin-enabled integrated satellite-terrestrial vehicle networks,''
  \emph{IEEE Transactions on Vehicular Technology}, 2021.

\bibitem{han2022dynamic}
Y.~Han, D.~Niyato, C.~Leung, D.~I. Kim, K.~Zhu, S.~Feng, S.~X. Shen, and
  C.~Miao, ``A dynamic hierarchical framework for iot-assisted digital twin
  synchronization in the metaverse,'' \emph{IEEE Internet of Things Journal},
  2022.

\bibitem{greitzer2008combating}
F.~L. Greitzer, A.~P. Moore, D.~M. Cappelli, D.~H. Andrews, L.~A. Carroll, and
  T.~D. Hull, ``Combating the insider cyber threat,'' \emph{IEEE Security \&
  Privacy}, vol.~6, no.~1, pp. 61--64, 2008.

\bibitem{hendaoui20083d}
A.~Hendaoui, M.~Limayem, and C.~W. Thompson, ``3d social virtual worlds:
  research issues and challenges,'' \emph{IEEE internet computing}, vol.~12,
  no.~1, pp. 88--92, 2008.

\bibitem{liang20162015}
G.~Liang, S.~R. Weller, J.~Zhao, F.~Luo, and Z.~Y. Dong, ``The 2015 ukraine
  blackout: Implications for false data injection attacks,'' \emph{IEEE
  Transactions on Power Systems}, vol.~32, no.~4, pp. 3317--3318, 2016.

\bibitem{su2020lvbs}
Z.~Su, Y.~Wang, Q.~Xu, and N.~Zhang, ``Lvbs: Lightweight vehicular blockchain
  for secure data sharing in disaster rescue,'' \emph{IEEE Transactions on
  dependable and secure computing}, 2020.

\bibitem{guo2017availability}
H.~Guo, Y.~Yu, T.~Xiang, H.~Li, and D.~Zhang, ``The availability of
  wearable-device-based physical data for the measurement of construction
  workers' psychological status on site: From the perspective of safety
  management,'' \emph{Automation in Construction}, vol.~82, pp. 207--217, 2017.

\bibitem{kumar2008second}
S.~Kumar, J.~Chhugani, C.~Kim, D.~Kim, A.~Nguyen, P.~Dubey, C.~Bienia, and
  Y.~Kim, ``Second life and the new generation of virtual worlds,''
  \emph{Computer}, vol.~41, no.~9, pp. 46--53, 2008.

\bibitem{liang2017provchain}
X.~Liang, S.~Shetty, D.~Tosh, C.~Kamhoua, K.~Kwiat, and L.~Njilla, ``Provchain:
  A blockchain-based data provenance architecture in cloud environment with
  enhanced privacy and availability,'' in \emph{2017 17th IEEE/ACM
  International Symposium on Cluster, Cloud and Grid Computing (CCGRID)}.\hskip
  1em plus 0.5em minus 0.4em\relax IEEE, 2017, pp. 468--477.

\bibitem{shang2020arspy}
J.~Shang, S.~Chen, J.~Wu, and S.~Yin, ``Arspy: Breaking location-based
  multi-player augmented reality application for user location tracking,''
  \emph{IEEE Transactions on Mobile Computing}, 2020.

\bibitem{ometov2016facilitating}
A.~Ometov, S.~V. Bezzateev, J.~Kannisto, J.~Harju, S.~Andreev, and
  Y.~Koucheryavy, ``Facilitating the delegation of use for private devices in
  the era of the internet of wearable things,'' \emph{IEEE Internet of Things
  Journal}, vol.~4, no.~4, pp. 843--854, 2016.

\bibitem{wasserkrug2008inference}
S.~Wasserkrug, A.~Gal, and O.~Etzion, ``Inference of security hazards from
  event composition based on incomplete or uncertain information,'' \emph{IEEE
  transactions on knowledge and data engineering}, vol.~20, no.~8, pp.
  1111--1114, 2008.

\bibitem{wei2020ldp}
J.~Wei, J.~Li, Y.~Lin, and J.~Zhang, ``Ldp-based social content protection for
  trending topic recommendation,'' \emph{IEEE Internet of Things Journal},
  vol.~8, no.~6, pp. 4353--4372, 2020.

\bibitem{li2021verifiable}
X.~Li, J.~He, P.~Vijayakumar, X.~Zhang, and V.~Chang, ``A verifiable
  privacy-preserving machine learning prediction scheme for edge-enhanced
  hcpss,'' \emph{IEEE Transactions on Industrial Informatics}, vol.~18, no.~8,
  pp. 5494--5503, 2021.

\bibitem{bertino2017botnets}
E.~Bertino and N.~Islam, ``Botnets and internet of things security,''
  \emph{Computer}, vol.~50, no.~2, pp. 76--79, 2017.

\bibitem{kirkpatrick2006metaverse}
M.~Kirkpatrick, ``Metaverse breached: Second life customer database hacked,''
  \emph{Retrieved January}, vol.~11, p. 2011, 2006.

\bibitem{yu2018leveraging}
J.~Yu, Z.~Kuang, B.~Zhang, W.~Zhang, D.~Lin, and J.~Fan, ``Leveraging content
  sensitiveness and user trustworthiness to recommend fine-grained privacy
  settings for social image sharing,'' \emph{IEEE transactions on information
  forensics and security}, vol.~13, no.~5, pp. 1317--1332, 2018.

\bibitem{ali2018applications}
M.~S. Ali, M.~Vecchio, M.~Pincheira, K.~Dolui, F.~Antonelli, and M.~H. Rehmani,
  ``Applications of blockchains in the internet of things: A comprehensive
  survey,'' \emph{IEEE Communications Surveys \& Tutorials}, vol.~21, no.~2,
  pp. 1676--1717, 2018.

\bibitem{zhang2014sybil}
K.~Zhang, X.~Liang, R.~Lu, and X.~Shen, ``Sybil attacks and their defenses in
  the internet of things,'' \emph{IEEE Internet of Things Journal}, vol.~1,
  no.~5, pp. 372--383, 2014.

\bibitem{ritzdorf2018toward}
H.~Ritzdorf, C.~Soriente, G.~O. Karame, S.~Marinovic, D.~Gruber, and S.~Capkun,
  ``Toward shared ownership in the cloud,'' \emph{IEEE Transactions on
  Information Forensics and Security}, vol.~13, no.~12, pp. 3019--3034, 2018.

\bibitem{de2018swarm}
L.~C.~C. De~Biase, P.~C. Calcina-Ccori, G.~Fedrecheski, G.~M. Duarte, P.~S.~S.
  Rangel, and M.~K. Zuffo, ``Swarm economy: a model for transactions in a
  distributed and organic iot platform,'' \emph{IEEE Internet of Things
  Journal}, vol.~6, no.~3, pp. 4561--4572, 2018.

\bibitem{suhail2021trustworthy}
S.~Suhail, R.~Hussain, R.~Jurdak, and C.~S. Hong, ``Trustworthy digital twins
  in the industrial internet of things with blockchain,'' \emph{IEEE Internet
  Computing}, 2021.

\bibitem{zhang2021privacy}
M.~Zhang, L.~Yang, S.~He, M.~Li, and J.~Zhang, ``Privacy-preserving data
  aggregation for mobile crowdsensing with externality: an auction approach,''
  \emph{IEEE/ACM Transactions on Networking}, vol.~29, no.~3, pp. 1046--1059,
  2021.

\bibitem{xu2014collusion}
Z.~Xu and W.~Liang, ``Collusion-resistant repeated double auctions for relay
  assignment in cooperative networks,'' \emph{IEEE Transactions on Wireless
  Communications}, vol.~13, no.~3, pp. 1196--1207, 2014.

\bibitem{li2008free}
M.~Li, J.~Yu, and J.~Wu, ``Free-riding on bittorrent-like peer-to-peer file
  sharing systems: Modeling analysis and improvement,'' \emph{IEEE Transactions
  on Parallel and Distributed Systems}, vol.~19, no.~7, pp. 954--966, 2008.

\bibitem{zhou2019cyber}
Y.~Zhou, F.~R. Yu, J.~Chen, and Y.~Kuo, ``Cyber-physical-social systems: A
  state-of-the-art survey, challenges and opportunities,'' \emph{IEEE
  Communications Surveys \& Tutorials}, vol.~22, no.~1, pp. 389--425, 2019.

\bibitem{casey2019immersive}
P.~Casey, I.~Baggili, and A.~Yarramreddy, ``Immersive virtual reality attacks
  and the human joystick,'' \emph{IEEE Transactions on Dependable and Secure
  Computing}, vol.~18, no.~2, pp. 550--562, 2019.

\bibitem{d2022terrorist}
E.~d'Argenlieu, ``Terrorist use of the metaverse: new opportunities and new
  challenges,'' \emph{Technology}, 2022.

\bibitem{zhu2020activity}
J.~Zhu, P.~Ni, and G.~Wang, ``Activity minimization of misinformation influence
  in online social networks,'' \emph{IEEE Transactions on Computational Social
  Systems}, vol.~7, no.~4, pp. 897--906, 2020.

\bibitem{valluripally2021modeling}
S.~Valluripally, A.~Gulhane, K.~A. Hoque, and P.~Calyam, ``Modeling and defense
  of social virtual reality attacks inducing cybersickness,'' \emph{IEEE
  Transactions on Dependable and Secure Computing}, 2021.

\bibitem{vellaithurai2014cpindex}
C.~Vellaithurai, A.~Srivastava, S.~Zonouz, and R.~Berthier, ``Cpindex:
  Cyber-physical vulnerability assessment for power-grid infrastructures,''
  \emph{IEEE Transactions on Smart Grid}, vol.~6, no.~2, pp. 566--575, 2014.

\bibitem{almeida2021ecosystem}
V.~Almeida, F.~Filgueiras, and D.~Doneda, ``The ecosystem of digital content
  governance,'' \emph{IEEE Internet Computing}, vol.~25, no.~3, pp. 13--17,
  2021.

\bibitem{bai2021public}
Y.~Bai, Q.~Hu, S.-H. Seo, K.~Kang, and J.~J. Lee, ``Public participation
  consortium blockchain for smart city governance,'' \emph{IEEE Internet of
  Things Journal}, 2021.

\bibitem{sayeed2020smart}
S.~Sayeed, H.~Marco-Gisbert, and T.~Caira, ``Smart contract: Attacks and
  protections,'' \emph{IEEE Access}, vol.~8, pp. 24\,416--24\,427, 2020.

\bibitem{li2021towards}
M.~Li, J.~Weng, J.-N. Liu, X.~Lin, and C.~Obimbo, ``Towards vehicular digital
  forensics from decentralized trust: an accountable, privacy-preserving, and
  secure realization,'' \emph{IEEE Internet of Things Journal}, 2021.

\bibitem{li2017secret}
Z.~Li, Q.~Pei, I.~Markwood, Y.~Liu, and H.~Zhu, ``Secret key establishment via
  rss trajectory matching between wearable devices,'' \emph{IEEE Transactions
  on Information Forensics and security}, vol.~13, no.~3, pp. 802--817, 2017.

\bibitem{liu2018cooperative}
H.~Liu, X.~Yao, T.~Yang, and H.~Ning, ``Cooperative privacy preservation for
  wearable devices in hybrid computing-based smart health,'' \emph{IEEE
  Internet of Things Journal}, vol.~6, no.~2, pp. 1352--1362, 2018.

\bibitem{shen2020blockchain}
M.~Shen, H.~Liu, L.~Zhu, K.~Xu, H.~Yu, X.~Du, and M.~Guizani,
  ``Blockchain-assisted secure device authentication for cross-domain
  industrial iot,'' \emph{IEEE Journal on Selected Areas in Communications},
  vol.~38, no.~5, pp. 942--954, 2020.

\bibitem{chen2021xauth}
J.~Chen, Z.~Zhan, K.~He, R.~Du, D.~Wang, and F.~Liu, ``Xauth: Efficient
  privacy-preserving cross-domain authentication,'' \emph{IEEE Transactions on
  Dependable and Secure Computing}, 2021.

\bibitem{gehrmann2019digital}
C.~Gehrmann and M.~Gunnarsson, ``A digital twin based industrial automation and
  control system security architecture,'' \emph{IEEE Transactions on Industrial
  Informatics}, vol.~16, no.~1, pp. 669--680, 2019.

\bibitem{ruth2019secure}
K.~Ruth, T.~Kohno, and F.~Roesner, ``Secure $\{$Multi-User$\}$ content sharing
  for augmented reality applications,'' in \emph{28th USENIX Security Symposium
  (USENIX Security 19)}, 2019, pp. 141--158.

\bibitem{lv2020industrial}
Z.~Lv, D.~Chen, R.~Lou, and H.~Song, ``Industrial security solution for virtual
  reality,'' \emph{IEEE Internet of Things Journal}, vol.~8, no.~8, pp.
  6273--6281, 2020.

\bibitem{shahsavari2019situational}
A.~Shahsavari, M.~Farajollahi, E.~M. Stewart, E.~Cortez, and H.~Mohsenian-Rad,
  ``Situational awareness in distribution grid using micro-pmu data: A machine
  learning approach,'' \emph{IEEE Transactions on Smart Grid}, vol.~10, no.~6,
  pp. 6167--6177, 2019.

\bibitem{li2019toward}
M.~Li, J.~Weng, A.~Yang, J.-N. Liu, and X.~Lin, ``Toward blockchain-based fair
  and anonymous ad dissemination in vehicular networks,'' \emph{IEEE
  Transactions on Vehicular Technology}, vol.~68, no.~11, pp. 11\,248--11\,259,
  2019.

\bibitem{jiang2021cooperative}
L.~Jiang, H.~Zheng, H.~Tian, S.~Xie, and Y.~Zhang, ``Cooperative federated
  learning and model update verification in blockchain empowered digital twin
  edge networks,'' \emph{IEEE Internet of Things Journal}, 2021.

\bibitem{lau2021coalitional}
P.~Lau, L.~Wang, Z.~Liu, W.~Wei, and C.-W. Ten, ``A coalitional cyber-insurance
  design considering power system reliability and cyber vulnerability,''
  \emph{IEEE Transactions on Power Systems}, vol.~36, no.~6, pp. 5512--5524,
  2021.

\bibitem{he2021datingsec}
X.~He, Q.~Gong, Y.~Chen, Y.~Zhang, X.~Wang, and X.~Fu, ``Datingsec: Detecting
  malicious accounts in dating apps using a content-based attention network,''
  \emph{IEEE Transactions on Dependable and Secure Computing}, vol.~18, no.~5,
  pp. 2193--2208, 2021.

\bibitem{zou2018multigranularity}
D.~Zou, J.~Zhao, W.~Li, Y.~Wu, W.~Qiang, H.~Jin, Y.~Wu, and Y.~Yang, ``A
  multigranularity forensics and analysis method on privacy leakage in cloud
  environment,'' \emph{IEEE Internet of Things Journal}, vol.~6, no.~2, pp.
  1484--1494, 2018.

\bibitem{wang2022metasocieties}
F.-Y. Wang, R.~Qin, X.~Wang, and B.~Hu, ``Metasocieties in metaverse:
  Metaeconomics and metamanagement for metaenterprises and metacities,''
  \emph{IEEE Transactions on Computational Social Systems}, vol.~9, no.~1, pp.
  2--7, 2022.

\bibitem{tao2018secured}
H.~Tao, M.~Z.~A. Bhuiyan, A.~N. Abdalla, M.~M. Hassan, J.~M. Zain, and
  T.~Hayajneh, ``Secured data collection with hardware-based ciphers for
  iot-based healthcare,'' \emph{IEEE Internet of Things Journal}, vol.~6,
  no.~1, pp. 410--420, 2018.

\bibitem{islam2019buav}
A.~Islam and S.~Y. Shin, ``Buav: A blockchain based secure uav-assisted data
  acquisition scheme in internet of things,'' \emph{Journal of Communications
  and Networks}, vol.~21, no.~5, pp. 491--502, 2019.

\bibitem{deepa2022survey}
N.~Deepa, Q.-V. Pham, D.~C. Nguyen, S.~Bhattacharya, B.~Prabadevi, T.~R.
  Gadekallu, P.~K.~R. Maddikunta, F.~Fang, and P.~N. Pathirana, ``A survey on
  blockchain for big data: approaches, opportunities, and future directions,''
  \emph{Future Generation Computer Systems}, 2022.

\bibitem{xu2021light}
C.~Xu, Y.~Qu, T.~H. Luan, P.~W. Eklund, Y.~Xiang, and L.~Gao, ``A light-weight
  and attack-proof bidirectional blockchain paradigm for internet of things,''
  \emph{IEEE Internet of Things Journal}, 2021.

\bibitem{bouraga2021taxonomy}
S.~Bouraga, ``A taxonomy of blockchain consensus protocols: A survey and
  classification framework,'' \emph{Expert Systems with Applications}, vol.
  168, p. 114384, 2021.

\bibitem{lashkari2021comprehensive}
B.~Lashkari and P.~Musilek, ``A comprehensive review of blockchain consensus
  mechanisms,'' \emph{IEEE Access}, vol.~9, pp. 43\,620--43\,652, 2021.

\bibitem{zhang2021research}
L.~Zhang, Z.~Zhang, W.~Wang, Z.~Jin, Y.~Su, and H.~Chen, ``Research on a covert
  communication model realized by using smart contracts in blockchain
  environment,'' \emph{IEEE Systems Journal}, 2021.

\bibitem{guo2021reliable}
J.~Guo, X.~Ding, and W.~Wu, ``Reliable traffic monitoring mechanisms based on
  blockchain in vehicular networks,'' \emph{IEEE Transactions on Reliability},
  2021.

\bibitem{xu2021latency}
X.~Xu, G.~Sun, L.~Luo, H.~Cao, H.~Yu, and A.~V. Vasilakos, ``Latency
  performance modeling and analysis for hyperledger fabric blockchain
  network,'' \emph{Information Processing \& Management}, vol.~58, no.~1, p.
  102436, 2021.

\bibitem{alrubei2020latency}
S.~M. Alrubei, E.~A. Ball, J.~M. Rigelsford, and C.~A. Willis, ``Latency and
  performance analyses of real-world wireless iot-blockchain application,''
  \emph{IEEE Sensors Journal}, vol.~20, no.~13, pp. 7372--7383, 2020.

\bibitem{chen2022blockchain}
L.~Chen, Q.~Fu, Y.~Mu, L.~Zeng, F.~Rezaeibagha, and M.-S. Hwang,
  ``Blockchain-based random auditor committee for integrity verification,''
  \emph{Future Generation Computer Systems}, vol. 131, pp. 183--193, 2022.

\bibitem{bian2021demystifying}
Y.~Bian, J.~Leng, and J.~L. Zhao, ``Demystifying metaverse as a new paradigm of
  enterprise digitization,'' in \emph{International Conference on Big
  Data}.\hskip 1em plus 0.5em minus 0.4em\relax Springer, 2021, pp. 109--119.

\bibitem{xie2019survey}
J.~Xie, F.~R. Yu, T.~Huang, R.~Xie, J.~Liu, and Y.~Liu, ``A survey on the
  scalability of blockchain systems,'' \emph{IEEE Network}, vol.~33, no.~5, pp.
  166--173, 2019.

\bibitem{kraus2022facebook}
S.~Kraus, D.~K. Kanbach, P.~M. Krysta, M.~M. Steinhoff, and N.~Tomini,
  ``Facebook and the creation of the metaverse: radical business model
  innovation or incremental transformation?'' \emph{International Journal of
  Entrepreneurial Behavior \& Research}, 2022.

\bibitem{liu2020blockchain}
L.~Liu, J.~Feng, Q.~Pei, C.~Chen, Y.~Ming, B.~Shang, and M.~Dong,
  ``Blockchain-enabled secure data sharing scheme in mobile-edge computing: an
  asynchronous advantage actor--critic learning approach,'' \emph{IEEE Internet
  of Things Journal}, vol.~8, no.~4, pp. 2342--2353, 2020.

\bibitem{egliston2021critical}
B.~Egliston and M.~Carter, ``Critical questions for facebook’s virtual
  reality: data, power and the metaverse,'' \emph{Internet Policy Review},
  vol.~10, no.~4, 2021.

\bibitem{yu2021blockchain}
K.~Yu, L.~Tan, M.~Aloqaily, H.~Yang, and Y.~Jararweh, ``Blockchain-enhanced
  data sharing with traceable and direct revocation in iiot,'' \emph{IEEE
  transactions on industrial informatics}, vol.~17, no.~11, pp. 7669--7678,
  2021.

\bibitem{rashid2021blockchain}
A.~Rashid, A.~Masood, H.~Abbas, and Y.~Zhang, ``Blockchain-based public key
  infrastructure: A transparent digital certification mechanism for secure
  communication,'' \emph{IEEE Network}, vol.~35, no.~5, pp. 220--225, 2021.

\bibitem{vashistha2021echain}
N.~Vashistha, M.~M. Hossain, M.~R. Shahriar, F.~Farahmandi, F.~Rahman, and
  M.~Tehranipoor, ``echain: A blockchain-enabled ecosystem for electronic
  device authenticity verification,'' \emph{IEEE Transactions on Consumer
  Electronics}, 2021.

\bibitem{min2022portrait}
T.~Min and W.~Cai, ``Portrait of decentralized application users: an overview
  based on large-scale ethereum data,'' \emph{CCF Transactions on Pervasive
  Computing and Interaction}, pp. 1--18, 2022.

\bibitem{ali2021comparative}
O.~Ali, A.~Jaradat, A.~Kulakli, and A.~Abuhalimeh, ``A comparative study:
  Blockchain technology utilization benefits, challenges and functionalities,''
  \emph{IEEE Access}, vol.~9, pp. 12\,730--12\,749, 2021.

\bibitem{luo2019blockchain}
Y.~Luo, H.~Jin, and P.~Li, ``A blockchain future for secure clinical data
  sharing: A position paper,'' in \emph{Proceedings of the ACM International
  Workshop on Security in Software Defined Networks \& Network Function
  Virtualization}, 2019, pp. 23--27.

\bibitem{gao2021b}
Y.~Gao, W.~Wu, P.~Si, Z.~Yang, and F.~R. Yu, ``B-rest: Blockchain-enabled
  resource sharing and transactions in fog computing,'' \emph{IEEE Wireless
  Communications}, vol.~28, no.~2, pp. 172--180, 2021.

\bibitem{belchior2021survey}
R.~Belchior, A.~Vasconcelos, S.~Guerreiro, and M.~Correia, ``A survey on
  blockchain interoperability: Past, present, and future trends,'' \emph{ACM
  Computing Surveys (CSUR)}, vol.~54, no.~8, pp. 1--41, 2021.

\bibitem{madine2021appxchain}
M.~Madine, K.~Salah, R.~Jayaraman, Y.~Al-Hammadi, J.~Arshad, and I.~Yaqoob,
  ``appxchain: Application-level interoperability for blockchain networks,''
  \emph{IEEE Access}, vol.~9, pp. 87\,777--87\,791, 2021.

\bibitem{jabbar2020blockchain}
R.~Jabbar, N.~Fetais, M.~Krichen, and K.~Barkaoui, ``Blockchain technology for
  healthcare: Enhancing shared electronic health record interoperability and
  integrity,'' in \emph{2020 IEEE International Conference on Informatics, IoT,
  and Enabling Technologies (ICIoT)}.\hskip 1em plus 0.5em minus 0.4em\relax
  IEEE, 2020, pp. 310--317.

\bibitem{siyaev2021towards}
A.~Siyaev and G.-S. Jo, ``Towards aircraft maintenance metaverse using speech
  interactions with virtual objects in mixed reality,'' \emph{Sensors},
  vol.~21, no.~6, p. 2066, 2021.

\bibitem{arvas2022gutenberg}
{\.I}.~S. ARVAS, ``Gutenberg galaksisinden meta evrenine:
  {\"U}{\c{c}}{\"u}nc{\"u} ku{\c{s}}ak {\.i}nternet, web 3.0,'' \emph{AJIT-e:
  Bili{\c{s}}im Teknolojileri Online Dergisi}, vol.~13, no.~48, pp. 53--70,
  2022.

\bibitem{kostenko2022electronic}
O.~Kostenko, ``Electronic jurisdiction, metaverse, artificial intelligence,
  digital personality, digital avatar, neural networks: theory, practice,
  perspective,'' \emph{World Science}, no. 1 (73), 2022.

\bibitem{kumar2021ppsf}
P.~Kumar, R.~Kumar, G.~Srivastava, G.~P. Gupta, R.~Tripathi, T.~R. Gadekallu,
  and N.~N. Xiong, ``Ppsf: a privacy-preserving and secure framework using
  blockchain-based machine-learning for iot-driven smart cities,'' \emph{IEEE
  Transactions on Network Science and Engineering}, vol.~8, no.~3, pp.
  2326--2341, 2021.

\bibitem{hassan2019privacy}
M.~U. Hassan, M.~H. Rehmani, and J.~Chen, ``Privacy preservation in blockchain
  based iot systems: Integration issues, prospects, challenges, and future
  research directions,'' \emph{Future Generation Computer Systems}, vol.~97,
  pp. 512--529, 2019.

\bibitem{ramu2022federated}
S.~P. Ramu, P.~Boopalan, Q.-V. Pham, P.~K.~R. Maddikunta, T.-H. The, M.~Alazab,
  T.~T. Nguyen, and T.~R. Gadekallu, ``Federated learning enabled digital twins
  for smart cities: Concepts, recent advances, and future directions,''
  \emph{Sustainable Cities and Society}, p. 103663, 2022.

\bibitem{chen2021research}
Q.~Chen and S.-J. Lee, ``Research status and trend of digital twin: Visual
  knowledge mapping analysis,'' \emph{International journal of advanced smart
  convergence}, vol.~10, no.~4, pp. 84--97, 2021.

\bibitem{zhuang2021digital}
C.~Zhuang, J.~Gong, and J.~Liu, ``Digital twin-based assembly data management
  and process traceability for complex products,'' \emph{Journal of
  manufacturing systems}, vol.~58, pp. 118--131, 2021.

\bibitem{lee2021integrated}
D.~Lee, S.~H. Lee, N.~Masoud, M.~Krishnan, and V.~C. Li, ``Integrated digital
  twin and blockchain framework to support accountable information sharing in
  construction projects,'' \emph{Automation in construction}, vol. 127, p.
  103688, 2021.

\bibitem{shen2021secure}
W.~Shen, T.~Hu, C.~Zhang, and S.~Ma, ``Secure sharing of big digital twin data
  for smart manufacturing based on blockchain,'' \emph{Journal of Manufacturing
  Systems}, vol.~61, pp. 338--350, 2021.

\bibitem{wang2021explainable}
S.~Wang, M.~A. Qureshi, L.~Miralles-Pechua{\'a}n, T.~Huynh-The, T.~R.
  Gadekallu, and M.~Liyanage, ``Explainable ai for b5g/6g: Technical aspects,
  use cases, and research challenges,'' \emph{arXiv preprint arXiv:2112.04698},
  2021.

\bibitem{song2021build}
Y.~Song and S.~Hong, ``Build a secure smart city by using blockchain and
  digital twin,'' \emph{Int. J. Adv. Sci. Converg}, vol.~3, pp. 9--13, 2021.

\bibitem{kumar2022virtual}
S.~Kumar, ``Virtual power plants: Metaverse for the power sector,'' \emph{.},
  2022.

\bibitem{chen2022exploring}
D.~Chen and R.~Zhang, ``Exploring research trends of emerging technologies in
  health metaverse: A bibliometric analysis,'' \emph{Available at SSRN
  3998068}, 2022.

\bibitem{owens2011empirical}
D.~Owens, A.~Mitchell, D.~Khazanchi, and I.~Zigurs, ``An empirical
  investigation of virtual world projects and metaverse technology
  capabilities,'' \emph{ACM SIGMIS Database: the DATABASE for Advances in
  Information Systems}, vol.~42, no.~1, pp. 74--101, 2011.

\bibitem{furstenau2019process}
D.~F{\"u}rstenau, C.~Auschra, S.~Klein, and M.~Gersch, ``A process perspective
  on platform design and management: evidence from a digital platform in health
  care,'' \emph{Electronic Markets}, vol.~29, no.~4, pp. 581--596, 2019.

\bibitem{cheung2020disambiguating}
S.~Cheung, ``Disambiguating the benefits and risks from public health data in
  the digital economy,'' \emph{Big Data \& Society}, vol.~7, no.~1, p.
  2053951720933924, 2020.

\bibitem{dhelim2022edge}
S.~Dhelim, T.~Kechadi, L.~Chen, N.~Aung, H.~Ning, and L.~Atzori, ``Edge-enabled
  metaverse: The convergence of metaverse and mobile edge computing,''
  \emph{arXiv preprint arXiv:2205.02764}, 2022.

\bibitem{gupta2017ifogsim}
H.~Gupta, A.~Vahid~Dastjerdi, S.~K. Ghosh, and R.~Buyya, ``ifogsim: A toolkit
  for modeling and simulation of resource management techniques in the internet
  of things, edge and fog computing environments,'' \emph{Software: Practice
  and Experience}, vol.~47, no.~9, pp. 1275--1296, 2017.

\bibitem{24}
L.~Evans, J.~Frith, and M.~Saker, ``Entertainment worlds,'' in \emph{From
  Microverse to Metaverse}.\hskip 1em plus 0.5em minus 0.4em\relax Emerald
  Publishing Limited, 2022, pp. 65--73.

\bibitem{6}
F.~Richter, ``Infographic: Gaming: The most lucrative entertainment industry by
  far,'' \emph{Retrieved July}, vol.~18, p. 2021, 2020.

\bibitem{33}
J.~N. Njoku, C.~I. Nwakanma, G.~C. Amaizu, and D.-S. Kim, ``Prospects and
  challenges of metaverse application in data-driven intelligent transportation
  systems,'' \emph{IET Intelligent Transport Systems}, 2022.

\bibitem{m44}
L.~Petrigna and G.~Musumeci, ``The metaverse: A new challenge for the
  healthcare system: A scoping review,'' \emph{Journal of Functional Morphology
  and Kinesiology}, vol.~7, no.~3, p.~63, 2022.

\bibitem{42}
M.~Xu, W.~C. Ng, W.~Y.~B. Lim, J.~Kang, Z.~Xiong, D.~Niyato, Q.~Yang, X.~S.
  Shen, and C.~Miao, ``A full dive into realizing the edge-enabled metaverse:
  Visions, enabling technologies, and challenges,'' \emph{IEEE Communications
  Surveys \& Tutorials}, 2022.

\bibitem{gadekallu2022blockchain}
T.~R. Gadekallu, T.~Huynh-The, W.~Wang, G.~Yenduri, P.~Ranaweera, Q.-V. Pham,
  D.~B. da~Costa, and M.~Liyanage, ``Blockchain for the metaverse: A review,''
  \emph{arXiv preprint arXiv:2203.09738}, 2022.

\bibitem{chang20226g}
L.~Chang, Z.~Zhang, P.~Li, S.~Xi, W.~Guo, Y.~Shen, Z.~Xiong, J.~Kang,
  D.~Niyato, X.~Qiao \emph{et~al.}, ``6g-enabled edge ai for metaverse:
  Challenges, methods, and future research directions,'' \emph{arXiv preprint
  arXiv:2204.06192}, 2022.

\bibitem{slater2020ethics}
M.~Slater, C.~Gonzalez-Liencres, P.~Haggard, C.~Vinkers, R.~Gregory-Clarke,
  S.~Jelley, Z.~Watson, G.~Breen, R.~Schwarz, W.~Steptoe \emph{et~al.}, ``The
  ethics of realism in virtual and augmented reality,'' \emph{Frontiers in
  Virtual Reality}, vol.~1, p.~1, 2020.

\bibitem{Sean2021}
\BIBentryALTinterwordspacing
W.~Sean, ``The single biggest question that'll determine the future of the \$30
  trillion metaverse,'' December 2021. [Online]. Available:
  \url{https://www.fool.com/investing/2021/12/14/question-determine-future-of-30-trillion-metaverse}
\BIBentrySTDinterwordspacing

\bibitem{Kyle2022}
\BIBentryALTinterwordspacing
W.~Kyle, ``The environmental impact of the metaverse,'' January 2022. [Online].
  Available:
  \url{https://venturebeat.com/2022/01/26/the-environmental-impact-of-the-metaverse/}
\BIBentrySTDinterwordspacing

\bibitem{lim2022realizing}
W.~Y.~B. Lim, Z.~Xiong, D.~Niyato, X.~Cao, C.~Miao, S.~Sun, and Q.~Yang,
  ``Realizing the metaverse with edge intelligence: A match made in heaven,''
  \emph{arXiv preprint arXiv:2201.01634}, 2022.

\end{thebibliography}
%

\begin{IEEEbiography}{}

\end{IEEEbiography}

\begin{IEEEbiographynophoto}{}

\end{IEEEbiographynophoto}








\end{document}